\documentclass{amsproc}

\usepackage{enumerate}
\usepackage{amsmath,amssymb}
\usepackage{amsthm,graphicx}
\usepackage{graphicx}

\newcommand{\AAA}{\mathcal{A}}
\newcommand{\BB}{\mathcal{B}}
\newcommand{\C}{\mathbb{C}}
\newcommand{\D}{{\mathrm d}}
\newcommand{\e}{{\mathrm e}}
\newcommand{\HH}{\mathcal{H}}
\newcommand{\JJ}{\mathcal{J}}
\newcommand{\KK}{\mathcal{K}}
\newcommand{\MM}{\mathcal{M}}
\newcommand{\N}{\mathbb{N}}
\newcommand{\OO}{\mathcal{O}}
\newcommand{\PP}{\mathcal{P}}

\newcommand{\R}{\mathbb{R}}
\newcommand{\RR}{\mathcal{R}}
\newcommand{\SSS}{\mathcal{S}}
\newcommand{\UU}{\mathcal{U}}
\newcommand{\Z}{\mathbb{Z}}
\newcommand{\eps}{\varepsilon}
\newcommand{\im}{{\rm Im\,}}
\newcommand{\re}{{\rm Re\,}}
\newcommand{\slim}{{\rm s\mbox{\rule[.5ex]{.6ex}{.1ex}}\hspace{.15ex}lim}}
\newcommand{\Slim}{{\rm s}\mbox{\rule[.5ex]{.6ex}{.1ex}}\hspace{-.1ex}\lim}
\newcommand{\restr}{\vert\hskip -5.5pt \phantom{\vert}^{\scriptscriptstyle \backslash}}
\newcommand{\supp}{{\rm supp\,}}

\newcommand{\Hab}{H_{\alpha,\beta}}

\newtheorem{theorem}{Theorem}[section]

\newtheorem{proposition}[theorem]{Proposition}

\theoremstyle{definition}

\theoremstyle{remark}

\numberwithin{equation}{section}

\begin{document}

\title{Solvable models of resonances and decays}

\author{Pavel Exner}
\address{Department of Theoretical Physics, Nuclear Physics
Institute, Czech Academy of Sciences, 25068 \v{R}e\v{z} near Prague, and
Doppler Institute for Mathematical Physics and Applied
Mathematics, Czech Technical University, B\v{r}ehov\'{a}~7, 11519
Prague, Czechia}
\email{exner@ujf.cas.cz}
\thanks{The project was partially supported by the Czech Science Foundation project P203/11/0701.}

\date{January 1, 1994 and, in revised form, June 22, 1994.}

\dedicatory{}


\begin{abstract}
Resonance and decay phenomena are ubiquitous in the quantum world. To understand them in their complexity it is useful to study solvable models in a wide sense, that is, systems which can be treated by analytical means. The present review offers a survey of such models starting the classical Friedrichs result and carrying further to recent developments in the theory of quantum graphs. Our attention concentrates on dynamical mechanism underlying resonance effects and at time evolution of the related unstable systems.
\end{abstract}

\maketitle

\tableofcontents

\section{Introduction}

Any general physical theory deals not only with objects as they are but also has to ask how they emerge and disappear in the time evolution and what one can learn from their temporary existence. In the quantum realm such processes are even more important than in classical physics. With few notable exceptions the elementary particles are unstable and also among nuclei, atoms and molecules unstable systems widely outnumber stable ones, even if the lack of permanence is a relative notion --- it is enough to recall that the observed lifetime scale of particles and nuclei ranges from femtoseconds to geological times.

It is natural that the quantum theory had to deal with such temporarily existent objects already in its nascent period, and it did it using simple means suggested by the intuition of the founding fathers. As time went, of course, a need appeared for a better understanding of these phenomena even if there was no substantial doubt about their mechanism; one can cite. e.g., a critical discussion of the textbook derivation of the `Fermi golden rule' \cite[lecture~23]{Fe} in \cite[notes to Sec.~XII.6]{RS}, or the necessarily non-exponential character of decay laws \cite[notes to Sec.~I.3]{Ex} which surprisingly keeps to puzzle some people to this day.

The quest for mathematically consistent description of resonance effects brought many results. It is worth to mention that some of them were rather practical. Maybe the best example of the last claim is the method of determining resonance poles using the so-called complex scaling. It has distinctively mathematical roots, in particular, in the papers \cite{AC71, BC71, Si79}, however, its applications in molecular physics were so successful that people in this area refer typically to secondary sources such as \cite{Mo98} instead giving credit to the original authors.

Description of resonances and unstable system dynamics is a rich subject with many aspects. To grasp them in their full complexity it is useful to develop a variety of tools among which an important place belongs to various solvable models of such systems. Those are the main topic of the present review paper which summarizes results obtained in this area over a long time period by various people including the author and his collaborators. As a \emph{caveat}, however, one has say also that the subject has so many aspects that a review like this one cannot cover all of them; our ambition is to give just a reasonably complete picture. We also remain for the moment vague about what the adjective `solvable' could mean in the present context; we will return to this question in Section~\ref{s:point} below.

\section{Preliminaries} \label{s: prelim}

Before starting the review it is useful to recall some notions we will need frequently in the following. Let us start with \emph{resonances}. While from the physics point of view we usually have in mind a single phenomenon when speaking of a resonance in a quantum system, mathematically it may refer to different concepts.

We will describe two most important definitions starting from that of a \emph{resolvent resonance}. A conservative quantum system is characterized by a family of observables represented by self-adjoint operators on an appropriate state Hilbert space. A prominent role among them is played by its Hamiltonian $H$, or operator of total energy. As a self-adjoint operator it has the spectrum which is a subset of the real line while the rest of the complex plane belongs to its resolvent set $\varrho(H)$ and the resolvent $z\mapsto (H-z)^{-1}$ is an analytic function on it having thus no singularities. It may happen, however, that it has an analytic continuation, typically across the cut given by the continuous spectrum of $H$ --- one usually speaks in this connection about another sheet of the `energy surface' --- and that this continuation is meromorhic having pole singularities which we identify with resonances.

An alternative concept is to associate resonances with scattering. Given a pair $(H,H_0)$ of self-adjoint operators regarded as the full and free Hamiltonian of the system we can construct scattering theory in the standard way \cite{AJS, RS}, in particular, we can check existence of the scattering operator and demonstrate that it can be written in the form of a direct integral, the corresponding fiber operators being called \emph{on-shell scattering matrices}. The latter can be extended to meromorhic function and resonances are identified in this case with their poles.

While the resonances defined in the two above described ways often coincide, especially in the situations when $H=-\Delta +V$ is a Schr\"odinger operator and $H_0=-\Delta$ its free counterpart, there is no \emph{a priori} reason why it should be always true; it is enough to realize that resolvent resonances characterize a single operator while the scattering ones are given by a pair of them. Establishing equivalence between the two notions is usually one of the first tasks when investigating resonances.

In order to explain how resonances are related to temporarily existing objects we have to recall basic facts about unstable quantum systems. To describe such a system we must not regard it as isolated, rather as a part of a larger system including
its decay products. We associated with the latter a state space $\HH$ on which unitary evolution operator $U:\: U(t) = \e^{-iHt}$ related to a self-adjoint Hamiltonian $H$ acts. The unstable system corresponds to a proper subspace $\HH_\mathrm{u} \subset \HH$ associated with a projection $E_\mathrm{u}$. To get a nontrivial model we assume that $\HH_\mathrm{u}$ is not invariant w.r.t. $U(t)$ for any $t>0$; in that case we have $\|E_\mathrm{u} U(t)\psi\| < \|\psi\|$ for $\psi\in \HH_\mathrm{u}$ and the state which is at the initial instant $t=0$ represented by the vector $\psi$ evolves into a superposition containing a component in $\HH_\mathrm{u}^\perp$ describing the decay products. Evolution of the unstable system alone is determined by the \emph{reduced propagator}
$$
V:\:V(t)= E_\mathrm{u}U(t)\,\restr\,\HH_\mathrm{u}\,,
$$
which is a contraction satisfying $V(t)^*=V(-t)$ for any $t\in\R$, strongly continuous with respect to the time variable. For a unit vector
$\psi\in\HH_\mathrm{u}$ the {\em decay law}
 \begin{equation} \label{decaylaw}
P_{\psi}:\: P_{\psi}(t)= \Vert V(t)\psi\Vert^2 = \Vert
E_\mathrm{u}U(t)\psi\Vert^2
 \end{equation}
is a continuous function such that $0\leq P_{\psi}(t)\leq
P_{\psi}(0)=1$ meaning the probability that the system undisturbed by measurement with be found undecayed at time $t$.

Under our assumptions the reduced evolution cannot be a group, however, it is not excluded that it has the semigroup property, $V(s)V(t)=V(s+t)$ for all $s,t\in\R$. As a example consider the situation where $\HH_\mathrm{u}$ is one-dimensional being spanned by a unit vector $\psi\in\HH$ and the reduced propagator is a multiplication by
$$
v(t):= (\psi,U(t)\psi) = \int_{\R} \e^{-i\lambda t}
\D(\psi,E^H_{\lambda}\psi)\,,
$$
where $E^H_{\lambda} = E_H(-\infty,\lambda]$ is the spectral projection of $H$. If $\psi$ and $H$ are such that the measure has Breit-Wigner shape, $\D(\psi,E^H_{\lambda}\psi) = \frac{\Gamma}{2\pi} \big[ (\lambda-\lambda_0)^2 +\frac14 \Gamma^2 \big]^{-1} \D\lambda$ for some $\lambda_0\in\R$ and $\Gamma>0$, we get $v(t)= \e^{-i\lambda_0t -\Gamma\vert t\vert/2}$ giving exponential decay law. Note that the indicated choice of the measure requires $\sigma(H)=\R$; this conclusion is not restricted to the one-dimensional case but it holds generally.

\begin{theorem} \cite{Si72} \label{sinha}
Under the stated assumptions the reduced propagator can have the semigroup property only if $\sigma(H)=\R$.
\end{theorem}

At a glance, this seems to be a problem since the exponential character of the decay laws conforms with experimental evidence in most cases, and at the same time Hamiltonians are usually supposed to be below bounded. However, such a spectral restriction excludes only the \emph{exact} validity of semigroup reduced evolution allowing it to be an approximation, possibly a rather good one. To understand better its nature, let us express the reduced evolution by means of the \emph{reduced resolvent}, $R_H^\mathrm{u}(z):= E_\mathrm{u}R_H(z)\,\restr\,\HH_\mathrm{u}$. Using the spectral decomposition of $H$ we can write the reduced propagator as Fourier image,
\begin{equation} \label{reduced}
V(t)\psi= \int_{\R} \e^{-i\lambda t} \D F_{\lambda}\psi
\end{equation}
for any $\psi\in\HH_\mathrm{u}$, of the operator-valued measure on $\R$ determined by the relation $F(-\infty,\lambda]:= E_\mathrm{u}E_{\lambda}^{(H)}\restr \HH_\mathrm{u}\,$. By Stone formula, we can express the measure as
$$
\frac{1}{2}\big\{F[\lambda_1,\lambda_2]+
F(\lambda_1,\lambda_2)\big\}\, =\,\frac{1}{2\pi i}\:
\Slim_{\hspace{-.8em} \eta\to 0+}\,
\int_{\lambda_1}^{\lambda_2}\,\big\lbrack R_H^\mathrm{u}(\xi\!+\!i\eta)
-R_H^\mathrm{u}(\xi\!-\!i\eta)\big\rbrack\, \D\xi\,;
$$
the formula simplifies if the spectrum is purely absolutely continuous and the left-hand side can be simply written as $F(\lambda_1,\lambda_2)$. The support of $F(\cdot)\psi$ is obviously contained in the spectrum of $H$ and the same is true for $\supp F = \bigcup_{\psi\in\HH_\mathrm{u}} \supp F(\cdot)\psi$, if fact, the latter coincides with $\sigma(H)\,$ \cite{Ex76}.

In view of that the reduced resolvent makes no sense at the points $\xi\in\supp F$ but the limits $\slim_{\eta\to 0+}\, R_H^\mathrm{u}(\xi\pm i\eta)$ may exist; if they are bounded on the interval $(\lambda_1,\lambda_2)$ we may interchange the limit with the integral. Furthermore, since the resolvent is analytic in $\rho(H) = \C\setminus \sigma(H)$ the same is true for $R_H^\mathrm{u}(\cdot)$. At the points of $\supp F$ it has a singularity but it may have an \emph{analytic continuation} across it; the situation is particularly interesting when this continuation has a meromorphic structure, i.e. isolated poles in the lower halfplane. For the sake of simplicity consider again the situation with $\dim\HH_\mathrm{u} =1$ when the reduced resolvent acts as a multiplication by $r_H^\mathrm{u}(z)$ and suppose its continuation has a single pole,
\begin{equation} \label{poleapprox}
r_H^\mathrm{u}(z)= \frac{A}{z_\mathrm{p}-z}\,+f(z)
\end{equation}
for $\im z >0$, where $A\ne 0$, $\:f$ is holomorphic,
and $z_\mathrm{p}:=\lambda_\mathrm{p}\!-i\delta_\mathrm{p}$ is a point in the
lower halfplane. Since $r_H^\mathrm{u}(\lambda \!-\!i\eta)=
\overline{r_H^\mathrm{u}(\lambda\!+\!i\eta)}$, the measure in question is
$$
\D F_\lambda = \frac{A}{2\pi i} \left(
\frac{1}{\lambda\!-\!\overline z_\mathrm{p}}\,-\,
\frac{1}{\lambda\!-\!z_\mathrm{p}}\, \right)\, \D\lambda\,
+\,\frac{1}{\pi}\, \im f(\lambda)\, \D\lambda\,,
$$
and evaluating the reduced propagator using the residue theorem
we get
\begin{equation} \label{redapprox}
v(t) = A\,\e^{-i\lambda_\mathrm{p}t-\delta_\mathrm{p}\vert t\vert}+\,
\frac{1}{\pi}\, \int_{\R}\,\e^{-i\lambda t}\, \im f(\lambda)\,
d\lambda\,,
\end{equation}
which is close to a semigroup, giving an approximately exponential decay law with $\Gamma=2\delta_\mathrm{p}$, if the second term is small and $A$ does not differ much from one. At the same time, the presence of the pole in the analytic continuation provides a link to the concept of (resolvent) resonance quoted above.

The main question in investigation of resonances and decays is to analyze how such singularities can arise from the dynamics of the systems involved. A discussion of this question in a variety of models will be our main topic in the following sections.

\section{A progenitor: Friedrichs model} \label{s:friedrichs}

We start with the \emph{mother of all resonance models} for which we are indebted to Kurt O. Friedrichs who formulated it in his seminal paper \cite{Fr48}. This is not to say it was recognized as seminal immediately, quite the contrary. Only after T.D.~Lee six years later came with a caricature model of decay in quantum field theory, it was slowly recognized that its essence was already analyzed by Friedrichs; references to an early work on the model can be found in \cite[notes to Sec.~3.2]{Ex}.

The model exists in numerous modifications; we describe here the simplest one. We suppose that the state Hilbert space of the system has the form $\HH:=\C\oplus L^2(\R^+)$ where the one-dimensional subspace is identified with the space $\HH_\mathrm{u}$ mentioned above; the states are thus described by the pairs $\alpha \choose f$ with $\alpha\in\C$ and $f\in L^2(\R^+)$. The Hamiltonian is the self-adjoint operator on $\HH$, or rather the family of self-adjoint operators labelled by the \emph{coupling constant} $g\in\R$, defined by
\begin{equation} \label{FriedrichsH}
H_g = H_0+gV\,, \quad H_g{\alpha \choose f} := \left(\begin{array}{cc} \lambda_0\alpha & g(v,f) \\ g\alpha v & Qf \end{array}\right)\,,
\end{equation}
where $\lambda_0$ is a positive parameter, $v\in L^2(\R^+)$ is sometimes called \emph{form factor}, and $Q$ is the operator of multiplication, $(Qf)(\xi)=\xi f(\xi)$. This in particular means that the continuous spectrum of $H_0$ covers the positive real axis and the eigenvalue $\lambda_0$ is embedded in it; one expects that the perturbation $gV$ can move the corresponding resolvent pole from the real axis to the complex plane.

To see that it is indeed the case we have to find the reduced resolvent. The model is solvable in view of the \emph{Friedrichs condition}, $E_\mathrm{d} VE_\mathrm{d}=0$ where $E_\mathrm{d}$ is the projection to $\HH_\mathrm{d}:= L^2(\R^+)$, which means that the continuum states do not interact mutually. Using the second resolvent formula and the commutativity of operators $E_\mathrm{u}$ and
$R_{H_0}(z)$ together with $E_\mathrm{u}+E_\mathrm{d}=I$ we can write $E_\mathrm{u}R_{H_g}(z)E_\mathrm{u}$ as
$$
E_\mathrm{u}R_{H_0}(z)E_\mathrm{u} - gE_\mathrm{u}R_{H_0}(z)E_\mathrm{u}VE_\mathrm{u}R_{H_g}(z)E_\mathrm{u}-
gE_\mathrm{u}R_{H_0}(z)E_\mathrm{u}VE_\mathrm{d}R_{H_g}(z)E_\mathrm{u}\,;
$$
in a similar way we can express the `off-diagonal' part of the resolvent as
$$
E_\mathrm{d}R_{H_g}(z)E_\mathrm{u} = -gE_\mathrm{d}R_{H_0}(z)E_\mathrm{d}VE_\mathrm{u}R_{H_g}(z)E_\mathrm{u}\,,
$$
where we have also employed the Friedrichs condition. Substituting from the last relation to the previous one and using $(H_0\!-z)E_\mathrm{u}R_{H_0}(z) =E_\mathrm{u}\,$ together with the explicit form of the operators $H_0,\,V\,$, we find that $R^\mathrm{u}_{H_g}(z)$ acts for $\im z \neq 0$ on $\HH_\mathrm{u}=\C$ as multiplication by the function
 \begin{equation} \label{Fresolv}
r_g^\mathrm{u}:\: r_g^\mathrm{u}(z) := \left(-z+\lambda_0+g^2\int_0^{\infty}\,
\frac{\vert v(\lambda)\vert^2}{z-\lambda}\,\D\lambda
\right)^{-1}\,.
 \end{equation}
To make use of this result we need an assumption about the form factor, for instance

\medskip

\noindent (a) there is an entire $f:\,\C\to\C$ such
that $\vert v(\lambda)\vert^2= f(\lambda)$ holds for all
$\lambda\in (0,\infty)\,$;

\medskip

\noindent for the sake of notational simplicity one usually writes $f(z)=\vert v(z)\vert^2$ for nonreal $z$ too keeping in mind that it is a complex quantity. This allows us to construct analytic continuation of $r_g^\mathrm{u}(\cdot)$ over $\sigma_\mathrm{c}(H_g) = \R^+$ to the lower complex halfplane in the form $r(z)= [-z+w(z,g)]^{-1}$, where
\begin{eqnarray}
w(\lambda,g) := \lambda_0+ g^2I(\lambda)-\pi ig^2\vert
v(\lambda)\vert^2 \; &\dots& \;\; \lambda>0 \nonumber \\ [-.5em]
\label{Fcontin} \\ [-.5em]
w(z,g) :=\ \lambda_0+ g^2 \int_0^{\infty}\, \frac{\vert
v(\xi)\vert^2}{z-\xi}\,\D\xi - 2\pi ig^2\vert v(z)\vert^2
\;  &\dots& \; \im z < 0 \nonumber
\end{eqnarray}
and $I(\lambda)$ is defined as the principal value of the
integral,
$$
I(\lambda) := \PP\int_0^{\infty}\, \frac{\vert v(\xi)\vert^2}
{\lambda-\xi}\, \D\xi \,:=\,\lim_{\eps\to 0+}\, \left(\,
\int_0^{\lambda-\eps}\, +\,\int_{\lambda+\eps}^{\infty}\,\right)\,
\frac{\vert v(\xi)\vert^2}{\lambda-\xi}\, \D\xi\,;
$$
the analyticity can be checked using the edge-of-the-wedge theorem .

These properties of the reduced resolvent make it possible to
prove the meromorphic structure of its analytic continuation. Resonances in the model are then given by zeros of the function $z\mapsto w(z,g)-z$. An argument using the implicit-function theorem \cite[sec.~3.2]{Ex} leads to the following conclusion:

\begin{theorem}
Assume (a) and $v(\lambda_0)\neq 0$, then $r(\cdot)$ has for all sufficiently small $|g|$ exactly one simple pole $z_\mathrm{p}(g):=\lambda_\mathrm{p}(g)-i\delta_\mathrm{p}(g)$. The function
$z_\mathrm{p}(\cdot)$ is infinitely differentiable and the expansions
\begin{equation} \label{Friedrichs pole}
\lambda_P(g) = \lambda_0+ g^2I(\lambda_0)+ \OO(g^4)\,, \quad
\delta_\mathrm{p}(g) = \pi g^2\vert v(\lambda_0)\vert^2+ \OO(g^4)\,,
\end{equation}
are valid in the vicinity of the point $g=0$ referring to the unperturbed Hamiltonian.
\end{theorem}

To summarize the above reasoning we have seen that resonance poles can arise from perturbation of eigenvalues embedded in the continuous spectrum and that, at least locally, their distance from the real axis is the smaller the weaker is the perturbation. Moreover, one observes here the phenomenon called \emph{spectral concentration}: it is not difficult to check that the spectral projections of $H_g$ to the intervals $I_g:= (\lambda_0 -\beta g, \lambda_0 -\beta g)$ with a fixed $\beta>0$ satisfy the relation
$$
\Slim_{g\to 0} E_{H_g}(I_g) = E_\mathrm{u} \,.
$$
Friedrichs model also allows us to illustrate other typical features of resonant systems. We have mentioned already the deep insight contained in the \emph{Fermi golden rule}, which in the present context can be written as
$$
\Gamma_\mathrm{F}(g)= 2\pi g^2 \left. \frac{d}{d\lambda}\,
\left(V\psi_\mathrm{u}, E_{\lambda}^{(0)} P_c(H_0)V\psi_\mathrm{u}\right)\,
\right\vert_{\lambda=\lambda_0} \,,
$$
where $\{E_{\lambda}^{(0)}\}$ is the spectral decomposition
of $H_0$ and $P_\mathrm{c}(H_0)$ the projection to the continuous spectral subspace of this operator. To realize that this is indeed what we known from quantum-mechanical textbooks, it is enough to realize that we use the convention $\hbar=1$ and formally it holds $\frac{\D}{\D\lambda} E_{\lambda}^{(0)} P_\mathrm{c}(H_0)= \vert\lambda\rangle\langle\lambda\vert$. Using the explicit form of the operators involved we find
$$
\Gamma_\mathrm{F}(g)= 2\pi g^2\, \frac{d}{d\lambda}\,
\int_0^{\lambda} \vert v(\xi)\vert^2\,d\xi\, \Big\vert
_{\lambda=\lambda_0} = 2\pi g^2\,\vert v(\lambda_0)\vert^2\,,
$$
which is nothing else than the first nonzero term
in the Taylor expansion (\ref{Friedrichs pole}). On the other hand, a formal use of the rule may turn its gold into brass: a warning example concerning the situation when the unperturbed eigenvalue is situated at the threshold of $\sigma_\mathrm{c} (H_0)$ is due to J.~Howland \cite{Ho74}, see also \cite[Example~3.2.5]{Ex}. Recent analysis of near-threshold effects in a generalized Friedrichs model together with a rich bibliography can be found in \cite{DJN11}.

Resonances discussed so far have been resolvent resonances. One can also consider the pair $(H_g,H_0)$ as a scattering system. Existence and completeness of the wave operators is easy to establish since the perturbation $gV$ has rank two. What we are interested in is the on-shell S-matrix: if $v$ is piecewise continuous and bounded in $\R^+$ one can check \cite[Prop.~3.2.6]{Ex} that it acts as multiplication by
$$
S(\lambda) = 1 + 2\pi i g^2 \lim_{\epsilon\to 0+} |v(\lambda)|^2\, r_\mathrm{u}(\lambda +i\epsilon)\,.
$$
If $v$ satisfies in addition the assumption (a) above, the function $S(\cdot)$ can be analytically continued across $\R^+$. It is obvious that if such a continuation has a pole at a point $z_\mathrm{p}$ of the lower complex halfplane, the same is true for $r(\cdot)$, on the other hand, it may happen that a resolvent resonance is not a scattering resonance, namely if the continuation of $|v(\cdot)|^2$ has a zero at the point $z_\mathrm{p}$.

Finally, the model can also describe a decaying system if we suppose that at the initial instant $t=0$ the state is described by the vector ${1 \choose 0}$ which span the one-dimensional subspace $\HH_\mathrm{u}$. The main question here is whether one can approximate the reduced evolution by a semigroup in the sense of (\ref{redapprox}); a natural guess is that it should be possible in case of a weak coupling. Since the reduced resolvent is of the form (\ref{poleapprox}) we can express the corresponding measure, calculate the reduced propagator according to (\ref{reduced}) and estimate the influence of the second term in (\ref{redapprox}). This leads to the following conclusion, essentially due to \cite{De76}:

\begin{theorem} \label{thm: poleapprox}
Under the stated assumptions there are positive $C,\,g_0$ such that
$$
\left| v(t) - A\,\e^{-iz_\mathrm{p}t} \right| < \frac{C g^2}{t}
$$
holds for all $t>0$ and $|g|<g_0$ with $A:= [1-g^2 I'(z_\mathrm{p})]^{-1}$, where $I(z)$ is the integral appearing in the second one of the formul{\ae} (\ref{Fcontin}).
\end{theorem}

\medskip

The simple Friedrichs model described here has many extensions and in no way we intend to review and discuss them here limiting ourselves to a few brief remarks:

\begin{enumerate}[(a)]
\setlength{\itemsep}{2pt}

\item Some generalization of the model cast it into a more abstract setting, cf. for example \cite{Mo96, DR07, DJN11}. Others are more `realistic' regarding it as a description of a system interacting with a field, either a caricature one-mode one \cite{DE87-89} or considerably closer to physical reality \cite{BFS98, HHH08} in a sense returning the model to its Lee version which stimulated interest to it.

\item Friedrichs model clones typically use the simple procedure --- attributed to Schur or Feshbach, and sometimes also to other people --- we employed to get relation (\ref{Fresolv}) expressing projection of the resolvent to the subspace $\HH_\mathrm{u}$; sometimes it is combined with a complex scaling.

\item While most Friedrichs-type models concern perturbations of embedded eigenvalues some go further. As an example, let us mention a caricature model of a crystal interacting with a field \cite{DEH04} in which the unperturbed Hamiltonian has a spectral band embedded in the continuous spectrum halfline referring to states of a lower band plus a field quantum. The perturbation turns the embedded band into a curve-shaped singularity in the lower complex halfplane with endpoints at the real axis. One can investigate in this framework decay of `valence-band' states analogous to Theorem~\ref{thm: poleapprox}, etc.

\item The weak-coupling behavior described in Theorem~\ref{thm: poleapprox} can be viewed also from a different point of view, namely that the decay law converges to a fixed exponential function as $g\to 0$ when we pass to the rescaled time $t'=g^{-2}t$. This is usually referred to as \emph{van Hove limit} in recognition of the paper \cite{vH55}; the first rigorous treatment of the limit belongs to E.B.~Davies, cf.~ \cite{Da}.

\end{enumerate}

\section{Resonances from perturbed symmetry}

The previous section illustrates the most common mechanism of resonance emergence, namely perturbations of eigenvalues embedded in the continuum. A typical source of embedded eigenvalues is a symmetry of the system which prevents transitions from the corresponding localized state into a continuum one. Once such a symmetry is violated, resonances usually occur. Let us demonstrate that in a model describing a Schr\"odinger particle in a straight waveguide, perturbed by a potential or by a magnetic field, the idea of which belongs to J.~N\"ockel \cite{No92}.

  \subsection{N\"ockel model} \label{ss:noeckel}

We consider two-dimensional `electrons' moving in a channel with a potential well. The guide is supposed to be either a hard-wall strip $\Omega:= \R\times S$ with $S=(-a,a)$, or alternatively the transverse confinement can be modelled by a potential in which case we have $S=\R$. The full Hamiltonian acting on $\HH:=L^2(\Omega)$ is given by
 \begin{equation} \label{NockelH}
H(B,\lambda):= \left( -i\partial_x -By\right)^2 +V(x) -\partial_y^2 +W(y) +\lambda U(x,y)\,;
 \end{equation}
if $\Omega$ is a strip of width $2a$ the transverse potential $W$ may be absent and we impose Dirichlet conditions at the boundary, $|y|=a$. The real-valued functions $V$ describing the well in the waveguide --- or a caricature quantum dot if you wish --- and $W$ are measurable, and the same is true for the potential perturbation $U$; further hypotheses will be given below. The number $B$ is the intensity of the homogeneous magnetic field perpendicular to $\Omega$ to which the system is exposed.

The unperturbed Hamiltonian $H(0):=H(0,0)$ can be written in the form $h^V\otimes I +I\otimes h^W$ which means its spectrum is the `sum' of the corresponding component spectra. If the spectrum of the transverse part $h^W$ is discrete the embedded eigenvalues can naturally occur; we are going to see what happens with them under influence of the potential perturbation $\lambda U$ and/or the magnetic field. Let us first list the assumptions using the common notation $\langle x\rangle:= \sqrt{1\!+\!x^2}$.

\begin{enumerate}[(a)]
\setlength{\itemsep}{2pt}

\item $\,\lim_{|x|\to\infty} W(x)=+\infty$ holds if $S=\R\,$,

\item $\,V\ne 0$ and $\,|V(x)|\le \mathrm{const\,}
\langle x\rangle^{-2-\eps}$ for some $\eps>0$, with $\int_{\R} V(x)\, \D x \le 0\,$,

\item the potential $V$ extends to a function analytic
in $\MM_{\alpha_0}:= \{ z\in\C:\: |\arg
z|\le\alpha_0\}$ for some $\alpha_0>0$ and obeys there the
bound of assumption (b),

\item $\,|U(x,y)|\le \mathrm{const\,}\langle x\rangle^{-2-\eps}$ holds  for some $\eps>0$ and all $(x,y)\in \Omega$. In addition, it does not factorize, $U(x,y)\ne U_1(x)\!+\! U_2(y)$, and $U(\cdot,y)$ extends for each fixed $y\in S$ to an analytic function in $\MM_{\alpha_0}$ satisfying there the same bound.

\end{enumerate}

The assumption (a) ensures that the spectrum of $h^{W}:= -\partial^2_y+W(y)$, denoted as $\{\nu_j\}_{j=1} ^\infty$, is discrete and simple, $\nu_{j+1}> \nu_j$. The same is true if $S=(-a,a)$ when we impose Dirichlet condition at $y=\pm a$, naturally except the case when $W$ grows fast enough as $y\to\pm a$ to make the operator essentially self-adjoint. The assumption (b) says, in particular, that the local perturbation responsible for the occurrence of localized states is short-ranged and non-repulsive in the mean; it is well known that in this situation the longitudinal part $h^V:= -\partial^2_x+V(x)$ of the unperturbed Hamiltonian has a
nonempty discrete spectrum,
 $$ 
\mu_1< \mu_2< \cdots \mu_N <0\,,
 $$ 
which is simple and finite \cite{Si76, BGS77}; the
corresponding normalized eigenfunctions $\phi_n,\:
n=1,\dots,N\,$, are exponentially decaying.

To be able to treat the resonances we need to adopt  analyticity hypotheses stated in assumptions (c) and (d). Note that in addition to the matter of our interest the system can also have `intrinsic' resonances associated with the operator $h^V$, however, the corresponding poles do not approach the real axis as the perturbation is switched off. In addition, they do not accumulate except possibly at the threshold \cite{AC71}, and if $V$ decays exponentially even that is excluded \cite[Lemma~3.4]{Je78}.

Since $\sigma_\mathrm{c}(h^V)=[0,\infty)$, the spectrum of the unperturbed Hamiltonian consists of the continuous part, $\sigma_\mathrm{c}(H(0))= \sigma_\mathrm{ess}(H(0)) =[\nu_1,\infty)$, and the infinite family of eigenvalues
 $$ 
\sigma_\mathrm{p}(H(0))\,=\,\left\lbrace\, \mu_n\!+\!\nu_j :\;
n=1,\dots,N\,,\; j=1,2,\dots\,\right\rbrace\,.
 $$ 
A finite number of them are isolated, while the remaining ones satisfying the condition $\mu_n+\nu_j> \nu_1$ are embedded in the continuum; let us suppose for simplicity that they  coincide with none of the thresholds, $\mu_n+\nu_j\ne \nu_k$ for any $k$.

To analyze the resonance problem it is useful to employ the transverse-mode decomposition, in other words, to replace the original PDE problem by a matrix ODE one. Using the transverse eigenfunctions, $h^W\chi_j=\nu_j\chi_j\,$, we introduce the embeddings $\JJ_j$ and their adjoints acting as projections by
\begin{eqnarray*}
\JJ_j &\!:\!& L^2(\R) \to L^2(\Omega)\,, \qquad \JJ_j f =f\otimes
\chi_j\,, \\ \JJ_j^* &\!:\!& L^2(\Omega) \to L^2(\R)\,, \qquad
(\JJ_j^*g)(x) = \left(\chi_j, g(x,\cdot)\right)_{L^2(S)}\,;
\end{eqnarray*}
then we replace $H(B,\lambda)$ by the matrix differential operator $\{H_{jk}(B,\lambda)\} _{j,k=1}^{\infty}$ with
\begin{eqnarray*}
H_{jk}(B,\lambda) &\!:=\!& \JJ^*_j H(B,\lambda) \JJ_k \,=\,
\left( -\partial_x^2 +V(x) +\nu_j \right) \delta_{jk}
+\UU_{jk}(B,\lambda) \,, \label{matrix H} \\
\nonumber \\
\UU_{jk}(B,\lambda) &\!:=\!& 2iB\, m_{jk}^{(1)} \partial_x
+B^2 m_{jk}^{(2)} + \lambda U_{jk}(x) \,, \label{matrix U}
\end{eqnarray*}
where $m_{jk}^{(r)}:= \int_S y^r \overline\chi_j(y) \chi_k(y) \,\D y$ and $U_{jk}(x):=\int_S U(x,y) \overline\chi_j(y)
\chi_k(y) \,\D y$.

  \subsection{Resonances by complex scaling}

N\"ockel model gives us an opportunity to illustrate how the complex scaling method mentioned in the introduction can be used in a concrete situation. We apply here the scaling transformation to the longitudinal variable starting from the unitary operator
 $$ 
\SSS_{\theta}:\; (\SSS_{\theta}\psi)(x,y)= \e^{\theta/2}
\psi(\e^{\theta}x,y)\,, \quad \theta\in\R\,,
 $$ 
and extending this map analytically to $\MM_{\alpha_0}$ which is possible since the transformed Hamiltonians are of the form
$H_{\theta}(B,\lambda):= \SSS_{\theta}H(B,\lambda)
{\SSS_{\theta}}^{-1} =H_{\theta}(0)+ \UU_{\theta}(B,\lambda)$
with
\begin{equation} \label{scaled free}
H_{\theta}(0)\,:=\, -\e^{-2\theta} \partial_x^2 -\partial_y^2
+V_{\theta}(x) +W(y)\,,
\end{equation}
where $V_{\theta}(x):= V(\e^{\theta x})$ and the interaction part
 $$ 
\UU_{\theta}(B,\lambda) := 2i\, \e^{-\theta}By\, \partial_x
+B^2 y^2 + \lambda U_{\theta}(x,y)
 $$ 
with $U_{\theta}(x,y):= U(\e^{\theta x},y)$. Thus in view of the assumptions (c) and (d) they constitute a type (A) analytic family of $m$-sectorial operators in the sense of \cite{Ka} for $|\im\theta |<\min\{\alpha_0,\pi/4\}\,$. Denoting $R_{\theta}(z):= (H_{\theta}(0)-z)^{-1}$ one can check \cite{DEM01} that
 \begin{equation}  \label{eq:type A estimate}
        \|\UU_{\theta}(B,\lambda) R_{\theta}(\nu_{1}+\mu_{1}-1)\|\leq
        c(|B|+|B|^2+|\lambda|)
 \end{equation}
holds for $|\im\theta|<\min\{\alpha_0,\pi/4\}$, and consequently, the operators $H_{\theta}(B,\lambda)$ also form a type (A) analytic family for $B$ and $\lambda$ small enough. The free part (\ref{scaled free}) of the transformed operator separates variables, hence its spectrum is
\begin{equation} \label{transf spec1}
\sigma\left( H_{\theta}(0)\right) = \bigcup_{j=1}^{\infty}
\left\lbrace\, \nu_j+ \sigma\left( h^V_{\theta} \right)
\right\rbrace\,,
\end{equation}
where $h^V_{\theta}:= -\e^{-2\theta}\partial_x^2 +V_{\theta}(x)$. Since the potential is dilation-analytic by assumption, we have a typical picture: the essential spectrum is rotated into the lower halfplane revealing (fully or partly) the discrete spectrum of the non-selfadjoint operator $h^V_{\theta}$ which is independent of $\theta$; we have
\begin{equation} \label{transf spec2}
\sigma\left( h^V_{\theta}\right) = \e^{-2\theta} \R^+ \,\cup\,
\{\mu_1, \dots,\mu_N\} \,\cup\, \{\rho_1,\rho_2\,\dots\}\,,
\end{equation}
where $\rho_r$ are the `intrinsic' resonances of $h^V$.
In view of the assumptions (c) and $\mu_n+\nu_j\ne \nu_k$ for no $k$ the supremum of $\im \rho_k$ over any finite region of the complex plane which does not contain any of the points $\nu_k$ is negative, hence each eigenvalue $\mu_n+\nu_j$ has a neighbourhood containing none of the points $\rho_k+\nu_{j'}$. Consequently, the eigenvalues of $H_{\theta}(0)$ become isolated once $\im\theta>0$. Using the relative
boundedness (\ref{eq:type A estimate}) we can draw a contour
around an unperturbed eigenvalue and apply perturbation theory; for simplicity we shall consider only the non-degenerate
case when $\mu_n+\nu_j\ne \mu_{n'}+\nu_{j'}$ for different
pairs of indices.

We fix an unperturbed eigenvalue $e_0=\mu_n+\nu_j$ and choose $\theta=i\beta$  with a  $\beta>0$; then in view of (\ref{transf spec1}) and (\ref{transf spec2}) we may chose a contour $\Gamma$ in the resolvent set of $H_{\theta}(0)$ which
encircles just the eigenvalue $e_{0}$. We use the symbol  $P_{\theta}$ for the eigenprojection of $H_{\theta}(0)$ referring to $e_{0}$ and set
 $$
S^{(p)}_{\theta}:=\frac{1}{2\pi i}\int_{\Gamma}
\frac{R_{\theta}(z)}{(e_{0}-z)^{p}}\, \D z
 $$
for $p=0,1,\dots\,$, in particular, $S^{(0)}_{\theta} = -P_{\theta}$ and $\hat R_{\theta}(z):= S^{(1)}_{\theta}$ is the reduced resolvent of $H_{\theta}(0)$ at the point $z$. The bound (\ref{eq:type A estimate}) implies easily
 $$
\big\|\UU_\theta(B,\lambda) S^{(p)}_{\theta}\big\|\leq
    c\, \frac{|\Gamma|}{2\pi}\, \big(\mathrm{dist} (\Gamma,e_{0})\big)^{-p}
    (|B|+|B|^2+|\lambda|)
 $$
with some constant $c$ for all $\im\theta \in (0,\alpha_{0})$ and $p\ge 0$. It allows us to write the perturbation expansion. Since $\, e_{0}= \mu_{n}+\nu_{j}$ holds by assumption for a unique pair of the indices, we obtain using \cite[Sec.~II.2]{Ka} the following convergent series
\begin{equation} \label{perturbed ev}
e(B,\lambda) = \mu_n+\nu_j+ \sum_{m=1}^{\infty} e_m(B,\lambda)
\,,
\end{equation}
where $e_m(B,\lambda) = \sum_{p_1+\cdots+p_m=m-1} {(-1)^m \over m}\: \mathrm{Tr}\, \prod_{i=1}^m\, \UU_{\theta}(B,\lambda) S^{(p_i)}_{\theta}$. Using the above estimate we can estimate the order of each term with respect to the parameters. We find $e_m(B,\lambda) = \sum_{l=0}^m \OO\left(B^l \lambda^{m-l}\right)$, in particular, we have $e_m(B)= \OO(B^m),$ and $e_m(\lambda)= \OO(\lambda^m)$ for pure magnetic and pure potential perturbations, respectively.

The lowest-order terms in the expansion (\ref{perturbed ev}) can be computed explicitly. In the non-degenerate case, $\dim
P_{\theta}=1$, we have $e_1^{j,n}(B,\lambda) = \mathrm{Tr} \left(\UU_{\theta}(B,\lambda) P_{\theta} \right)$. After a short calculation we can rewrite the expression at the right-hand side in the form $2i B m_{jj}^{(1)} \left(\phi_n, \phi'_n \right) + B^2 m_{jj}^{(2)} +\, \lambda \left(\phi_n, U_{jj} \phi_n \right)$. Moreover, $i\left(\phi_n, \phi'_n\right)= \left(\phi_n, i\partial_x \phi_n \right)$ is (up to a sign) the group velocity of the wavepacket, which is zero in a stationary state; recall that eigenfunction $\phi_n$ of $h^V$ is real-valued up to a phase factor. In
other words,
 \begin{equation} \label{Nfirst}
e_1^{j,n}(B,\lambda) = B^2 \int_S y^2\left| \chi_j(y)\right|^2\, \D y\,+\, \lambda \int_{\R\times S} U(x,y) \left|\phi_n(x) \chi_j(y)\right|^2\, \D x\, \D y
 \end{equation}
with the magnetic part independent of $n$. As usual in such
situations the first-order correction is real-valued and thus does not contribute to the resonance width.

The second term in the expansion (\ref{perturbed ev}) can be computed in the standard way \cite[Sec.XII.6]{RS}; taking the limit $\im\theta \to 0$ in the obtained expression we get
 \begin{equation} \label{Nsecond}
e_2^{j,n}(B,\lambda) = -\sum_{k=1}^{\infty}
\left(\UU_{jk}(B,\lambda) \phi_n, \left(\left( h^V\! -\!e_{0}
+\nu_k\! -\!i0 \right)^{-1} \right) \!\!\hat{\phantom{|}}\,
\UU_{jk}(B,\lambda) \phi_n \right)\,.
 \end{equation}
We shall calculate the imaginary part which determines the resonance width in the leading order. First we note that it can be in fact expressed as a finite sum. Indeed, $k_{e_{0}}:= \max\{ k:\,e_{0}-\nu_{k}>0\}$ is finite and nonzero if the eigenvalue $e_0$ is embedded, otherwise we set it equal to zero. It is obvious that $\RR_{k}:= \big(\left( h^V\! -\!e_{0} +\nu_k\! -\! i0 \right)^{-1} \! \big) \!\hat{\phantom{|}}$ is Hermitean for $k>k_{e_{0}}$, hence the corresponding terms in (\ref{Nsecond}) are real and
 $$ 
    \im e_{2}^{j,n}(B,\lambda) =
    \sum_{k=1}^{k_{e_{0}}} \left(\UU_{jk}(B,\lambda) \phi_n,
    (\im\RR_{k}) \UU_{jk}(B,\lambda) \phi_n \right)\,.
 $$ 
The operators $\im\RR_{k}$ can be expressed by a straightforward computation \cite{DEM01}. To write the result we need $\omega(z):= \big[ I+|V|^{1/2}(-\partial^{2}_{x} -z)^{-1} |V|^{1/2} \mathrm{sgn\,}(V) \big]^{-1}$, in other words, the inverse to the operator acting as
 $$
\big( \omega^{-1}(z) f\big)(x)= f(x)
+\frac{i|V(x)|^{1/2}}{2\sqrt{z}} \int_{\R} \e^{i\sqrt{z}
|x-x'|}|V(x')|^{1/2}\mathrm{sgn\,}V(x')\,f(x')\, \D x'\,.
 $$
We also need the trace operator $\tau_E^\sigma:\,\HH^1\to\C$  which acts on the first Sobolev space $W^{1,2}$ as $\tau^{\sigma}_{E} \phi := \hat\phi(\sigma \sqrt{E})$ for $\sigma=\pm$ and $E>0$ where $\hat\phi$ is the Fourier transform of $\phi$. Armed with these notions we can write the imaginary part of the resonance pole position up to higher-order terms as
\begin{eqnarray} \label{eq:long Im e2}
    \lefteqn{\phantom{\quad} \im e_{2}^{j,n}(B,\lambda) = \sum_{k=1}^{k_{e_{0}}}\sum_{\sigma=\pm}^{}
    \frac{\pi}{2\sqrt{e_{0}\!-\!\nu_{k}}}\big| \tau^{\sigma}_{e_{0}\!-\!\nu_{k}}
    \omega(e_{0}\!-\!\nu_{k}\!+\! i0)\,
    \UU_{jk}(B,\lambda) \phi_{n} \big|^{2}}
    \\[.1em] &&
    = \sum_{k=1}^{k_{e_{0}}}\sum_{\sigma=\pm}^{}
    \frac{\pi}{\sqrt{e_{0}\!-\!\nu_{k}}}
    \Big\{ -2 B^{2}\, |m^{(1)}_{jk}|^{2}\left|\tau^{\sigma}_{e_{0}\!-\!\nu_{k}}
    \omega(e_{0}\!-\!\nu_{k}\!+\! i0) \phi'_{n} \right|^{2}
    \nonumber \\[.3em] &&
    \quad +2\lambda B\,  m^{(1)}_{jk}\, \im
    \Big(\tau^{\sigma}_{e_{0}\!-\!\nu_{k}}
    \omega(e_{0}\!-\!\nu_{k}\!+\! i0) \phi'_{n}\,,
    \tau^{\sigma}_{e_{0}\!-\!\nu_{k}}
    \omega(e_{0}\!-\!\nu_{k}\!+\! i0) U_{jk}\phi_{n}\Big)
    \nonumber \\[.3em] &&
    \quad -\frac{\lambda^{2}}{2} \left|\tau^{\sigma}_{e_{0}\!-\!\nu_{k}}
    \omega(e_{0}\!-\!\nu_{k}\!+\! i0)\,U_{jk}\phi_{n}\right|^{2}
    \Big\} + \OO(B^{3})+ \OO(B^{2}\lambda)\,,
    \nonumber
\end{eqnarray}
where as usual $f(E+i0)= \lim_{\eps\to 0+} f(E+i\eps)$.  Let us summarize the results:

 \begin{theorem} \label{Lthm}
Assume (a)--(d) and suppose that an unperturbed
eigenvalue $e_{0}= \mu_{n}+\nu_{j} > \nu_1$ is simple and coincides with no threshold $\nu_k$. For small enough $B$ and $\lambda$ the N\"ockel model Hamiltonian (\ref{NockelH}) has a simple resonance pole the position of which is given by the relations (\ref{perturbed ev})--(\ref{Nsecond}). The leading order in the expansion obtained by neglecting the error terms in (\ref{eq:long Im e2}) is the Fermi golden rule in this case.
 \end{theorem}

The symmetry in this example is somewhat hidden; it consists of the factorized form of the unperturbed Hamiltonian $H(0)$ which makes it reducible by projections to subspaces associated with the transverse modes. It is obvious that both the potential perturbation --- recall that we assumed $U(x,y)\ne U_1(x)\!+\! U_2(y)$ --- and the magnetic field destroy this symmetry turning thus embedded eigenvalues coming from higher transverse modes into resonances. At the same time, the described decomposition may include other, more obvious symmetries. For instance, if the potential $W$ is even with respect to the strip axis --- including the case when $S=(-a,a)$ and $W=0$ --- the unperturbed Hamiltonian commutes with the transverse parity operator, $\psi(x,y) \mapsto \psi(x,-y)$, and the transversally odd states are orthogonal to the even ones so embedded eigenvalues arise.

N\"ockel model is by far not the only example of this type. We limit ourselves here to quoting one more. Consider an \emph{acoustic waveguide} in the form of a planar strip of width $2a$ into which we place an axially symmetric obstacle; the corresponding Hamiltonian acts as Laplacian with Neumann condition at the boundary, both of the strip and the obstacle. Due to the axial symmetry the odd part of the operator gives rise to at least one eigenvalue in the interval $\big(0, \frac14(\frac{\pi}{a})^2 \big)$ which is embedded into the continuous spectrum covering the whole positive real axis \cite{ELV94}. If the obstacle is shifted by $\eps$ in the direction perpendicular to the axis, such an eigenvalue turns again into a resonance for the position of which one can derive an expansion in powers of $\eps$ analogous to Theorem~\ref{Lthm}, cf.~\cite{APV00}.

\section{Point contacts} \label{s:point}

The resonance models discussed in the previous two sections show that we should be more precise speaking about \emph{solvable} models. The question naturally is what we have finally to \emph{solve} when trying to get conclusions such as formul{\ae} for resonance pole positions. In both cases we have been able to derive weak-coupling expansions with explicit leading terms which could be regarded as confirmation of the Fermi golden rule for the particular model. One have to look, however, into which sort of problem the search for resonances was turned. For the Friedrichs model it was the functional equation $w(z,g)=z$ with the left-hand side given by (\ref{Fcontin}), and a similar claim is true for its clones, while in the N\"ockel model case we had to perform spectral analysis of the non-selfadjoint operator\footnote{The same is true also for most `realistic' descriptions of resonances using complex scaling, in particular, in the area of atomic and molecular physics --- see, e.g., \cite[Sec.~XII.6]{RS} or \cite{Mo98}.} $H_\theta(B,\lambda)$.

Not only the latter has been more difficult in the above discussion, the difference becomes even more apparent if we try to go beyond the weak-coupling approximation. Following the pole trajectory over a large interval of coupling parameters may not be easy even if its position is determined by a functional equation and one have to resort usually to numerical methods, however, it is still much easier than to analyze a modification of the original spectral problem. Recall that for the Friedrichs model pole trajectories were investigated already in \cite{Ho58} where it had been shown, in particular, that for strong enough coupling the pole may return to the (negative part of the) real axis becoming again a bound state.

In the rest of this review we will deal with models which are `solvable' at least in the sense of the Friedrichs model, that is, their resonances are found as roots of a functional --- sometimes even algebraic --- equation. In this section we will give examples showing that this is often the case in situations where the interaction responsible for occurrence of the resonances is of point or contact type.

  \subsection{A simple two-channel model} \label{ss: 2channel}

The first model to consider here will describe a system the state space of which has two subspaces corresponding to two internal states; the coupling between them is of a \emph{contact nature}. To be specific, one can think of a system consisting of a neutron and a nucleus having just two states, the ground state and an excited one. Their relative motion can be described in the Hilbert space $L^2(\R^3)\oplus L^2(\R^3)$; we suppose that the reduced masses is the two channels are the same and equal to $\frac12$ so that the Hamiltonian acts on functions supported away from the origin of the coordinates as $-\Delta$ and $-\Delta+E$, respectively, where $E>0$ is the energy difference between the ground and the excited states.

Before proceeding further, let us note that the above physical interpretation of the model coming from \cite{Ex91} is not the only possible. The two channels can be alternatively associated, for instance, with two spin states; this version of the model was worked out in \cite{CCF09}, also in dimensions one and two.

To construct the Hamiltonian we start from the direct sum $A_0=A_{0,1}\oplus A_{0,2}$ where the component operators act as $A_{0,1}:=-\Delta$ and $A_{0,2}:=-\Delta+E$, respectively, being defined on $W^{2,2}(\R^3\setminus \{0\})$. It is not difficult to check that $A_0$ is a symmetric operator with deficiency indices $(2,2)$; we will choose the model Hamiltonian among its self-adjoint extensions. The analysis can be simplified using the rotational symmetry, since the components of $A_0$ referring to nonzero values of the angular momentum are essentially self-adjoint, and therefore a nontrivial coupling is possible in the s-wave only. As usual we pass to reduced radial wave functions $f:\, f(r):=r\psi(r)\,$; we take $\HH=\HH_1 \oplus \HH_2$ with $\HH_j:=L^2(\R^+)$ as the state space of the problem. The construction starts from the operator $H_0=H_{0,1}\oplus H_{0,2}$, where
 $$
H_{0,1} := -\,\frac{\D^2}{\D r^2}\,, \quad H_{0,2} := -\,\frac{\D^2}{\D r^2}+E\,, \quad D(H_{0,j}) = W^{2,2}_0(\R^+)\,,
 $$
which has again deficiency indices $(2,2)$ and thus a four-parameter family of self-adjoint extensions.
They can be characterized by means of boundary conditions:
for each matrix $\AAA= {a\;\; c \choose \bar{c} \;\; b}$ with $a,b\in \R\,$ and $\,c\in \C$ we denote by $H_\AAA$ the operator given by the same differential expression as $H_0$ with the domain $D(H_\AAA)\subset D(H_0^*) = W^{2,2}(\R^+) \oplus W^{2,2}(\R^+)$ specified by the conditions
\begin{equation} \label{2channel_bc}
f_1'(0)=af_1(0)+cf_2(0)\,,\;\quad f_2'(0)=\bar cf_1(0)+bf_2(0)\,;
\end{equation}
it is easy to check that any such $H_\AAA$ is a self-adjoint extension of $H_0$. There may be other extensions, say, with decoupled channels corresponding to $a=\infty$ or $b=\infty$ but it is enough for us to consider `most part' of them given by (\ref{2channel_bc}).

If the matrix $\AAA$ is real the operator $H_\AAA$ is invariant with respect to time reversal. The channels are not coupled if $c=0\,$; in that case $H_\AAA=H_a\oplus H_b$ where the two operators correspond to the s-wave parts of the point-interaction Hamiltonians $H_{\alpha,0}$ and $H_{\beta,0}$ in the two channels \cite{AGHH} with the interaction strengths
$\alpha:=\frac{a}{4\pi}$ and $\beta:=\frac{b}{4\pi}$, respectively, and its spectrum is easily found. To determine $\sigma(H_\AAA)$ in the coupled case, we have to know its resolvent which can be determined by means of Krein's formula using the integral kernel $G_\mathrm{D}(r,r';k) = \mathrm{diag}\big( \frac{\e^{ik\vert r+r'\vert}-\e^{ik\vert r-r'\vert}}{2ik},\, \frac{\e^{i\kappa\vert r+r'\vert}-\e^{i\kappa\vert r-r'\vert}}
{2i\kappa} \big)$, where $\kappa:=\sqrt{k^2\!-\!E}$, of the operator $H_\mathrm{D}$ with Dirichlet decoupled channels. The kernel of $(H_\AAA\!-\!z)^{-1}$ for $z\in\rho(H_\AAA)$ equals
 $$
G_\AAA(r,r';k) = G_\mathrm{D}(r,r';k) +D(k)^{-1}\,\left(\begin{array}{cc}
(b\!-\!i\kappa)\,\e^{ik(r+r')} & -\,c\,\e^{(ikr+\kappa r')} \\
-\,\bar c\, \e^{i(\kappa r+kr')} & (a\!-\!ik)\,\e^{i\kappa(r+r')} \end{array} \right )\,,
 $$
where as usual $k:=\sqrt z$ and  $D(k) :=(a\!-\!ik)(b\!-\!i\kappa)-\vert c\vert^2$.

It is straightforward to check that pole singularities of the above the resolvent can come only from zeros of the `discriminant' $D(k)$. In the decoupled case, i.e. if $c=0$ and $\AAA_0= {a\;\; 0 \choose 0 \;\; b}$, the expression factorizes, and consequently, it vanishes \emph{iff} $k=-ia$ or $\kappa=-ib$. Several different situations may arise:

\medskip

\noindent $\bullet$ If $a<0$ the operator $H_{\AAA_0}$ has eigenvalue $-a^2$ corresponding to the eigenfunction $f(r) =\sqrt{-2a}\, {\e^{ar} \choose 0}$ while for $a\geq 0$ the pole now corresponds to a zero-energy resonance or an antibound state

\medskip

\noindent $\bullet$ If $b<0$, then $H_{\AAA_0}$ has eigenvalue $E\!-\!b^2$ corresponding to $f(r)=\sqrt{-2b}\, {0 \choose \e^{br}}$, otherwise it has a zero-energy resonance or an antibound state.

\medskip

\noindent The continuous spectrum of the decoupled operator covers the positive real axis being simple in
$[0,E)$ and of multiplicity two in $[E,\infty)$. We are interested mainly in the case when both $a,b$ are negative and  $b^2<E\,$; under the last condition the eigenvalue of $H_b$ is embedded in the continuous spectrum of $H_a$.

Let us next turn to the interacting case, $c\neq 0$. Since the deficiency indices of $H_0$ are finite, the essential spectrum is not affected by the coupling. To find the eigenvalues and/or resonances of $H_\AAA$, we have to solve the equation
\begin {equation} \label{2model spectrum}
\left( a\!-\!ik\right) \left(b\!-\!i\sqrt{k^2\!-\!E}\right) = \vert c\vert^2.
\end{equation}
It reduces to a quartic equation, and can therefore be solved in terms of radicals; for simplicity we limit ourselves to the \emph{weak-coupling case} when one can make the following conclusion \cite{Ex91}.

\begin{theorem} \label{2chann_spec}
(a) Let $\sigma_\mathrm{p}\big(H_{\AAA_0}\big)$ be simple,
$-a^2\neq E\!-\!b^2$, then the perturbed first-channel
bound/antibound state behaves for small $|c|$ as
$$
e_1(c) = -a^2+\frac{2a|c|^2}{b+\sqrt{a^2\!+E}}+
\frac{a^2-E-b\sqrt{a^2\!+E}}{\sqrt{a^2\!+E}\left(b+\sqrt{a^2\!
+E}\right)^3} |c|^4+\OO(|c|^6)\,.
$$
In particular, zero-energy resonance corresponding to
$a=0$ turns into an antibound state if $H_{\AAA_0}$ has an
isolated eigenvalue in the second channel, $b<-\sqrt{E}$, and
into a bound state otherwise. \\
(b) Under the same simplicity assumption, if $H_{\AAA_0}$ has
isolated eigenvalue in the second channel, $b<-\sqrt{E}\,$,
the perturbation shifts it as follows
$$
e_2(c)=E\,-\,b^2+\frac{2b\vert c\vert^2}{a+\sqrt{b^2\!
-\!E}}+\frac{b^2+E-a\sqrt{b^2\!-\!E}}{\sqrt{b^2\!-\!E}
\left(a+\sqrt{b^2\!-\!E}\right)^3} \,\vert c\vert^4 +
\OO(\vert c\vert^6)\,.
$$
On the other hand, if $H_{\AAA_0}\,$ has embedded eigenvalue,
$-\sqrt{E}<b<0$, it turns under the perturbation into a pole of the analytically continued resolvent with
\begin{eqnarray*}
\re e_2(c) \!&=&\! E\,-\,b^2+\frac{2ab\vert c\vert^2}{a^2\!
-\!b^2\!+E}+ \OO(\vert c\vert^4)\,, \\[.5em]
\im e_2(c) \!&=&\! \frac{2b\vert c\vert^2\sqrt{E\!-\!b^2}}
{a^2\!-\!b^2\!+E}+ \OO(\vert c\vert^4)\,.
\end{eqnarray*}
(c) Finally, let $H_{\AAA_0}$ have an isolated eigenvalue of multiplicity two, $b=-\sqrt{a^2\!+E},$; then under the perturbation it splits into
$$
e_{1,2}(c)=-a^2\,\mp\,2\sqrt{-a}\root 4 \of {a^2\!+E}\, \vert
c\vert+ \frac{2a^4\!+4a^2E\!+\!E^2}{2a(a^2\!+E)^{3/2}}\, \vert
c\vert^2+ \OO(\vert c\vert^3)\,.
$$
\end{theorem}

The model can be investigated also from the scattering point of view. Since the couplings is a rank-two perturbation of the free resolvent, the existence and completeness of the wave operators $\Omega_{\pm} (H_\AAA,H_{\AAA_0})$ follow
from Birman-Kuroda theorem \cite[Sec.~XI.3]{RS}. It is also easy to check that the scattering is asymptotically complete, what is more interesting is the explicit form of the S-matrix. To find it we look for generalized eigenfunctions of the form
$f(r)= \big( e^{-ikr}\!-\!A \e^{ikr}, B \e^{i\kappa r} \big)^\mathrm{T}$ which belong \emph{locally} to the domain of $H_\AAA$. Using boundary conditions (\ref{2channel_bc}) we find
$$
A= S_0(k) = \frac{(a\!+\!ik)(b\!-\!i\kappa)-\vert
c\vert^2}{D(k)}\,, \quad B = \frac{2ik\bar c}{D(k)}\,.
$$
If the second channel is closed, $k^2\leq E$, the scattering is elastic, $|A|=1$. We are interested primarily in the case when $H_{\AAA_0}\,$ has an embedded eigenvalue which turns under the perturbation into a resonant state whose  lifetime is
$$
T(c)\,:=\,-\,\frac{a^2\!-\!b^2+E}{4b\vert
c\vert^2\sqrt{E\!-\!b^2}}\, \left(1+\OO(\vert c\vert^2)\right)\,;
$$
inspecting the phase shift we see that it has a jump by $\pi$ in the interval of width of the order $2\,\im e_2(c)$ around $\re e_2(c)$. More specifically, writing the on-shell S-matrix
conventionally through the phase shift as
$S_0(k)=e^{2i\delta_0(k)}$ we have
$$
\delta_0(k)=\arctan\frac{k(b\!+\!\sqrt{E\!-\!k^2})}
{a(b+\sqrt{E\!-\!k^2}) -\vert c\vert^2}\;\pmod{\pi}\,.
$$
The resonance is then seen as a local change of the transmission probability (and related quantities such as the scattering cross section), the sharper it is the closer the pole is to the real axis. This is probably the most common way in which resonances are manifested, employed in papers too numerous to be quoted here.

On the other hand, if the second channel is open, $k^2>E$, the reflection and transmission amplitudes given above
satisfy $\vert A\vert^2+\frac{\kappa}{k}\,\vert B\vert^2 = 1$.
The elastic scattering is now non-unitary since $B\neq 0$ which means that the `nucleus' may now leave the interaction
region in the excited state. The said relation between the amplitudes can alternatively be written as $\vert S_{0,1\rightarrow 1}(k)\vert^2+ \vert S_{0,1\rightarrow2}(k)\vert^2=1$ which is a part of the full two-channel S-matrix unitarity condition.

The model also allows us to follow the time evolution of the resonant state, in particular, to analyze the pole approximation (\ref{redapprox}) in this particular case. The natural choice for the `compound nucleus' wave function is the eigenstate of the unperturbed Hamiltonian,
\begin{equation} \label{2resonant state}
f\,:\:f(r)=\sqrt{-2b}\,\left(
\begin{array}{l} 0 \\ \e^{br} \end{array}\!\right)\,.
\end{equation}
Using the explicit expression of the resolvent we find
$$
(f,(H_\AAA\!-\!k^2)^{-1}f) = \frac{\vert c\vert^2+(a\!-\!ik)
(b\!+\!i\kappa)}{(b\!+\!i\kappa)^2\bigl\lbrack\vert c\vert^2
-(a\!-\!ik)(b\!-\!i\kappa)\bigr\rbrack}\,.
$$
The reduced evolution is given by (\ref{reduced}); using the last formula, evaluating the integral by means of the residue theorem and estimating the remainder we arrive after a straightforward computation to the following conclusion \cite{Ex91}.

\begin{theorem} \label{2chann_evolution}
Assume $a\neq 0$ and $-\sqrt{E}<b<0$. The reduced propagator of the resonant state (\ref{2resonant state}) is given by
\begin{eqnarray*}
\lefteqn{v_\AAA(t) = \Biggl\{\,\e^{-ik_2^2t}\,-\, \vert c\vert^2 \Biggl\lbrack \,\frac{2(\vert a\vert \!-\!a)b}
{(a^2\!-\!b^2\!+E)^2}\, \e^{-ik_1^2t}\, +
\frac{ib}{\sqrt{E\!-\!b^2} (a-i\sqrt{E\!-\!b^2})^2}\, \e^{-ik_2^2t}
}
\\[.3em] && \phantom{AAAAAA}
+\frac{4b}{\pi} \,\e^{-\pi i/4} \,\int_{0}^{\infty}\, \frac {z^2 \e^{-z^2 t}\,\D z}{(z^2\!+ia^2)(z^2\!-i(E\!-\!b^2))^2}
\,\Biggr\rbrack \Biggr\}\, \left(\,1+\OO(\vert c\vert^2) \right) \phantom{AAAA}
\end{eqnarray*}
and the decay law is
\begin{eqnarray*}
\lefteqn{P_\AAA(t) = \Biggl\{\,\e^{2({\rm Im}\,e_2)t} - 2\vert
c\vert^2 {\rm Re}\,\Biggl\lbrack\,\frac{2(\vert a\vert \!-\!a)b}
{(a^2\!-\!b^2\!+E)^2}\, \e^{-i(k_1^2\!-\bar k^2_2)t}} \\[.3em]
&& \phantom{AAAAAA} +\,\frac{ib}{\sqrt{E\!-\!b^2}
(a-i\sqrt{E\!-\!b^2})^2}\, e^{2({\rm Im}\,e_2)t}
\\ [.3em] && \phantom{AAAAAA}
+\frac{4b}{\pi}\,e^{i(\bar k_2^2t-\pi/4)}\,\int_{0}^{\infty}\,
\frac {z^2 e^{-z^2 t}\,\D z}{(z^2\!+ia^2)(z^2-i(E\!-\!b^2))^2}
\,\Biggr\rbrack \Biggr\}\, \left(\,1+\OO(\vert c\vert^2)
\right)\,,
\end{eqnarray*}
where $e_j= e_j(c)=: k_j^2\,,\: j=1,2\,,$ are specified in Theorem~\ref{2chann_spec}.
\end{theorem}

Hence we have explicit formula for deviations from the exponential decay law. Some of its properties, however, may not be fully obvious. For instance, the initial decay rate vanishes, $\dot P_\AAA(0+)=0$, since $\im(f,(H_\AAA\!-\!\lambda)^{-1}f) = \OO(\lambda^{-5/2})$ as $\lambda\to\infty$, cf. Proposition~\ref{zero_initial} below. On the other hand, the long-time behavior depends substantially on the spectrum of
the unperturbed Hamiltonian. If $H_{\AAA_0}$ has an eigenvalue in the first channel, then the decay law contains a term of order of $|c|^4$, which does not vanish as $t\to\infty$; it comes from the component of the first-channel bound state contained in the resonant state (\ref{2resonant state}).

The last name fact it is useful to keep in mind when we speak about an unstable state \emph{lifetime}. It is a common habit, motiveted by the approximation (\ref{redapprox}), to identify the latter with the inverse distance of the pole from the real axis as we did above when writing $T(c)$. If the decay law is differentiable, however, $-\dot P_\AAA(t)$ expresses the probability density of decay at the instant $t$ and a simple integration by parts allows us to express the average time for which the initial state survives as $T_\AAA = \int_0^\infty P_\AAA(t)\,\D t$; this quantity naturally diverges if $\lim_{t\to\infty} P_\AAA(t) \ne 0$.

  \subsection{K-shell capture model: comparison to stochastic mechanics}

The above model has many modifications, we will describe briefly two of them. The first describes a $\beta$-decay process in which an atomic electron is absorbed by the nucleus and decays through the reaction $e\!+\!p\to n\!+\!\nu$ with a neutrino emitted. One usually speaks about a  \emph{K-shell capture} because the electron comes most often from the lowest energy orbital, however, from the theoretical point any orbital mode can be considered. We assume again spherical symmetry and take $\HH=\HH_1 \oplus \HH_2$ with $\HH_j:=L^2(\R^+)$ as the state space. The first component refers to the (s-wave part of) atomic wave function, the other is a caricature description of the decayed states; we neglect the fact that neutrino is a relativistic particle.

The departing point of the construction is again a non-selfadjoint operator of the form $H_0=H_{0,1}\oplus H_{0,2}$, the components of which act as
 \begin{equation} \label{Kstart}
H_{0,1}:=- \frac{1}{2m}\,\frac{\D^2}{\D r^2}+V_{\ell}(r)\,, \quad
H_{0,2}:=-{1\over 2M}\, \frac{\D^2}{\D x^2}-E
 \end{equation}
with the domains $D(H_{0,1})=\{f\in W^{2,2}(\R^+):\:u(R_j-)=u(R_j+)=0\}$ for fixed $0<R_1<R_2<\dots<R_N\:$ --- we add the requirement $f(0+)=0$ if the angular momentum $\ell=0\:$ --- and $D(H_{0,2})= \{W^{2,2}(\R^+):\: f(0+)=f'(0+)=0\}$. Here as usual $V_{\ell}(r)= V(r)+ \frac{\ell(\ell\!+\!1)}{2mr^2}$ and the potential is supposed to satisfy the conditions
 $$ 
\lim_{r \to \infty} V(r)=0 \,,\quad
\limsup_{r \to 0}\, rV(r)=\gamma < \infty \,,
 $$ 
under which the operator $H_0$ is symmetric with deficiency indices $(N+1,N+1)$. Of all its self-adjoint extensions we choose a subclass that (i) allows us to switch off the coupling and (ii) couples each sphere \emph{locally} to the other space. The adjoint operator $H_0^*$ acts again as (\ref{Kstart}); the extensions $H(a)$ described by $a=(a_1,\dots,a_N)$ are specified by the boundary conditions
 $$ 
u'(R_j+)-u'(R_j-) = a_jf'(0+)\,,\quad j=1,\dots ,N \quad\mathrm{and}\quad
\sum_{j=1}^N \bar{a}_ju(R_j) = -\,{m\over M}\,f(0+)\,;
 $$ 
it is easy to see that under them the appropriate boundary form vanishes, the channels are decoupled for $a=0$, and the Hamiltonians $H_a$ are time-reversal invariant.

To solve the resonance problem in the model we have to find the resolvent of $H(a)$ which can be again done using Krein's formula. We will describe the resolvent projection to the `atomic' channel referring for the full expression and the proof to \cite{ET92}. We introduce the kernel
 $$ 
G_1(r,s;k^2)= \frac{1}{W(v_k,u_k)}\, u_k(r_<)v_k(r_>)\,,
 $$ 
where as usual $r_<:= \min\{r,s\},\: r_>:=\max\{r,s\}$, and the functions
$u_k,v_k$ are solutions to $-\frac{1}{2m}\,u''+Vu=\frac{k^2}{2m}\,u\,$ such that $u_k(0+)=0\,$ and $\,v_k$ is $L^2$ around $\,\infty$, and furthermore, $W(v_k,u_k) :=v_k(r)u'_k(r)-v'_k(r)u_k(r)$ is their Wronskian.

Before stating the result, let us mention that the model can also cover the situation when the electron can be absorbed anywhere within the volume of the nucleus approximating this behavior by a family of equidistant spheres with $R_j:=jR/N,\: j=1,\dots,N$, where $R$ is the nucleon radius. Let $a:\,[0,R]\to\R$ be a bounded piecewise continuous function understood as `decay density', and take $H(a^N)$ corresponding to $a_j^N:= \frac{R}{N}\,a \big(\frac{jR}{N}\big)$. On a formal level, the limit $N\to\infty$ leads to an operator describing the two channels coupled through the boundary conditions
 $$ 
u''(r)= a(r)f'(0+)\,, \quad \int_0^R a(s)u(s)\, \D s= -\,{m\over M}\, f(0+)\,,
 $$ 
however, we limit ourselves to checking the strong resolvent convergence \cite{ET92}.

\begin{theorem} \label{atomic kernel}
The projection of the resolvent $(H(a)-z)^{-1}$ to the `atomic' channel is an integral operator with the kernel
$$
G_1(r,s;z)+\,\sum_{j,k=1}^{N} \frac{i\kappa M a_j{\bar a}_k
G_1(r,R_j;z)G_1(R_k,s;z)} {m-i\kappa M\sum_{i,l=1}^{N}a_i{\bar a}_l
G_1(R_l,R_i;z)}\,.
$$
The projections of $(H(a^N)-z)^{-1}$ converge as $N\to\infty$ to  operator with the kernel
$$
G_1(r,s;z)+\,\frac{i\kappa M\phi_k(r)\phi_k(s)}
{m-i\kappa M\int_0^R\int_0^R a(r)a(s)G_1(r,s;z)\,\D r\D s}\,,
$$
where $\phi_k:=\int_0^R a(s) G_1(\cdot,s;z)\,\D s$, in the strong resolvent sense.
\end{theorem}

The singularities correspond to zeros of the denominators in the above expression. As an example, consider the `atom' with Coulomb potential,
$$
V_{\ell}(r)= \frac{\gamma}{r}+ \frac{\ell(\ell\!+\!1)}{2mr^2}\,,\quad \gamma<0\,,
$$
which has in the decoupled case, $a=0$, poles at $ k_n= -\frac{im \gamma}{n},\: n=1,2,\dots\,$. The Green function $G_1$ can be expressed in terms of the standard Coulomb wave functions $\psi_{n\ell m}(r, \vartheta,\varphi) = R_{n\ell}(r) Y_{\ell m}(\vartheta,\varphi)$. To analyze weak-coupling behavior of the poles, we restrict ourselves to two cases, a \emph{surface-supported decay} when $N=1$ and a \emph{volume-supported decay} with a constant `density' when $a(r)=a$ is a constant function on $[0,R]$. We put $\kappa_n := \sqrt{2mE - \left( \frac{m\gamma}{n} \right)^2}$ and introduce the form factor
$$
B_n(r):= \left\{ \begin{array}{lcl} R\,R_{n\ell}(r) &\quad \dots \quad & \text{surface-supported decay} \\[.3em] \int_0^R r\,R_{n\ell}(r)\, \D r &\quad \dots \quad & \text{volume-supported decay} \end{array} \right.
$$
A straightforward calculation \cite{ET92} then yields the shifted pole positions,
$$
\frac{k_n^2(a)}{2m} = -\,\frac{m\gamma^2}{2n} \,-\,\frac{i}{4}\, a^2m \kappa_n \gamma^3 (2\ell+1)! \, B_n(R)^2 + \OO(a^4)\,.
$$
We are interested particularly in the situation where the unperturbed eigenvalue is embedded, $n> \sqrt{\frac{m\gamma^2}{2E}}$, when $\kappa_n$ is real and the coupling shifts the pole into the lower complex halfplane giving rise to the resonant state with the lifetime
$$
T_n(a) = \frac{8a^2 (B_n(R))^{-2}}{m \kappa_n \gamma^3 (2\ell+1)!} + \OO(a^0)\,.
$$
In the case of the real decay, of course, all the unperturbed eigenvalues are embedded and the K-shell contribution is dominating. It has the shortest lifetime since $R_{n\ell}(0)$ is nonzero for $\ell=0$ only and $m\gamma R\ll 1$, typically of order $10^{-4}$, so the form factor value is essentially determined by the wave function value at the origin.

The K-shell capture model allows us to make an important reflection concerning relations between quantum and stochastic mechanics. The two theories are sometimes claim to lead to the same results \cite{Ne} and there are cases when such a claim can be verified. The present model shows that in general there is a principal difference between the two. One can model such a decay in stochastic mechanics too considering random electron trajectories and summing the decay probabilities for their parts situated within the nucleus. The formula is given in \cite{ET92} and we are not going to reproduce it here; what is important that the total probability is expressed as the sum of probabilities of all the contributing processes. In the quantum-mechanical model discussed here, on the other hand, one adds the \emph{amplitudes} --- it is obvious from the form factor expression in case of a volume-supported decay --- and the total probability is the squared modulus of the sum.

  \subsection{A model of heavy quarkonia decay}

Let us finally mention one more modification of the model, this time aiming at description of decays of charmonium or bottomium, which are bound states of heavy quark-antiquark pairs, into a meson-antimeson pair. Such processes are known to be essentially non-relativistic; as an example one can take the decay $\psi''(3770)\to D\bar D$ where the $D$ meson mass is $\approx\, 1865\,\mathrm{MeV/c}^2\,$, thus rest energy of the meson pair is two orders of magnitude larger than the kinetic one released in the decay.

If the interaction responsible for the decay is switched off the quark and meson pairs are described by the operators
 $$
\hat H_{0,j}:= -\,\frac{1}{2m_j}\Delta_{j1}
-\,\frac{1}{2m_j}\Delta_{j2} + V_j(\vert\vec x_{j1}\!
-\!\vec x_{j2}\vert) + 2m_jc^2\,,
 $$
where $m_1$ and $m_2$ are the quark and meson masses, respectively. As before we separate the center-of-mass motion and use the rotational invariance. Adjusting the energy threshold to $2m_2c^2$ we can reduce the problem to investigation of self-adjoint extensions of the
operator $H_0^{(\ell)}:= H_{0,1}^{(\ell)} \oplus H_{0,2}^{(\ell)}$ on
$L^2(R_1,\infty)\oplus L^2(R_2,\infty)$ defined by
 $$
H_{0,j}^{(\ell)}:= -\, \frac{1}{m_j}\, \frac{\D^2}{\D r_j^2} +
V_j(r_j) + \frac{\ell(\ell+1)}{m_jr_j^2}+ 2(m_j-m_2)c^2
 $$
with $D(H_{0,j}^{(\ell)}):= C_0^{\infty}(R_j,\infty)$. Let us list the assumptions. We suppose that the quarks can annihilate only they `hit each other', $R_1=0$, while for mesons we allow existence of a hard core, $R_2=R\ge 0$. One the other hand, the mesons are supposed to be non-interacting, $V_2=0$; this may not be realistic if they are charged but it simplifies the treatment. In contrast to that, the interquark potential is \emph{confining}, $\lim_{r\to\infty}V_1(r)=\infty$; we also assume that $V_1\in L^1_\mathrm{loc}$ and a finite $\lim_{r\to 0+} V_1(r)$ exists.

To couple the two channels the deficiency indices of $H_0^{(\ell)}$ have to be $(2,2)$; since we have put $R_1=0$ it happens only if $\ell=0$ and we drop thus the index $\ell$ in the following. We will not strive again to describe all the extensions and choose a particular one-parameter family: the domain of the extension $H_a$ will consist of functions $f\in W^{2,2}(\R^+) \oplus W^{2,2}(R,\infty)$ satisfying the conditions
\begin{equation} \label{quark_bc}
f_1(0)=a\,f'_2(R)\,, \quad f_2(R)= \frac{m_2}{m_1}\,
\bar a\, f'_1(0)\,,
\end{equation}
with $a\in\C$. In the decoupled case, $a=0$, we get Dirichlet boundary condition in both channels as expected; for $a\in\R$ the Hamiltonian is time-reversal invariant.

As the first thing we have to find the resolvent of $H_a$, in particular its projection to the quark channel. In analogy with the previous section we can write its integral kernel $G_0(r,s;z)$ in terms of two solutions of the equation
 $$
\Big(-\, \frac{1}{m_j}\, \frac{\D^2}{\D r^2} +
V_1(r) + 2(m_1-m_2)c^2 \Big) f(r) = zf(r)
 $$
for $z\not\in\R$ such that $u(0)=0$ and $v$ is $L^2$ at infinity. Krein's formula helps again; by a straightforward computation \cite{AES94} we get
 $$
G_a(r,s;z) = G_0(r,s;z) + \frac{-ikm_2\vert a\vert^2 v(r;z) v(s;z)} {m_1 v(0;z) D(v,a;z)}\,,
 $$
where the denominator is given by
 $$
D(v,a;z):= v(0;z)-ik\vert a\vert^2 \frac{m_2}{m_1}\,v'(0;z)\,.
 $$
The singularities are again determined by zeros of the last expression. One can work out examples such a natural confining potential, $V_1(r)= \alpha r+ V_0+ 2(m_2\!-m_1)c^2$, and its modifications, in which the resonance width can be expressed through the value of the quark wave function at the origin. This appears to be the case generally.

\begin{theorem} \cite{AES94} \label{quark_dec_width}
Under the stated assumptions, the quarkonium decay width is given for the $n$-th s-wave state by
\begin{equation} \label{decay width}
\Gamma_n(a)= 8\pi k_n\, \frac{m_2}{m_1^2}\, |a|^2 |\psi_n(0)|^2+ \OO(|a|^6)\,,\quad
k_n:=\sqrt{m_1E_n}\,,
\end{equation}
provided the bound-state energy $E_n$, adjusted by the difference of the rest energies, is positive; $\psi_n(0)$ is the value of the corresponding wave function at the origin.
\end{theorem}

Note that while we have assumed the quark potential to be below bounded at the origin, the assumption can be relaxed. The theorem holds also for potentials with sufficiently weak singularity, in particular, for the physically interesting case of a linear confinement combined with a Coulomb potential.

\section{More about the decay laws}

Let us return to the time evolution of unstable systems, in particular, to properties of the decay laws. In addition to the elementary properties mentioned together with the definition (\ref{decaylaw}) we know so far only that in the weak-coupling situation they do not differ much from an exponential function coming from the leading term of the pole approximation. This says nothing about local properties of the decay laws which is the topic we are going to investigate in this section.

Historically the first consequence of non-exponentiality associated with the below bounded energy spectrum concerned the long-time behavior of the decay laws; already in \cite{Kh57} it was observed that a sharp energy cut-off leads to the $\OO(t^{-3/2})$ behavior as $t\to\infty$, and other examples of that type followed. Moreover, it is even possible that a part of the initial state survives the decay; we have seen a simple example at the end of Sec.~\ref{ss: 2channel} and another one will be given in Sec.~\ref{s: leaky} below. Here we concentrate on two other local properties of decay laws.

  \subsection{Initial decay rate and its implications}

The first one concerns the behavior of the system immediately after its preparation. Exponential decay has a constant decay rate which, in particular, means it is nonzero at $t=0$. This may not be true for other decay laws. We note, for example, that $P_\psi$ is by definition an even function of $t$, hence if the (two-sided) derivative $\dot P_\psi(0)$ exists it has to be zero. This happens for vectors from the form domain of the Hamiltonian: we have $|(\psi, \e^{-iHt}\psi)|^2 \le P_\psi(t) \le 1$ which leads easily to the following conclusion \cite{EH73}.

\begin{proposition} \label{zero_initial}
If $\psi\in Q(H)$ the decay law satisfies $\dot P_\psi(0+)=0$.
\end{proposition}

The importance of this result stems from the peculiar behavior of unstable systems subject to frequently repeated measurements knows as \emph{quantum Zeno effect}, namely that in the limit of permanent measurement the system cannot decay. This fact was known essentially already to von Neumann and Turing, in the context of unstable particle decay it was first described by Beskow and Nilsson \cite{BN67} followed by a serious mathematical work \cite{Fr72, Ch} which elucidated the mechanism. It became truly popular, however, only after the flashy name referring to Zeno's aporia about a flying arrow was proposed in \cite{MS77}. Since then the effect was a subject of numerous investigations, in part because it became interesting also from experimental and application points of view. However, since Zeno-type problems are not the subject of this survey we limit ourselves to quoting the review papers \cite{Sch04, FP08} as a guide to further reading, and will discuss the topic only inasmuch it concerns the initial decay rate.

Suppose that we perform on an unstable system a series of measurements at times $t/n,\, 2t/n\,\, \dots,\, t$, in which we ascertain that it is still undecayed. If the outcome of each of them is positive, the state reduction returns the state vector into the subspace $\HH_\mathrm{u}$ and the resulting non-decay probability is
 $$ 
  M_n(t)= P_{\psi}(t/n) P_{\psi_1}(t/n) \cdots
  P_{\psi_{n-1}}(t/n)\,,
 $$ 
where $\psi_{j+1}$ is the normalized projection of $\e^{-iHt/n} \psi_j$ on $\HH_\mathrm{u}$ and $\psi_0:= \psi$, in particular, $M_n(t)= (P_{\psi}(t/n))^n$ if $\dim\HH_\mathrm{u}=1$ and all the vectors $\psi_j$ coincide with $\psi$ up to a phase factor. Since $\lim_{n\to\infty} (f(t/n)^n= \exp\{{-\dot f(0+)t}\}$ holds whenever $f(0)=1$ and the one-sided derivative $\dot f(0+)$ exists, we see that $\dot P_\psi(0+)=0$ implies the Zeno effect, $M(t):=\lim_{n\to\infty} M_n(t)=1$ for all $t>0$, and the same is true if $\dim\HH_\mathrm{u}>1$ provided the
derivative $\dot P_\psi(0+)$ has such a property for \emph{any} $\psi\in \HH_\mathrm{u}$.

At the same time the above simple argument suggests that an opposite situation, an \emph{anti-Zeno effect}, is possible when $\dot P_\psi(0+)$ is negative infinite; then $M(t)=0$ for any $t>0$ which means that the decay is accelerated and the unstable system disappears once the measurement started. The possibility of such a behavior was mentioned early \cite{CSM77}, however, the attention to it is of a recent date only --- we refer again to the review work quoted above.
Before proceeding further we have to say that the two effects are understood differently in different communities. For experimental physicists the important question is the change of the observed lifetime when the measurement are performed with a certain frequency, on the other hand a theoretical or mathematical physicist typically asks what happens if the period between two successive measurements tends to zero.

Let us return to the initial decay rate. It is clear we have to estimate $1-P_\psi(t)$ for small values of $t$, which we can  write as $2\,\mathrm{Re\,} (\psi, E_\mathrm{u}(I\!-\!\e^{-iHt}) \psi) -\| E_\mathrm{u}(I\!-\!\e^{-iHt})\psi\|^2$, or alternatively cast it using spectral theorem into the form
 $$ 
  4\int_{-\infty}^\infty \!\sin^2 \frac{\lambda t}{2}\,
  \D\|E^H_\lambda \psi\|^2 -4 \left\|\int_{-\infty}^\infty
  \!\e^{-i\lambda t/2}\,\sin \frac{\lambda t}{2}
  \, \D E_\mathrm{u} E^H_\lambda \psi\, \right\|^2,
 $$ 
where the non-decreasing projection-valued function $\lambda \mapsto E^H_\lambda: = E_H((-\infty,\lambda])$ generates spectral measure $E_H$ of the Hamiltonian $H$. By Schwarz inequality the above expression is non-negative; we want to find tighter upper and lower bounds.

To this aim we choose an orthonormal basis $\{\chi_j\}$ in the unstable system subspace $\HH_\mathrm{u}$ and expand the initial state vector as $\psi= \sum_j c_j\chi_j$ with $\sum_j |c_j|^2 =1$. The second term in the above expression can be then written as
$$ 
  -4 \sum_m \bigg| \sum_j c_j\int_{-\infty}^\infty
  \!\e^{-i\lambda t/2}\, \sin \frac{\lambda t}{2}
  \, \D(\chi_m, E^H_\lambda \chi_j) \bigg|^2,
$$ 
where $\D\omega_{jk}(\lambda):= \D(\chi_j, E^H_\lambda \chi_k)$ are real-valued measures symmetric with respect to
interchange of the indices. Since the measure appearing in
the first term can be written as $\D\|E^H_\lambda\psi\|^2 = \sum_{jk} \bar c_j c_k \D\omega_{jk}(\lambda)$, the decay probability becomes
\begin{eqnarray*}\label{diff5}
  \lefteqn{1-P_\psi(t)= 4\sum_{jk} \bar c_j c_k \bigg\{
  \int_{-\infty}^\infty \!\sin^2 \frac{\lambda t}{2}\,
  d\omega_{jk}(\lambda) } \nonumber \\ &&
  - \sum_m \int_{-\infty}^\infty
  \! e^{-i\lambda t/2}\,\sin \frac{\lambda t}{2}
  \, d\omega_{jm}(\lambda)\, \int_{-\infty}^\infty
  \! e^{i\mu t/2}\,\sin \frac{\mu t}{2} \, d\omega_{km}(\mu) \bigg\}\,;
\end{eqnarray*}
if $\dim\HH_\mathrm{u}=\infty$ the involved series can easily be seen to converge using Parseval relation. Using next the normalization $\int_{-\infty}^\infty \D\omega_{jk}(\lambda) =
\delta_{jk}$ we arrive after a simple calculation \cite{Ex05} at the formula
 \begin{equation} \label{diff6}
1-P_\psi(t)= 2\sum_{jkm} \bar c_j c_k
  \int_{-\infty}^\infty \int_{-\infty}^\infty
  \sin^2 \frac{(\lambda-\mu) t}{2}\,
  \D\omega_{jm}(\lambda) \D\omega_{km}(\mu)\,.
 \end{equation}
Consider first an upper bound. We fix $\alpha\in (0,2]$ and use the inequalities $|x|^\alpha \ge |\sin x|^\alpha \ge \sin^2 x$ together with $|\lambda-\mu|^\alpha \le 2^\alpha (|\lambda|^\alpha +|\mu|^\alpha)$ to infer that
\begin{eqnarray*}
  \lefteqn{ \frac{1-P_\psi(t)}{t^\alpha} \le 2^{1-\alpha}
  \sum_{jkm} \bar c_j c_k \int_{-\infty}^\infty \int_{-\infty}^\infty
  |\lambda-\mu|^\alpha\, \D\omega_{jm}(\lambda) \D\omega_{km}(\mu) } \\ &&
  \le 2 \sum_{jkm} \bar c_j c_k \int_{-\infty}^\infty \int_{-\infty}^\infty
  (|\lambda|^\alpha +|\mu|^\alpha)\, \D\omega_{jm}(\lambda) \D\omega_{km}(\mu) \le 4 \langle |H|^\alpha \rangle_\psi\,,
\end{eqnarray*}
which means that $1-P_\psi(t)= \OO(t^\alpha)$ if $\psi\in \mathrm{Dom\,} (|H|^{\alpha/2})$. If this is true for some $\alpha>1$ we get $\dot P_\psi(0+)=0$ which a slightly weaker result than Proposition~\ref{zero_initial}. Note also that if $\dim\HH_\mathrm{u}=1$ and the spectrum of $H$ is absolutely continuous there is an alternative way to justify the claim using Lipschitz regularity, since $P(t)= |\hat\omega(t)|^2$ in this case and $\hat\omega$ is bounded and uniformly
$\alpha$-Lipschitz \emph{iff} $\int_\mathbb{R} \omega(\lambda) (1+|\lambda|^\alpha)\,\D\lambda < \infty$.

A lower bound is more subtle. We use the inequality $\big| \sin \frac{(\lambda-\mu) t}{2} \big| \ge C|\lambda-\mu|t$ which holds with a suitable $C>0$ for $\lambda, \mu\in [-1/t,1/t]$ to estimate (\ref{diff6}) as follows
\begin{eqnarray*}
  \lefteqn{1-P_\psi(t)\ge 2C^2t^2
  \int_{-1/t}^{1/t} \int_{-1/t}^{1/t}
  (\lambda-\mu)^2\,
  (\psi, \D E^H_\lambda E_\mathrm{u} \D E^H_\mu \psi)}
  \\ && =4C^2t^2 \bigg\{
  \int_{-1/t}^{1/t} \int_{-1/t}^{1/t}
  (\lambda^2-\lambda\mu)\, (\psi, \D E^H_\lambda E_\mathrm{u} \D E^H_\mu \psi)
  \bigg\} \\ &&
  =4C^2t^2 \left\{ (\psi, H^2_{1/t} E_\mathrm{u} I_{1/t} \psi)
  - \|PH_{1/t}\psi\|^2 \right\}\,, \phantom{AAAAAAA}
\end{eqnarray*}
where $H_N$ denotes the cut-off Hamiltonian, $HE_H(\Delta_N)$ with $\Delta_N:= (-N,N)$. Dividing the expression at the right-hand side by $t$ and choosing $t=N^{-1}$, we arrive at the following conclusion.

\begin{proposition} \label{inf initial}
The initial decay rate of $\psi\in \HH_\mathrm{u}$ satisfies $\dot P_\psi(0+)=-\infty$ provided $\left(\langle H^2_N E_\mathrm{u} E_H(\Delta_N)\rangle_\psi - \|PH_N\psi\|^2\right)^{-1} = o(N)$ holds as $N\to\infty$.
\end{proposition}

To illustrate how does the initial decay rate depend on spectral properties of the decaying state, consider an
\emph{example} in which $\dim\HH_\mathrm{u}=1$, the
Hamiltonian is bounded from below and $\psi$ from its absolutely continuous spectral subspace is such
that $\D(\psi,E^H_\lambda\psi) = \omega(\lambda)\,\D\lambda$ where $\omega(\lambda) \approx c\lambda^{-\beta}$ as $\lambda\to +\infty$ for some $c>0$ and $\beta>1$. If $\beta>2$, Proposition~\ref{zero_initial} implies $\dot P_\psi(0+)=0$. On the other hand, one can easily find the asymptotic behavior of the quantity appearing in Proposition~\ref{inf initial}: $\int_{-N}^N \omega(\lambda)\, \D\lambda$ tends to one, while the other two integrals diverge giving
$$ 
  \int_{-N}^N \lambda^2\, \D\omega(\lambda)
  \int_{-N}^N \D\omega(\lambda)
  -\left( \int_{-N}^N \lambda\, \D\omega(\lambda)
  \right)^2 \approx \frac{c}{3-\beta}\,N^{3-\beta}-
  \left(\frac{c}{2-\beta}\right)^2 N^{4-2\beta}
$$ 
as $N\to +\infty$, and consequently, $\dot P(0+)=-\infty$ holds for $\beta\in (1,2)$. This shows that the exponential decay --- which requires, of course, $\sigma(H)=\R$ by Theorem~\ref{sinha} --- walks a thin line between the two extreme initial-decay-rate possibilities. Let us remark finally that while `Zeno' limit is trivial for the exponential decay, it may not exist in other cases with $\beta=2$; in \cite[Rem.~2.4.9]{Ex} the reader can find an example of such a distribution with a sharp cut-off leading to rapid oscillations of the function $t\mapsto(\psi, \e^{-iHt}\psi)$ which obscure existence of the limit.

  \subsection{Irregular decay: example of Winter model}

Now we turn to another decay law property. In the literature it is usually tacitly assumed that $P_\psi(\cdot)$ is a `nice', i.e. sufficiently regular function, typically by dealing with its derivatives. Our aim is to show that this property cannot be taken for granted which we are going to illustrate on another well-known solvable model of decay.

An inspiration comes from the striking behavior of some wave functions in a one-dimensional hard-wall potential well observed in \cite{Be96, Th}. The simplest example concerns the situation when the initial function is constant (and thus not belonging to the domain of the Dirichlet Laplacian): it evolves into a steplike $\psi(x,t)$ for times  which are rational
multiples of the period, $t=qT$ with $q=N/M$, and the number of steps increases with growing $M$, while for an irrational $q$ the function $\psi(x,t)$ is fractal with respect to the variable $x$. One may expect that such a behavior will not disappear completely if the hard wall is replaced by a singular potential barrier. It was illustrated in a `double well' system \cite{VDS02}; here we instead let the initial state decay into continuum through the tunneling.

Decay due to a barrier tunneling is among the core problem of quantum mechanics which can be traced back to Gamow's paper \cite{Ga28}. The model in which the barrier is a spherical $\delta$-shell is usually referred to as \emph{Winter model} after the paper \cite{Wi61} where it was introduced. A thorough analysis of this model can be found in \cite{AGS87}; it has also various generalizations, we refer to \cite{AGHH} for a
bibliography. The Hamiltonian acting in $L^2(\R^3)$ is of the form
 $$ 
 H_\alpha = -\Delta + \alpha \delta(|\vec{r}|-R)\,, \quad \alpha >0\,,
 $$ 
with a fixed $R>0$; as usual we employ rational units, $\hbar = 2m = 1$. For simplicity we restrict our attention to the s-wave part of the problem, using the reduced wave functions $\psi(\vec{r},t) =\frac{1}{\sqrt{4\pi}} r^{-1} \phi(r,\,t)$ and the corresponding Hamiltonian part,
 $$ 
 h_\alpha = -\frac{\D^2}{\D r^2} + \alpha \delta(r - R)\,;
 $$ 
we are interested in the time evolution, $\psi(\vec{r},t) = \e^{-iH_\alpha t}\psi(\vec{r},0)$ for a fixed initial condition $\psi(\vec{r},0)$ with the support inside the
ball of radius $R$, and the corresponding decay law $P_\psi(t) = \int_0^R |\phi(r,\,t)|^2\,\D r\,$ referring to $\HH_\mathrm{u} = L^2(B_R(0))$.

It is straightforward to check \cite{AGS87} that $H_\alpha$ has no bound states, on the other hand, it has infinitely many resonances with the widths increasing logarithmically with respect to the resonance index \cite{EF06}; a natural idea is to employ them as a tool to expand the quantities of interest \cite{GMM95}. In order to express reduced evolution in the way described in Sec.~\ref{s: prelim} we need to know Green's function of the Hamiltonian $h_\alpha$ which can be obtained
from Krein's formula,
 $$
 (h_\alpha - k^2)^{-1}(r,r') = (h_0 - k^2)^{-1}(r,r') +
 \lambda(k) \Phi_k(r)\Phi_k(r')\,,
 $$
where $\Phi_k(r):= G_0(r,R)$ is the free Green function with one argument fixed, in particular, $\Phi_k(r) = \frac{1}{k}\,
\sin(kr)\, \e^{ikR}$ holds for $r<R$, and $\lambda(k)$ is
determined by $\delta$-interaction matching conditions at $r=R$; by a direct calculation one finds
 $$ 
\lambda(k) = -\frac{\alpha}{1 + \frac{i \alpha}{2 k}(1 -
\e^{2ikR})}\,.
 $$ 
Using it we can write the integral kernel of $\e^{-ih_\alpha t}$ as Fourier transformation, $u(t,r,r') = \int_0^\infty p(k,r,r') \e^{-ik^2t}\, 2 k\, \D k$, where the explicit form of the resolvent gives
 $$
 p(k,\,r,\,r') = \frac{2k\sin(kr)\sin(kr')}{\pi(2 k^2 +
 2\alpha^2 \sin^2 kR + 2 k \alpha \sin 2kR)}\,.
 $$
The resonances understood as poles of the resolvent continued to the lower halfplane appear in pairs, those in the fourth
quadrant, denoted as $k_n$ in the increasing order of their real parts, and $-\bar{k}_n$; we denote $S = \{k_n,\,-k_n,\, \bar{k}_n,\,-\bar{k}_n :\:n\in\mathbb{N}\}$. In the vicinity of $k_n$ the function $p(\cdot,r,r')$ can be written as
 $$
p(k,r,r') = \frac{i}{2 \pi} \frac{v_n(r) v_n(r')}{k^2 - k_n^2} + \chi(k,r,r')\,,
 $$
where $v_n(r)$ solves the differential equation $h_\alpha
v_n(r) = k_n^2 v_n(r) $ and $\chi$ is locally
analytic. It is not difficult to see that the function
$p(\cdot,r,r')$ decreases in every direction of the $k$-plane, hence it can be expressed as the sum over the pole singularities,
 $$
p(k,r,r') = \sum_{\tilde{k} \in S}\,
\frac{1}{k-\tilde{k}}\:
\mathrm{Res}_{\tilde{k}}\, p(k,r,r')
 $$
and the residue theorem implies $\sum_{\tilde{k} \in S}\, \mathrm{Res}_{\tilde{k}}\, p(k,r,r') = 0$. Using these relations and denoting $k_{-n} := -\bar{k}_{n}$ with $v_{-n}$ being the associated solution of the equation $H_\alpha v_{-n}(r) = k_{-n}^2 v_{-n}(r)$, we arrive after a short computation \cite{EF07} at
 $$
u(t,r,r') = \sum_{n \in \mathbb{Z}}\, M(k_n,t) v_n(r) v_n(r')
 $$
with $M(k_n,t) = \frac{1}{2}\, \e^{u_n^2}\, \mathrm{erfc}(u_n)$ and $u_n:= -\e^{-i\pi/4} k_n \sqrt{t}$, leading to the decay law
 $$
P_\psi(t) = \sum_{n,l} C_n \bar{C}_l I_{nl} M(k_n,t)
\overline{M(k_l,t)}
 $$
with $C_n := \int_0^R \phi(r,0) v_n(r)\, \D r$ and $I_{nl} := \int_0^R v_n(r) \bar{v}_l(r)\, \D r\,$; in our particular
case we have $v_n(r) = \sqrt{2} Q_n\sin(k_n r)$ with the coefficient $Q_n$ equal to
 $$
\left(\frac{-2i k_n^2}{2k_n + \alpha^2 R \sin 2 k_nR + \alpha\sin 2k_nR + 2 k_n \alpha R \cos 2k_nR}\right)^{1/2}.
 $$
These explicit formul{\ae} allow us to find $P_\psi(t)$ numerically. Let us quote an example worked out in \cite{EF07} in which $R=1$ and $\alpha = 500$; the initial wave function is chosen to be constant, i.e. the ground state of the \emph{Neumann} Laplacian in $L^2(B_R(0))$ which corresponds to $\phi(r,0) = R^{-3/2}\sqrt{3} r\, \chi_{[0,R]}(r)\,$.

\begin{figure}
\includegraphics[width=10cm,keepaspectratio]{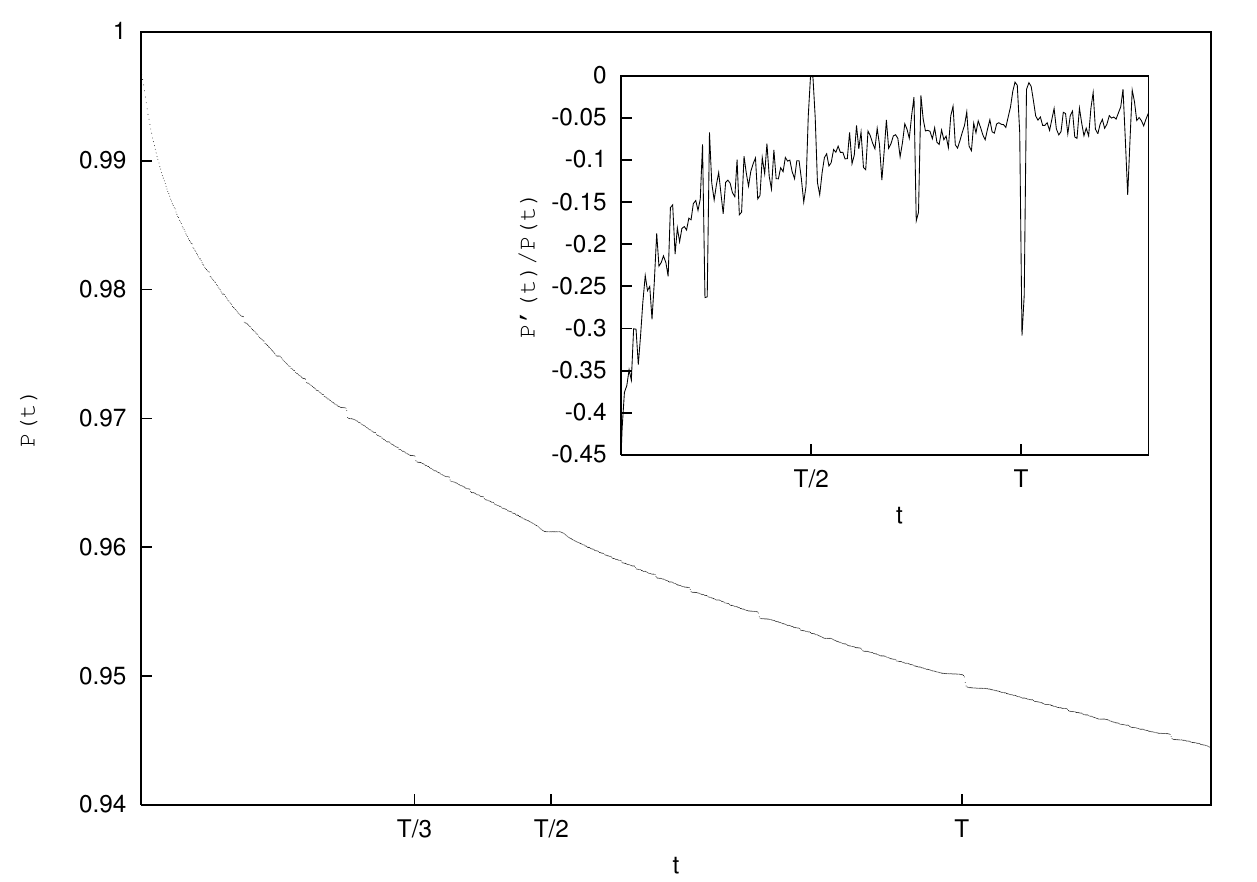}
\caption{\label{fig:1} Decay law for the initial state
$\phi(r,0) = R^{-3/2}\sqrt{3} r$; the inset shows its
logarithmic derivative averaged over intervals of the length
approximately $T/200$.}
\end{figure}

The respective decay law is plotted in Figure~\ref{fig:1}; we see that it is irregular having `steps', the most pronounced at the period $T=2R^2/\pi$ and its simple rational multiples. This is made even more visible from the plot of its logarithmic derivative (for numerical reasons it is locally smeared, otherwise the picture would be a fuzzy band). It is reasonable to conjecture that the function is in fact \emph{fractal}.

Let us add a few heuristic considerations in favor of this conjecture concerning the behavior of the derivative in the limit $\alpha \to\infty$. We can write the wave function as $\phi(r,t) \approx \sum_{n} c_n \exp(-i k_n^2 t) v_n(r)$ where resonance position expands for a fixed $n$ around $k_{n,0}:=n\pi/R$ as $k_n \approx k_{n,0} - k_{n,0} (\alpha R)^{-1}+ k_{n,0}(\alpha R)^{-2} -ik^2_{n,0}(\alpha^2 R)^{-1}$. In the leading order we have $v_n(r) \approx \sqrt{\frac{2}{R}} \sin(k_n r)$ and the substantial contribution to the expansion of $\phi(r,t)$ comes from terms with $ n \lesssim [\alpha^{1-\varepsilon} \frac{R}{\pi}]$ for some
$0<\varepsilon<1/3$.

The derivative of the decay law $P_{\psi,\alpha}(\cdot)$ can be identified with the probability current through the sphere, $\dot{P}_{\psi,\alpha}(t) = -2 \im (\phi'(R,t) \bar{\phi}(R,t))$. To use it we have to know the expressions on the right-hand side; using the above expansion we find
 $$
\phi(R,t) \approx \sqrt{\frac{2}{R}} \sum_{n=1}^{\infty} (-1)^n c_n \e^{-i k^2_{n,0} t\left(1-\frac{2}{\alpha R} \right)} \e^{-\frac{2k^3_{n,0}}{\alpha^2 R}t} \bigg(-\frac{k_{n,0}} {\alpha} - i\frac{k^2_{n,0}}{\alpha^2} \bigg)
 $$
and a similar expansion for $\phi'(R,t)$ with the last bracket
replaced by $k_{n,0}$. We observe that $\sum_{n=1}^{\infty} \exp\left(-2\frac{k^3_{n,0}}{\alpha^2 R}t \right) k^j_{n,0} \approx \frac{R}{\pi} \left(\frac{R}{2t} \right)^{(j+1)/3} \alpha^{2(j+1)/3} I_j$ holds for $j>-1$ where on the right-hand side we have denoted $I_j := \int_0^\infty \e^{-x^3}\,x^j\, \D x = \frac{1}{3} \Gamma\left(\frac{j+1}{3} \right)$.

Using this result we can argue that the decay law regularity depends on the asymptotic behavior of the coefficients $c_n$. Suppose for simplicity that it is power-like, $c_n = \OO(k^{-p}_{n,0})$ as $n\to\infty$. If the decay is fast enough, $p>1$, we find that $|\dot{P}_{\psi,\alpha}(t)| \leq \mathrm{const}\, \alpha^{4/3 - 4/3p} \to 0$ holds as $\alpha\to\infty$ uniformly in the time variable. The situation is different if the decay is slow, $p \leq 1$. Consider the example mentioned above leading to the decay law featured in Figure~\ref{fig:1} where $c_n = (-1)^{n+1} \frac{\sqrt{6}}{R k_n}$. Since the real parts of the resonance poles change with
$\alpha$, it is natural to look at the limit of $\dot{P}_{\psi, \alpha}(t_\alpha)$ as $\alpha\to\infty$ at the moving time
value $t_\alpha := t (1+2/\alpha R)$.

For irrational multiples of $T$ we use the fact \cite{BG88} that the modulus of $\sum_{n=1}^{L} \e^{i\pi n^2
t}$ is for an irrational $t$ bound by $C\, L^{1-\varepsilon}$
where $C, \varepsilon$ depend on $t$ only. In combination with the estimate, $\sum_{n=1}^\infty a_n b_n \leq
\sum_{n=1}^\infty |\sum_{j=1}^n a_j|\, |b_{n} - b_{n+1}|$ we find
 $$
 \sum_{n=1}^\infty
 \e^{-ik^2_{n,0}t}\: \e^{-\frac{2k^2_{n,0}}
 {\alpha^2 R}t} k^j_{n,0} \,\lesssim\, \mathrm{const}\,
 \alpha^{2/3(j+1-\varepsilon)}\,,
 $$
and consequently, $\dot{P}_{\psi, \alpha}(t_\alpha)\to 0$ as $\alpha\to\infty$. Assume next that $t = \frac{p}{q}\, T$ with $p,q\in\N$. If $pq$ is odd then $S_L(t):=\sum_{n=1}^{L} \e^{i\pi n^2 t}$ repeatedly retraces according to \cite{BG88} the same pattern, hence $\dot{P}_{\psi, \alpha}(t_\alpha)
\to 0$ --- an illustration can be seen in Fig.~\ref{fig:1} at the half period. On the other hand, for $pq$ even $|S_L(t)|$ grows linearly with $L$, and consequently, $\lim_{\alpha \to \infty} \dot{P}_{\psi, \alpha}(t_\alpha) >0$. For instance, a direct computation \cite{EF07} yields the value at the period, $\lim_{\alpha \to \infty} \dot{P}_{\psi, \alpha}(T_\alpha) = -\frac{4}{3 \sqrt{3}} \approx -0.77$.

As the last remark in this section, we note that there is a relation between a lack of local regularity of the decay law and the `anti-Zeno' property of Proposition~\ref{inf initial}; both occur if the energy distribution of the decaying state has a slow enough decay at high energies. The connection is no doubt worth of further exploration.

\section{Quantum graphs} \label{s:qgraphs}

Many quantum systems, both spontaneously emerging in Nature and resulting from an experimentalist's design, no doubt intelligent one, have complicated geometrical and topological structure which can be conveniently modeled as a graph to which the particle motion is confined. Such a concept was first developed for the purpose of quantum chemistry \cite{RS53}, however, it became a subject of intense investigation only at the end of the 1980's when tiny graph-like structures of semiconductor and other materials gained a prominent position in experimental physics. The literature on quantum graphs is vast at present; we limit ourselves with referring to the proceedings volume \cite{EKKST} as a guide for further reading.

Quantum graphs are usually rich in resonances; the reason, as we see below, is that their spectra often exhibit embedded eigenvalues which, as we know, are typically sources of resonance effects. Before we turn to the review let us briefly mention that while describing real-world quantum system through graphs is certainly an idealization, they can be approximated by more realistic `fat-graph' structures in a well-defined mathematical sense; from our point of view here it is important than such approximations also include convergence of resonances \cite{EP07}.

  \subsection{Basic notions} \label{ss:graphs}

As a preliminary, let us recall some basic notions about quantum graph models we shall need in the following. For the purpose of this review, a graph $\Gamma$ consists of a set of vertices $\mathcal{V}=\{\mathcal{X}_j: j\in I\}$, a set of finite edges\footnote{We assume here implicitly that any two vertices are connected by not more than a single edge and that the graph has no loops, which is possible to do without loss of generality since we are always able to insert `dummy' vertices into `superfluous' edges.} $\mathcal{L} =\{\mathcal{L}_{jn}:  (\mathcal{X}_j, \mathcal{X}_n) \in I_\mathcal{L} \subset I\times I\}$, and a set of infinite edges, sometimes also called \emph{leads}, ${\mathcal L_\infty} = \{\mathcal{L}_{j\infty}:  \mathcal{X}_j \in I_\mathcal{C}\}$ attached to them. We consider \emph{metric} graphs which means that each edge of $\Gamma$ is isomorphic to a line segment; the notions of finiteness or (semi)infiniteness refer to the length of those segments. As indicated, we regard $\Gamma$ as the configuration space of a quantum system with the Hilbert space
 $$
   \mathcal{H} = \bigoplus_{\mathcal{L}_{j} \in \mathcal{L}} L^2([0,l_{j}])\oplus
\bigoplus_{\mathcal{L}_{j\infty} \in \mathcal{L_\infty}}
L^2([0,\infty))\,,
 $$
the elements of which are columns $\Psi = (\{f_{j}:
\mathcal{L}_j \in \mathcal{L}\},\, \{g_{j}: \mathcal{L}_{j\infty}\in \mathcal{L}_\infty\} )^\mathrm{T}$.

For most part of this section we will suppose that the motion on the graph edges is free, i.e. governed by the Hamiltonian which acts there as $-\frac{\mathrm{d}^2}{\mathrm{d} x^2}$ with respect to the arc-length variable parametrizing the particular edge. In order to make it a self-adjoint operator, we have to impose appropriate boundary conditions which couple the wave functions at the graph vertices. One of the possible general forms of such conditions \cite{GG, Ha00, KS00} is
 \begin{equation}
   (U_j -I) \Psi_j +i (U_j +I) \Psi_j' = 0\,,\label{rat-coup1}
 \end{equation}
where $U_j$ are unitary matrices, and $\Psi_j$ and $\Psi_j'$ are vectors of the functional values and of the (outward) derivatives at the vertex $\mathcal{X}_j$; in other words, the domain of the Hamiltonian consists of all functions on $\Gamma$ which are locally $W^{2,2}$ and satisfy
conditions (\ref{rat-coup1}). Note that coupling is local connecting boundary values in $\mathcal{X}_j$ only.

\begin{figure}
   \begin{center}
     \includegraphics{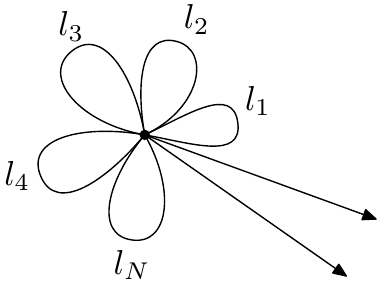}
     \caption{The model $\Gamma_0$ for a quantum graph $\Gamma$ with $N$ internal finite edges and $M$ external leads}\label{flower}
   \end{center}
\end{figure}

Since handling Hamiltonians of graphs with a complicated topology may be cumbersome, one can employ a trick proposed in \cite{Ku08} replacing $\Gamma$ with the graph $\Gamma_0$ in which all edge ends meet in a single vertex as sketched in Fig.~\ref{flower}; the actual topology of $\Gamma$ will be then encoded into the matrix which describes the coupling in the vertex. Denoting $N=\sharp\,\mathcal{L}$ and $M=\sharp\, \mathcal{L_\infty}$ we introduce the $(2N+M)$-di\-men\-sional vector of functional values by $\Psi = (\Psi_1^T, \dots ,\Psi_{\sharp\,\mathcal{V}}^T)^T$ and similarly the vector of derivatives $\Psi'$ at the vertex; the conditions (\ref{rat-coup1}) can be concisely rewritten as coupling on $\Gamma_0$ characterized by $(2 N+ M)\times (2 N+ M)$
unitary block-diagonal matrix $U$, consisting of the blocks $U_j$, in the form
 \begin{equation}
   (U -I) \Psi +i (U+I) \Psi' = 0\,;\label{rat-coup2}
 \end{equation}
it is obvious that one can treat the replacement as a unitary equivalence which does not alter spectral properties and preserves the system resonances (if there are any).

  \subsection{Equivalence of resonance notions}

In Section~\ref{s: prelim} we made it clear how important it is to establish connection between different objects labeled as resonances. Let us look now how this question looks like in the quantum graph setting. Let us begin with the resolvent resonances. One can write the resolvent of the graph Hamiltonian \cite{Pa10}, however, it is sufficient to inspect the spectral condition encoded in it and its behavior in the complex plane.

We employ an external complex scaling in which the external part are the semi-infinite leads where the functions are scaled as $g_{j\theta}(x) = \e^{\theta/2} g_j(x \e^\theta)$ with an imaginary~$\theta$; as usual this rotates the essential spectrum of the transformed (non-selfadjoint) Hamiltonian into the lower complex halfplane and reveals the second-sheet poles. In particular the `exterior' boundary values, to be inserted into (\ref{rat-coup2}), can be for $g_j(x)= c_j\,\e^{ikx}$ written as $g_j (0) = \mathrm{e}^{-\theta/2} g_{j\theta}$ and $g_j' (0) = ik \mathrm{e}^{-\theta/2} g_{j\theta}$ with an appropriate $g_{j\theta}$. On the other hand, the internal part of the graph is left unscaled. Choosing the solution on the $j$-th edge in the form $f_j (x) = a_j \sin{kx} + b_j \cos{kx}$ we easily find its boundary values; for $x=0$ it is trivial, for $x=l_j$ we use the standard transfer matrix. This allows us to express both $(f_j(0),f_j(l_j))^\mathrm{T}$ and $(f'_j(0),-f'_j(l_j))^\mathrm{T}$ through the coefficients $a_j,b_j$, cf.~\cite{EL10} and note the sign of $f'_j(l_j)$ reflecting the fact that the derivatives entering (\ref{rat-coup2}) are outward ones. Inserting these boundary values into the coupling condition we arrive at the system
 \begin{equation}
   (U-I)C_1(k)\left(\begin{array}{c}a_1\\b_1\\a_2\\\vdots\\b_N\\
   \mathrm{e}^{-\theta/2}g_{1\theta}\\\vdots\\
   \mathrm{e}^{-\theta/2}g_{M\theta}\end{array}\right)
   +ik(U+I)C_2(k)\left(\begin{array}{c}a_1\\b_1\\a_2\\\vdots\\b_N\\
   \mathrm{e}^{-\theta/2}g_{1\theta}\\\vdots\\
   \mathrm{e}^{-\theta/2}g_{M\theta}\end{array}\right)=0,\label{rat-scal}
 \end{equation}
where we have set $C_1 (k) = \mathrm{diag\,} (C_1^{(1)}(k), C_1^{(2)}(k), \dots,C_1^{(N)}(k), I_{M\times M})$ and $C_2  = \mathrm{diag\,}(C_2^{(1)}(k),C_2^{(2)}(k),\dots,C_2^{(N)}(k), i I_{M\times M})$ with
 $$
   C_1^{(j)} (k) =
   \left(\begin{array}{cc}
   0 & 1 \\
   \sin{kl_j} & \cos{kl_j}\end{array}\right)\,,\quad
   C_2^{(j)} (k) =
   \left(\begin{array}{cc}
   1 & 0  \\
   -\cos{kl_j} & \sin{kl_j}\end{array}\right)\,,
 $$
and $I_{M\times M}$ being the $M\times M$ unit matrix. The solvability condition of the system (\ref{rat-scal})
determines eigenvalues of the scaled non-selfadjoint operator, and \emph{mutatis mutandis}, poles of the analytically continued resolvent of the original Hamiltonian.

Looking at the same system from the scattering point of view we use the same solution as above on the internal edges while on the leads we take appropriate combinations of two planar
waves, $g_j = c_j \mathrm{e}^{-ikx}+ d_j \mathrm{e}^{ikx}$. We look for the on-shell S-matrix $S=S(k)$ which maps the vector of amplitudes of the incoming waves $c=\{c_n\}$ into the vector of amplitudes of the outgoing waves $d =\{d_n\}$, and ask about its complex singularities, $\mathrm{det}\,S^{-1}=0$. This leads to the system
 $$
   (U-I)C_1(k)\left(\begin{array}{c}a_1\\b_1\\a_2\\
     \vdots\\b_N\\c_1+d_1\\\vdots\\ c_M+d_M\end{array}\right)
   +ik(U+I)C_2(k)\left(\begin{array}{c}a_1\\b_1\\a_2\\
     \vdots\\b_N\\d_1-c_1\\\vdots\\ d_M-c_M\end{array}\right)=0\,;
 $$
eliminating the variables $a_j$, $b_j$ one can rewrite it as a
system of $M$ equations expressing the map $S^{-1}d =c$. The
condition under which the latter is not solvable, which is equivalent to our original question since $S$ is unitary, reads
 \begin{equation}
   \mathrm{det}\,\left[(U-I)\,C_1(k)+ik (U+I)\,C_2(k)\right]=0\,, \label{rat-res}
 \end{equation}
however, this is nothing else than the condition of solvability of the system (\ref{rat-scal}). Thus we can make the following conclusion \cite{EL10}.

\begin{theorem} \label{graph_res_equiv}
The notions of resolvent and scattering resonances coincide for quantum graph Hamiltonians described above.
\end{theorem}

Before proceeding further, let us mention one more way in which the resonance problem on a graph can be reformulated. To this purpose we rearrange the matrix $U$ permuting its rows and columns into the form $U= {U_1 \;\; U_2 \choose U_3 \;\; U4}$, where $U_1$ is the $2N\times 2N$ square matrix referring to the compact subgraph, $U_4$ is the $M\times M$ square matrix related to the exterior part, and $U_2$ and $U_3$
are rectangular matrices of the size $M \times 2N$ and $2N \times M$, respectively, connecting the two. The system (\ref{rat-scal}) can be then rewritten by elimination of the lead variables \cite{EL10} as
 \begin{equation} \label{eff-res-condition}
   (\tilde U(k)-I)F + i (\tilde U(k)+I)F' = 0\,,
 \end{equation}
where $F:=(f_1,\dots,f_{2N})^\mathrm{T}$, and similarly for $F'$, are the internal boundary values, and the effective coupling matrix appearing in this condition is given by
 \begin{equation} 
   \tilde U(k)=U_1 -(1-k) U_2 [(1-k) U_4 - (k+1) I]^{-1} U_3 \,. \label{rat-efu}
 \end{equation}
In other words, we have been able to cast the problem into the form of spectral question for the compact core of the graph with the effective coupling replacing the leads by the non-unitary and energy-dependent matrix (\ref{rat-efu}).

  \subsection{Line with a stub} \label{s:stub}

Next we will present several simple examples. In the first one $\Gamma$ is a line to which a segment is attached at the point $x=0$. The Hilbert space is thus $\HH:= L^2(\R)\oplus L^2(0,l)$ and we write its elements as columns $\psi= {f \choose u}$. To make the problem more interesting we suppose that the particle on the stub is exposed to a potential; the Hamiltonian acts
   $$ 
   (H\psi)_1(x)=-f''(x)\,, \quad (H\psi)_2(x)=(-u''+Vu)(x)\,,
   $$ 
outside the junction, where $V\in L^1_\mathrm{loc}(0,l)$ having finite limits at both endpoints of the segment so --- if the domain consists of functions vanishing in the vicinity of the junction --- the corresponding deficiency indices are $(3,3)$. The admissible Hamiltonians will be identified with self-adjoint extensions which, as before, can be conveniently characterized by appropriate boundary conditions. We will not explore all of them and restrict our attention to a subclass of those having the line component of the wave function continuous at the the junction, namely
   \begin{eqnarray} \label{stub-bc}
f(0+) = f(0-) =: f(0)\,, && u(0) = bf(0)+cu'(0) \,, \nonumber \\[-.7em] \\[-.7em]
f'(0+)-f'(0-) = df(0)-bu'(0) \,, &&
u(\ell) = 0 \,; \nonumber
   \end{eqnarray}
at the free end of the stub we fix Dirichlet condition. The coefficient matrix $\KK= { b \;\; c \choose d \; -b}$ is real; we restrict our attention to time-reversal invariant dynamics. The operator specified by the conditions (\ref{stub-bc}) will be denoted as $H_\KK$. The parameter $b$ controls the coupling; if $b=0$ the graph decomposes into the line with a point interaction at $x=0$ and the stub supporting Schr\"odinger operator $h_c:= -{\D^2\over \D x^2} +V$  with Robin condition $u(0)=cu'(0)$ at the junction referring again to $x=0$.

Let us begin with the scattering. To find the on-shell S-matrix we use the standard Ansatz, $f(x)=\e^{ikx}+r\,\e^{-ikx}$ and $t\,\e^{ikx}$ on the line for $x<0$ and $x>0$, respectively, while on the stub we take $u(x)= \beta u_l(x)$ where $u_l$ is a solution to $-u''+Vu= k^2u$ corresponding to the boundary conditions $u_l(l)=0$, unique up to a multiplicative constant. Using the coupling conditions (\ref{stub-bc}) we find
   $$
t(k) = {-2ik(cu'_\ell-u_\ell)(0) \over 2ik D(k)}\,, \quad r(k) = -\,{b^2u'_\ell(0)+d(cu'_\ell-u_\ell)(0) \over 2ik D(k)}\,;
  $$
where $2ik D(k):= b^2u'_\ell(0)+ (d-2ik)(cu'_\ell-u_\ell)(0)$; it is easy to check that these amplitudes satisfy $|t(k)|^2+ |r(k)|^2 =1$. We note that $H_\KK$ can have also an isolated eigenvalue; this happens if $D(i\kappa)=0$ with $\kappa>0$.
If $b=0$ such an eigenvalue exists provided $d<0$ and equals $-\,{1\over 4}d^2\,$; it remains isolated for $|b|$ small enough.

It is also not difficult to find the resolvent of $H_\KK$. The tool is as usual Krein's formula; we choose for comparison $H_{\KK_0}$ corresponding to $\KK_0=0$. In that case the operator decomposes, the kernel of line part being $G_1(x,y;z)={i\over 2k}\, \e^{ik|x-y|}$ where $k:=\sqrt z$ as usual; the stub part is $-\,{u_0(x_<) u_\ell(x_>)\over W(u_0,u_l)}$, where $u_l$ has been introduced above, $u_0$ is similarly a solution corresponding to the condition $u_0(0)=cu'_0(0)$, and $W(u_0,u_l)\,$ is the Wronskian of the two functions. The sought kernel then equals
 $$ 
(H_\KK-z)^{-1}(x,y)= (H_{\KK_0}-z)^{-1}(x,y)+ \sum_{j=1,2}\,
\lambda_{jk}(k) F_j(x)F_k(y)\,,
 $$ 
where the vectors $F_j$ can be chosen as $F_1(x):= {R_1(x,0) \choose 0}$ and $F_2(x):= {0 \choose u_l(x)}$; note that the stub part vanishes at $x=0$. The coefficients are obtained from the requirement that the resolvent must map any vector of $\HH$ into the domain of $H_\KK$; a straightforward computation \cite{ES94} gives
 $$
 \lambda_{11}(k) = \frac{b^2u'_\ell(0)+
d(cu'_\ell-u_\ell)(0)}{D(k)}\,, \quad \lambda_{22}(k) = {u_\ell(0)^{-1} \over D(k)}\,
\left( c+\,i\, {cd+b^2\over 2k} \right)\,,
 $$
together with $\lambda_{12}(k) = \lambda_{21}(k) = b D(k)^{-1}$. We see, in particular, that the coefficient denominator zeros in the complex plane coincide with those of the on-shell S-matrix as we expect based on Theorem~\ref{graph_res_equiv} proved above.

In the decoupled case, $b=0$, the expression for $D(k)$ factorizes giving rise to eigenvalues of the operator $h_c$ introduced above which are embedded in the continuous spectrum of the line Hamiltonian; the coupling turns them generally into resonances. In the case case of weak coupling, i.e. for small $|b|$ one can solve the condition $D(k)=0$ perturbatively arriving at the following conclusion \cite{ES94}.

 \begin{proposition}
Let $k_n$ refer to the $n$-th eigenvalue of $h_c$ and denote by $\chi_n$ the corresponding normalized eigenfunction; then for all sufficiently small $|b|$ there is a unique resolvent pole in the vicinity of $k_n$ given by
 $$ 
k_n(b)= k_n-\, {ib^2\chi'_n(0)^2 \over 2k_n(2k_n+id)} \,+ \OO(b^4)\,.
 $$ 
 \end{proposition}

This gives, in particular, the inverse value of resonance lifetime in the weak-coupling case, $\im z_n(b) = ib^2\chi'_n(0)^2 (2k_n+id)^{-1} + \OO(b^4)$. The simple form of the condition $D(k)=0$ allows us, however, to go beyond the weak coupling and to trace numerically the pole trajectories as the coupling constant $b$ runs over the reals. Examples are worked out in \cite{ES94} but we will not describe them here and limit ourselves with mentioning an important particular situation.

It concerns the case when the motion in the stub is free and the decoupled operator is specified by the Dirichlet boundary condition, i.e. $V=0$ and $c=d=0$. The condition $D(k)=0$ can be then solved analytically. Indeed, the embedded eigenvalues are $k_n^2$ with $k_n:= {n\pi\over\ell}$ and the equation reduces to $\tan k\ell =-{i\beta^2\over 2}$ solved by
 \begin{equation} \label{stubfreefall}
k_n(b) = \left\{ \begin{array}{lcl} {n\pi\over\ell}+ {i\over 2l}\, \ln {2-b^2\over 2+b^2} &\;\dots\; & |b|<\sqrt 2 \\[.5em]
{(2n-1)\pi\over 2\ell}+ {i\over 2l}\, \ln
{b^2-2\over b^2+2} &\;\dots\; & |b|>\sqrt 2 \end{array} \right.
 \end{equation}
Hence the poles move with the increasing $|b|$ vertically down in the $k$-plane and for $|b|>\sqrt 2$ they ascend, again vertically, returning to eigenvalues of Neumann version of $h_c$ as $|b|\to \infty$. An important conclusion from this example is that poles may disappear to infinite distance from the real axis and a quantum graph may have no resonances at all, as it happens here for $|b|=\sqrt 2$.

  \subsection{Regeneration in decay: a lasso graph} \label{s:lasso}

Let us next describe another simple example, now with a lasso-shaped $\Gamma$ consisting of a circular loop of radius
$R$ to which a halfline lead is attached. This time we shall suppose that the particle is charged and the graph is placed into a homogeneous magnetic field of intensity $B$ perpendicular to the loop plane\footnote{The assumptions of homogeneity and field direction are here for simplicity only, in fact the only thing which matters in the model is the magnetic flux through the loop.}. The vector potential can be then chosen tangent to the loop with the modulus $A= {1\over 2}\,BR= {\Phi\over L}$, where $\Phi$ is the flux through the loop and $L$ is its perimeter. With the convention we use, $e=c=2m=\hbar=1$, the natural flux unit is $\frac{hc}{e}= 2\pi$, so we can also write $A=\frac\phi R$ where $\phi$ is the flux value in these units. The Hilbert space of the lasso-graph model is $\HH:= L^2(0,L)\oplus L^2(\R^+)$; the wave functions are written as columns, $\psi= {u \choose f}$.

To construct the Hamiltonian we begin with the operator describing the free motion on the loop and the lead under the assumption that the graph vertex is `fully
disconnected', in other words $H_{\infty}= H_\mathrm{loop}(B) \oplus H_\mathrm{halfline}$, where
   $$ 
H_\mathrm{loop}(B)= \Big(-i\frac{\D}{\D x}+A \Big)^2\,, \quad H_\mathrm{halfline}= -\frac{\D^2}{\D x^2}
   $$ 
with Dirichlet condition, $u(0)=u(L)=f(0)=0$ at the junction. The spectrum of $H_\mathrm{loop}$ is discrete of multiplicity two; the eigenfunctions $\chi_n(x)= {\e^{-iAx}\over\sqrt{\pi R}}\, \sin\left( nx\over 2R\right)$ with $n=1,2,\dots\,$ correspond to the eigenvalues $\left(n\over 2R\right)^2$ which are embedded into the continuous spectrum of $H_\mathrm{halfline}$ covering the interval $[0,\infty)$; note that the effect of the magnetic field on the \emph{disconnected} loop amounts to a unitary equivalence, $H_\mathrm{loop}(B)= U_{-A} H_\mathrm{loop}(0) U_A$ where $(U_A u)(x):= \e^{iAx} u(x)$.

Restricting the domain of $H_\infty$ to functions vanishing in the vicinity of the junction we get a symmetric operator with deficiency indices $(3,3)$. We are going to consider a subclass of its self-adjoint extensions analogous to (\ref{stub-bc}) characterized by three real parameters; the Hamiltonian will act as
   $$ 
H_{\alpha,\mu,\omega}(B) {u \choose f} = {-u''\!-2iAu'\!+A^2u \choose -f''}
   $$ 
on functions from $W^{2,2}_\mathrm{loc}(\Gamma)$ continuous on the loop, $u(0)= u(L)$, which satisfy
   \begin{equation} \label{lasso_bc}
f(0) = \omega u(0)+ \mu f'(0)\,, \quad u'(0)-u'(L) = \alpha u(0)- \omega f'(0)\,,
   \end{equation}
for some $\alpha,\mu,\omega\in\R$ the latter being the coupling constant. This includes a particular case of \emph{$\delta$-coupling} corresponding to $\mu=0$ and
$\omega=1$ in which case the wave functions are fully continuous,
   \begin{equation} \label{lasso_delta}
u(0)= u(L)= f(0)\,, \quad u'(0)-u'(L)+f'(0)= \alpha f(0)\,;
   \end{equation}
in the fully decoupled case we have $\alpha=\infty$ as the notation suggests. For simplicity we will write $H_{\alpha,0,1}=H_\alpha$. Note that in general the vector potential enters the coupling conditions \cite{KS03} but here the outward tangent components of $\vec A$ at the junction have opposite signs so their contributions cancel mutually.

Let us start again with scattering, i.e. the reflection of the particle traveling along the halfline from the
magnetic-loop end. To find the generalized eigenvectors, $H_{\alpha,\mu,\omega}(B)\psi= k^2\psi$, we use $u(x)= \beta\, \e^{-iAx} \sin(kx+\gamma)$ and $f(x)= \e^{-ikx}\! +r\,\e^{ikx}$ as the Ansatz; using the coupling conditions (\ref{lasso_bc}) we get after a simple algebra
   $$ 
r(k)= -\,{(1+ik\mu)\left\lbrack \alpha- {2k\over \sin
kL}\,(\cos\Phi-\cos kL)\, \right\rbrack +i\omega^2k  \over
(1-ik\mu)\left\lbrack \alpha- {2k\over \sin
kL}\,(\cos\Phi-\cos kL)\, \right\rbrack -i\omega^2k}
   $$ 
for the reflection amplitude. The Hamiltonian $H_{\alpha,\mu,\omega}(B)$ can have also isolated eigenvalues but we shall skip this effect referring to \cite{Ex97}. On the other hand, it is important to mention that there may exist \emph{positive} eigenvalues embedded in the continuous spectrum even if $\omega\ne 0$. In view of (\ref{lasso_bc}) it is possible if $u(0)= u'(0)\!-u'(L)=0$, hence such bound states exist only at integer/halfinteger values of the magnetic flux (in the natural units) and the corresponding eigenfunctions are the $\chi_n$'s mentioned above with even $n$ for $\phi$ integer and odd $n$ for $\phi$ halfinteger.

Next we find the resolvent of $H_{\alpha,\mu,\omega}(B)$ using again Krein's formula to compare it to that of $H_\infty$ with the kernel $\:\mathrm{diag}\,\big(\e^{-iA(x-y)}\, {\sin
kx_< \sin k(x_>-L) \over k\,\sin kL},\,{\sin kx_<\, \e^{ikx_>} \over k}\big)$. The sought resolvent kernel can be then written as
   $$ 
G_{\alpha,\mu,\omega}(x,y;k)= G_{\infty}(x,y;k)+
\sum_{j,\ell=1}^2 \lambda_{j\ell}(k) F_j(x)F_{\ell}(y)\,,
   $$ 
where the deficiency subspaces involved are chosen in the form
   $$ 
F_1(x):= {w(x) \choose 0}\,, \qquad F_2(x):= { 0 \choose e^{ikx} }
   $$ 
with $w(x) := e^{iAx}\, {e^{-i\Phi} \sin kx -\sin k(x\!-\!L) \over \sin kL}$ and the coefficients $\lambda_{j\ell}(k)$ given by \cite{Ex97}
   $$ 
\lambda_{11} = -\, {1-i\mu k\over D(k)}\,, \quad \lambda_{22} = {\mu \left\lbrack 2k\, {\cos\Phi-\cos kL\over \sin kL}\,-\alpha
\right\rbrack - \omega^2 \over D(k)} \nonumber
   $$ 
together with $\lambda_{12}=\lambda_{21} = -{\omega\over D(k)}$, where
   $$ 
D(k)\equiv D(\alpha,\mu\,\omega;k):= (1-i\mu k) \left\lbrack
2k\, {\cos\Phi-\cos kL\over \sin kL}\,-\alpha \right\rbrack
-i\omega^2 k\,.
   $$ 
In the case of a $\delta$-coupling, in particular, the
coefficients acquire a simple form, $\,\lambda_{jl}(k)=
-D(k)^{-1}, \: j,l=1,2\,$. As expected, the denominator $D(k)$ determining the singularities is the same as for the on-shell S-matrix. A simple form of the condition $D(k)=0$ allows us to follow the pole trajectories, both with respect to the coupling parameters and the flux $\Phi$. At the same time, knowing the resolvent of $H_{\alpha,\mu,\omega}(B)$ we can express the decay law for states supported at the initial moment $t=0$ on the loop only; we will not go into details and refer the reader to \cite{Ex97} where the appropriate formul{\ae} and graphs are worked out.

Let us just mention one amusing feature of this model which can be regarded as an analogue of the effect known in particle physics as \emph{regeneration} in decay of neutral kaons and illustrates that intuition may misguide you when dealing with quantum systems.  Consider the lasso graph with the initial wave function $u$ on the loop such that $x\mapsto\e^{iAx}u(x)$ has no definite symmetry with respect to the connection point $x=0$. If the flux value $\phi$ is integer, the $A$-even component represents a superposition of embedded-eigenvalue bound states mentioned above, thus it survives, while the $A$-odd one dies out. Suppose that after a sufficiently long time we decouple the lead and attach it at a different point (or we may have a loop with two leads which may be switched on and off independently). For the decay of the surviving state the symmetry with respect to the new junction is important; from this point of view it is again a superposition of an $A$-even and an $A$-odd part, possibly even with same weights if the distance between the two junctions is $\frac14 L$.

  \subsection{Resonances from rationality violation}

The above simple examples illustrated that resonances are a frequent phenomenon in quantum graph models. To underline this point we shall describe in this section another mechanism giving rise to resonances, this time without need to change the coupling parameters. The observation behind this claim is that quantum-graph Hamiltonians may have embedded eigenvalues even if no edges are disconnected which is related to the fact that the unique continuation principle is generally not valid here and one can have compactly supported eigenfunctions. Indeed, eigenfunctions of a graph Laplacian are trigonometric functions, hence it may happen that the graph has a loop and the vertices on it have rationally related distances such that the eigenfunction has zeros there and the rest of the graph `does not know' about it. Let us present briefly two such examples referring to \cite{EL10} for more details.

\smallskip

\subsubsection{A loop with two leads} \label{loop2leads}

In this case $\Gamma$ consists of two internal edges of lengths~$l_1,\,l_2$ and one halfline attached at each of their endpoints, corresponding to the Hilbert space is $L^2(\mathbb{R}^+)\oplus L^2(\mathbb{R}^+)\oplus
L^2([0,l_1]) \oplus L^2([0,l_2])$; states of the system are
correspondingly described by columns $\psi=(g_1,g_2,f_1, f_2)^\mathrm{T}$. The Hamiltonian $H$ is supposed to act as negative Laplacian, $\psi \mapsto -\psi''$, separately on each edge. We consider coupling analogous to (\ref{stub-bc}) assuming that the functions of $\mathrm{Dom\,}H$ are continuous on the loop, $f_1(0)=f_2(0)$ and $f_1(l_1)=f_2(l_2)$, and satisfy
 \begin{eqnarray*}
    f_1(0) = \alpha_1^{-1} (f_1'(0)+f_2'(0))  + \gamma_1  g_1'(0)\,,&\!\!\!\!&
    f_1(l_1) = -\alpha_2^{-1} (f_1'(l_1)+f_2'(l_2))  + \gamma_2  g_2'(0)\,,\\
    g_1(0) = \gamma_1 (f_1'(0)+f_2'(0))  + \tilde\alpha_1^{-1}  g_1'(0)\,,&\!\!\!\!&
    g_2(0) = -\gamma_2 (f_1'(l_1)+f_2'(l_2))  + \tilde\alpha_2^{-1}   g_2'(0)\,,
 \end{eqnarray*}
for some $\alpha_j,\, \tilde\alpha_j,\, \gamma_j\in\R$. Since we want to examine behavior of the model with respect to the lengths of internal edges, let us parametrize them as $l_1 = l(1-\lambda),\: l_2 = l(1+\lambda)$ with $\lambda \in [0,1]$; changing $\lambda$ thus effectively means moving one of the connections points around the loop from the antipolar position for $\lambda=0$ to merging of the two vertices for $\lambda=1$. Due to the presence of the semi-infinite leads the essential (continuous) spectrum of $H$ is $[0,\infty)$. If we consider the loop itself, it has a discrete spectrum consisting of eigenvalues $k_n^2$ where $k_n= \frac{\pi n}{l}$ with $n\in\Z$. The corresponding eigenfunctions have nodes spaced by $\frac ln$ for $n\ne 0$, hence $H$ has embedded eigenvalues if the leads are attached to the loop at some of them.

\begin{figure}
   \begin{center}
     \includegraphics[width=8cm]{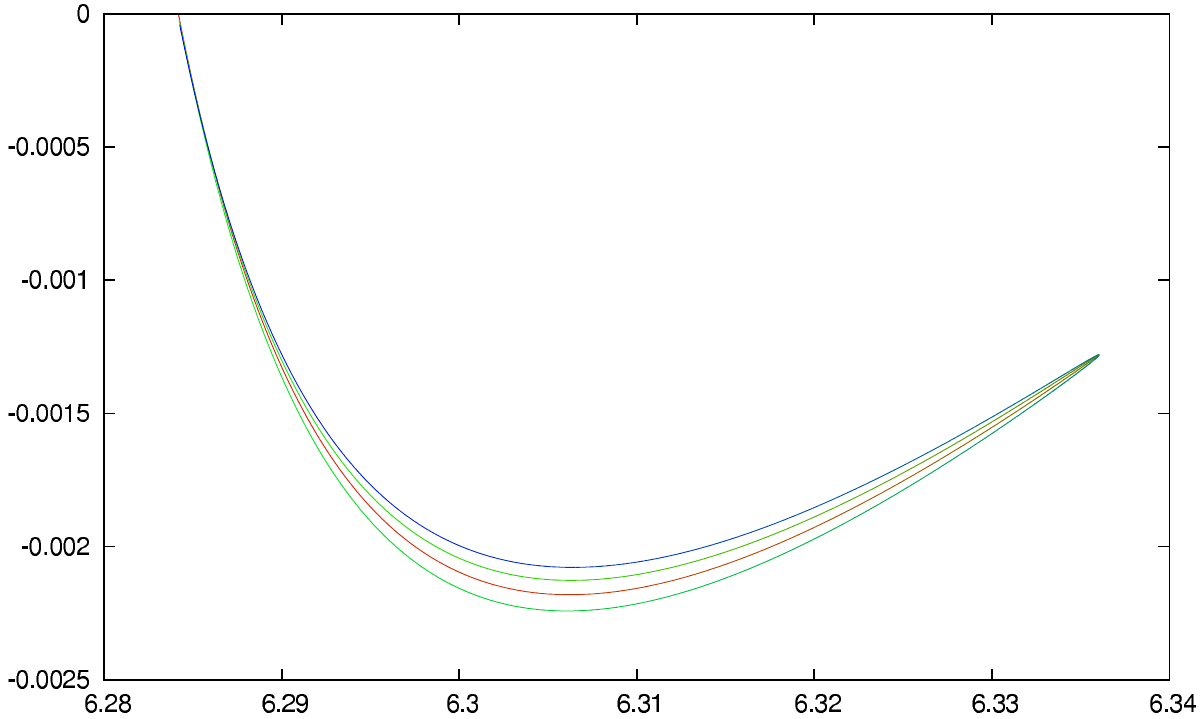}
     \caption{Trajectory of the resonance pole in the momentum
     plane starting from $k_0 = 2\pi$ corresponding to $\lambda=0$ for $l=1$ and the coefficients
     values $\alpha_1^{-1}=1$, $\tilde\alpha_1^{-1}=-2$, $\gamma_1=1$,
     $\alpha_2^{-1}=0$, $\tilde\alpha_2^{-1}=1$, $\gamma_2=1$, $n=2$.}
     \label{fig:looptrajectory}
   \end{center}
\end{figure}

If the rationality of the junction distances is violated these eigenvalues turn into resonances. The condition determining the singularities can be found in the same way as in the previous examples; it reads
 $$
   \sin{kl(1-\lambda)}\sin{kl(1+\lambda)}
   -\frac{4k^2}{\beta_1(k)\beta_2(k)}\sin^2{kl}
   +k\Big[\frac1{\beta_1(k)}+\frac1{\beta_2(k)}\Big]\sin{2kl}=0\,, \label{1-con}
 $$
where $\beta_i^{-1}(k) := \alpha_{i}^{-1}+ \frac{ik |\gamma_i|^2
}{1-ik \tilde\alpha_i^{-1}}$. One can solve it perturbatively but also to find numerically its solution describing pole trajectories as $\lambda$ runs through $[0,1].$ The analysis presented in \cite{EL10} shows that various situations may occur, for instance, a pole returning to the real axis after one or more loops in the complex plane --- an example is shown in Fig.~\ref{fig:looptrajectory} --- or a trajectory ending up in the lower halfplane at the endpoint of the parameter interval.

\smallskip

\subsubsection{A cross-shaped graph}

\begin{figure}
   \begin{center}
     \includegraphics[width=8cm]{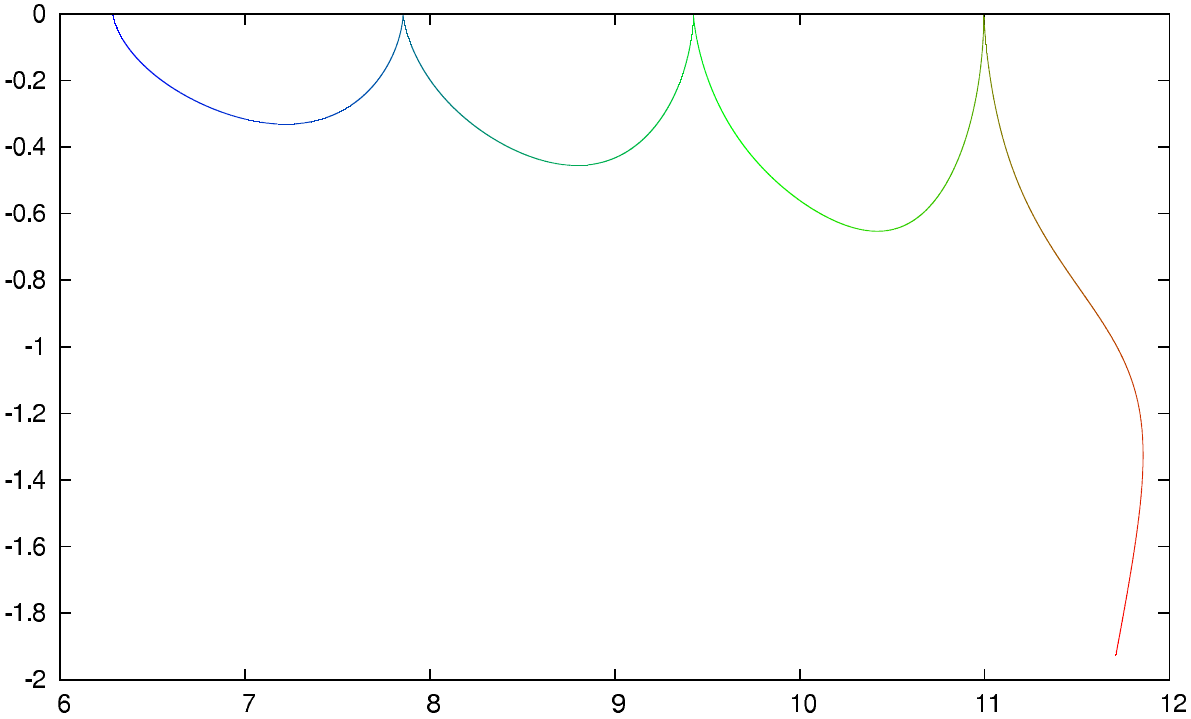}
     \caption{Resonance pole trajectory for $\alpha = 1$ and $n=2$.}
     \label{crosstrajectory}
   \end{center}
\end{figure}

We add one more simple example to illustrate that the same effect may occur even if the graph has no loops. Consider a cross-shaped $\Gamma$ consisting of two leads and two internal edges attached to the leads at one point; the lengths of the internal edges will be $l_1 = l(1-\lambda)$ and $l_2=l(1+\lambda)$. The Hamiltonian acts again as $-\mathrm{d}^2/\mathrm{d}x^2$ on the corresponding Hilbert space $L^2(\mathbb{R}^+)\oplus L^2(\mathbb{R}^+)\oplus L^2([0,l_1]) \oplus L^2([0,l_2])$ the elements of which are described by columns $\psi=(g_1,g_2,f_1,f_2)^\mathrm{T}$. For simplicity we restrict our attention to the $\delta$ coupling at the vertex and Dirichlet conditions at the loose ends, i.e. $f_1(0)=f_2(0)=g_1(0)=g_2(0)$ and $f_1(l_1)=f_2(l_2)=0$ together with the requirement
 $$
 f_1'(0)+f_2'(0)+ g_1'(0)+ g_2'(0) = \alpha f_1(0)
 $$
for $\alpha\in\R$. In the same way as above we can derive resonance condition in the form $k\sin{2kl}
+(\alpha-2ik)\sin{kl(1-\lambda)} \sin{kl(1+\lambda)}=0$, or equivalently
 $$
    2k\sin{2kl}+(\alpha-2ik)
    (\cos{2kl\lambda-\cos{2kl}})=0\,.\label{1-condresonator}
 $$
Asking when the solution is real we note that this happens if the real and imaginary parts of the left-hand side vanish. We find easily that it is the case if $\lambda = 1-2m/n\,$, $\:\mathbb{N}_0\ni m\leq n/2$, while if this rationality relation is violated the poles move into the lower halfplane. The condition can be again solved numerically giving pole trajectories for various parameter values; in addition to the possibilities mentioned above we can have trajectories returning to \emph{different} embedded eigenvalues --- an example shown in Fig.~\ref{crosstrajectory} calls to mind the effect of \emph{quantum anholonomy} \cite{Ch98} --- as well as avoided trajectory crossings, etc., see \cite{EL10} for more details.

\smallskip

\subsubsection{Local multiplicity preservation}

Let us turn from examples to the general case and consider an eigenvalue $k_0^2$ with multiplicity $d$ of a quantum-graph Hamiltonian $H$ which is embedded in the continuous spectrum due to rationality relations between the edges of $\Gamma$. We consider graphs $\Gamma_\varepsilon$ with modified edge lengths $l_j' = l_0 (n_j + \varepsilon_j)$ assuming that $n_j \in \mathbb{N}$ for $j \in \{1,\dots , n\}$ while $n_j$ may not be an integer for $j \in \{n+1,\dots ,N\}$ where $N:=\sharp \mathcal{L}$. The analysis of the perturbation is a bit involved, see \cite{EL10} for details, leading to the following conclusion.

 \begin{theorem}
Let $\Gamma$ be a quantum graph with $N$ finite edges of the
lengths $l_j$, $\:M$ infinite edges, and the coupling described by the matrix $U= {U_1 \;\; U_2 \choose U_3 \;\; U4}$, where $U_4$ corresponds to the coupling between the infinite edges. Let $k_0>0$ correspond to a pole of the resolvent $(H -
k_0^2)^{-1}$ of multiplicity $d$. Let $\Gamma_\varepsilon$ be a geometrically perturbed quantum graph with edge lengths $l_0 (n_j + \varepsilon_j)$ described above and the same coupling as $\Gamma$. Then there exists an $\varepsilon_0 >0$ such that for all $\vec\varepsilon \in \mathcal{U}_{\varepsilon_0}(0)$ the sum of multiplicities of the resolvent poles in the vicinity of $k_0$ is $d$.
 \end{theorem}

\section{High-energy behavior of quantum-graph resonances} \label{s: graphweyl}

Now we will look at quantum-graph resonances from a different point of view and ask about asymptotics of their numbers at high energies. Following the papers \cite{DP11, DEL10} we are going to show, in particular, that it may often happen that this asymptotics does not follow the usual Weyl's law. Following the standard convention we will count in this section embedded eigenvalues among resonances speaking about the poles in the open lower halfplane as of `true' resonances.

\subsection{Weyl asymptotics criterion}

It is useful for our purpose to rewrite the condition (\ref{eff-res-condition}) in terms of the exponentials $\mathrm{e}^{ikl_j}$ and $\mathrm{e}^{-ikl_j}$ using for brevity the symbols $e_j^\pm:=\mathrm{e}^{\pm ikl_j}$ and
$ e^\pm := \Pi_{j=1}^N e_j^\pm=e^{\pm ikV}$,
where $V=\sum_{j=1}^nl_j$ is the size of the finite part of $\Gamma$. The condition then becomes
\begin{eqnarray}
  \lefteqn{F(k):= \mathrm{det\,}\bigg\{\frac{1}{2}[(U\!-\!I)+k(U\!+\!I)]E_1(k)
   +\frac{1}{2}[(U\!-\!I)-k(U\!+\!I)]E_2(k)\nonumber} \\ && +k(U\!+\!I)E_3 +\,(U\!-\!I)E_4 +[(U\!-\!I)-k(U\!+\!I)]\,\mathrm{diag\,}
   \left(0,\dots,0,I_{M\times M}\right)\bigg\}\nonumber = 0 \,, \label{rese}
 \end{eqnarray}
where $E_i(k) = \mathrm{diag\,}\left(E_{i}^{(1)},E_{i}^{(2)},
\dots,E_{i}^{(N)},0,\dots,0\right)$, $\,i=1,2,3,4$, are matrices consisting of $N$ nontrivial $2\times 2$ blocks
 $$
   E_1^{(j)}= \left(\!\begin{array}{cc}0&0\\-i e_j^+&e_j^+\end{array}\!\right),
   \ E_2^{(j)} = \left(\!\begin{array}{cc}0&0\\i e_j^-&e_j^-\end{array}\!\right),
   \ E_3^{(j)} = {i\;\;0 \choose 0\;\;0}\,,
   \ E_4^{(j)} = {0\;\;1 \choose 0\;\;0}\,.
 $$
and a trivial $M\times M$ part. To analyze the asymptotics we employ the usual \emph{counting function} $N(R,F)$ defined for an entire function $F(\cdot)$ by
 $$ 
N(R,F)=\sharp\{ k:F(k)=0\mbox{ and } |k|<R\}\,,\label{NRF}
 $$ 
where the algebraic multiplicities of the zeros are taken into account. With the above spectral condition in mind we have to apply it to trigonometric polynomials with rational-function coefficients. We need the following result \cite{DEL10} which is a simple consequence of a classical theorem by Langer \cite{La31}.

 \begin{theorem}\label{exppol}
Let $F(k) = \sum_{r = 0}^{n} a_r(k) \,\e^{ik\sigma_r}$,
where $a_r(k)$ are rational functions of the complex variable $k$ with complex coefficients, and $\sigma_r\in \R$, $\sigma_0
<\sigma_1< \dots < \sigma_n$. Suppose that $\lim_{k\to\infty}
a_0(k) \ne 0$ and $\lim_{k\to\infty} a_n(k) \ne 0$. There
exists a compact set $\Omega\subset \C$, real numbers $m_r$ and positive $K_r$, $\:r =1,\dots,n$, such that the zeros of $F(k)$ outside $\Omega$ lie in one of $n$ logarithmic strips, each one bounded between the curves $-\mathrm{Im\,}k + m_r \log{|k|} =\pm K_r$. The counting function behaves in the limit $R\to\infty$ as
 $$
   N(R,F) = \frac{\sigma_n -\sigma_0}{\pi} R +\OO(1)\,. 
 $$
 \end{theorem}

To apply this result it is useful to pass to effective energy-dependent coupling (\ref{rat-efu}) which makes it possible to cast the spectral condition into a simpler form,
  \begin{eqnarray}
  \lefteqn{\hspace{4em}F(k)= \mathrm{det\,}\bigg\{\frac{1}{2}[(\tilde U(k)-I)+k(\tilde U(k)+I)]\tilde E_1(k) \label{endep}} \\&&
  +\frac{1}{2}[(\tilde U(k)-I)-k(\tilde U(k)+I)]\tilde E_2(k)
  + k(\tilde U(k)+I)\tilde E_3+(\tilde U(k)-I)\tilde E_4\bigg\} = 0\,, \nonumber
 \end{eqnarray}
where $\tilde E_j$ are the nontrivial $2N \times 2N$ parts of the matrices $E_j$, the first two of them being energy-dependent, and $I$ denotes the $2N \times 2N$ unit matrix. Then we have the following criterion \cite{DEL10} for the asymptotics to be of Weyl type.

 \begin{theorem}\label{thmeigenvalue}
Assume a quantum graph $(\Gamma,H_U)$ corresponding to
$\Gamma$ with finitely many edges and the coupling at vertices
$\mathcal{X}_j$ given by unitary matrices $U_j$. The asymptotics of the resonance counting function as $R\to \infty$ is of the form
 $$
   N(R,F) = \frac{2W}{\pi} R
   +\mathcal{O}(1)\,,
 $$
where the effective size of the graph $W$ satisfies $0\leq W\leq V:=\sum_{j=1}^{N}l_j$. More\-over, $W<V$ holds if and only if there exists a vertex where the corresponding energy-dependent coupling matrix $\tilde U_j(k)$ has an eigenvalue $\frac{1-k}{1+k}$ or $\frac{1+k}{1-k}$ for all $k$.
 \end{theorem}

To prove the theorem one has to realize that $\sigma_n=V$ and $\sigma_0=-V$, hence the asymptotics is not of Weyl type \emph{iff} either the senior or the junior coefficient in expression of $F(k)$, i.e. those of $e^\pm$, vanish. By a straightforward computation \cite{DEL10} we find that they equal $\left( \frac{i}{2}\right)^N\mathrm{det\,}[(\tilde U(k)-I)\pm k(\tilde U(k)+I)]$, respectively, and therefore they vanish under the condition stated in the theorem.

Before proceeding further, let us mention that the asymptotic number of resonances is not the only thing of interest. One can investigate other asymptotic properties such as the distribution of resonance pole spacings; quantum graphs are known to be a suitable laboratory to study quantum chaotic effects \cite{KoS03}.

 \subsection{Permutation-symmetric coupling}\label{s: symcoup}

Let us first look what the above criterion means in a particular class of vertex couplings which are invariant with respect to permutations of the edges connected at the vertex. It is easy to see that such couplings are described by matrices of the form $U_j = a_j J + b_j I$, where $a_j$, $b_j$ are complex numbers satisfying $|b_j|=1$ and $|b_j+a_j \mathrm{deg\,} \mathcal{X}_j| =1$; the symbol $J$ denotes the square matrix all of whose entries equal to one and $I$ stands for the unit matrix. Important examples are the
\emph{$\delta$-coupling} analogous to (\ref{lasso_delta}), with the functions continuous at the vertex and the sum of outward derivatives proportional to their common value, corresponding to $U_j = \frac{2}{d_j+i\alpha_j}J - I$,
where $d_j$ is the number of edges emanating from the vertex
$\mathcal{X}_j$ and $\alpha_j\in\mathbb{R}$ is the coupling
strength, and the \emph{$\delta'_\mathrm{s}$-coupling} corresponding to $U_j = -\frac{2}{d_j-i\beta_j}J + I$ with $\beta_j\in\mathbb{R}$ for which the roles of functions and derivatives are interchanged. The particular cases $\alpha_j = 0$ and $\beta_j=0$ are usually referred to as the
Kirchhoff and anti-Kirchhoff condition, respectively.

Consider a vertex which connects $p$ internal and $q$
external edges. For matrices of the form $U_j = a_j J + b_j I$ it is an easy exercise to invert them and to find the effective energy-dependent coupling; this allows us to make the following claim.

 \begin{theorem} \label{perm_result}
Let $(\Gamma,H_U)$ be a quantum graph with permutation-symmetric coupling conditions at the vertices, $U_j = a_j J + b_j I$. Then it has a non-Weyl asymptotics if and only if at least one of its vertices is \emph{balanced} in the sense that $p = q$, and the coupling at this vertex satisfies one the following conditions:
\begin{enumerate}
\item[(a)] $f_m = f_n,\quad \forall m,n\leq 2p$, \quad
$\sum_{m=1}^{2p} f_m' = 0$, \quad \textrm{i.e.}$\:\:U =
\frac{1}{p}J_{2p\times 2p}-I_{2p\times 2p}$\,,
\item[(b)] $f_m' = f_n',\quad \forall m,n\leq 2p$, \quad
$\sum_{m=1}^{2p} f_j = 0$, \quad \textrm{i.e.}$\:\:U =
-\frac{1}{p}J_{2p\times 2p}+I_{2p\times 2p}$\,.
\end{enumerate}
 \end{theorem}

\noindent In other words, if the graph has a balanced vertex there are exactly two situations when the asymptotics is non-Weyl, either if the coupling is Kirchhoff --- which is the case where the effect was first noted in \cite{DP11} --- or if it is anti-Kirchhoff.

  \smallskip

  \subsubsection{An example: a loop with two leads}

To illustrate the above claim let us return to the graph of
Example~\ref{loop2leads}. It is balanced if the two leads are attached at the same point. Changing slightly the notation we suppose that the loop length is $l$ and consider negative Laplacian on the Hilbert space is $L^2(0,l)\oplus L^2(\mathbb{R}^+)\oplus L^2(\mathbb{R}^+)$ with
its elements written as $(u,f_1,f_2)^\mathrm{T}$ defined on functions from $W^{2,2}_\mathrm{loc}(\Gamma)$ satisfying the
requirements $u(0) = f_1(0)$ and $u(l) = f_2(0)$ together with
 $$
  \alpha u(0) = u'(0)+ f_1'(0) +\beta (-u'(l)+f_2'(0))\,, \quad \alpha u(l) = \beta(u'(0)+ f_1'(0)) -u'(l)+f_2'(0)
 $$
with real parameters $\alpha$ and $\beta$; the choice $\beta = 1$ corresponds to the `overall' $\delta$-coupling of strength
$\alpha$, while $\beta = 0$ decouples two `inner-outer' pairs of mutually meeting edges turning $\Gamma$ into a line with two $\delta$-interactions at the distance $l$. In terms of the quatities $e^{\pm}= \e^{\pm ikl}$ the pole condition can be written \cite{DEL10} as
  $$
    8\: \frac{i\alpha^2 e^+
    + 4 k \alpha\beta -i[\alpha(\alpha-4ik)
    +4 k^2 (\beta^2-1)]\, e^-}{4(\beta^2-1)+\alpha(\alpha-4i)}=0\,.
  $$
The coefficient of $e^+$ vanishes \emph{iff} $\alpha = 0$, the one in the second term for $\beta = 0$ or if $|\beta| \not = 1$ and $\alpha = 0$, while the coefficient $e^-$ does not vanish for any combination of $\alpha$ and $\beta$. The graph has thus a non-Weyl asymptotics \emph{iff}\ $\alpha =0$. If, in addition, $|\beta|\not = 1$, then all resonances are confined to a circle, i.e. the graph has zero `effective size'. The only exceptions are the Kirchhoff condition, $\beta = 1$ and $\alpha=0$, and its anti-Kirhhoff counterpart, $\beta = - 1$ and $\alpha=0$, for which one half of the resonances is asymptotically preserved, in other words, the effective size of the graph is $\frac12 l$.

We can demonstrate how a `half' of the resonances disappears using the example of the $\delta$-coupling, $\beta = 1$. The resonance equation in this case becomes
  $$
    \frac{-\alpha\sin{kl}+2k(1 + i \sin{kl}-\cos{kl})}{\alpha-4i}=0\,.
  $$
A simple calculation shows that the graph Hamiltonian has a
sequence of embedded eigenvalues $k^2$ with $k = \frac{2\pi n}{l},\:n\in\Z$, and a family of resonances given by solutions to $\e^{ikl}=-1+\frac{4ik}{\alpha}$. The former do not depend
on $\alpha$, while the latter behave like
  $$
    \mathrm{Im\,}k = -\frac{1}{l} \ln \frac{1}{\alpha}
    +\mathcal{O}(1)\,, \quad  \mathrm{Re\,}k = n \pi +\mathcal{O}(\alpha)\,,
  $$
as $\alpha\to 0$, hence all the `true' resonances escape to the imaginary infinity in the limit, in analogy with the similar pole behavior described by relation (\ref{stubfreefall}).

   \subsection{The mechanism behind a non-Weyl asymptotics}

One naturally asks about reasons why graphs with balanced vertices and Kirchhoff/anti-Kirchhoff coupling have smaller than expected effective size. A simple observation is that if such a vertex has degree one, then Kirchhoff coupling between an external and internal edge is in fact no coupling at all, hence the internal edge can be regarded as a part of the lead
and the effective size is diminished by its length. We are going to show that this remains true in a sense also when the degree is larger than one.

 \smallskip

\subsubsection{Kirchhoff `size reduction'}\label{ss:reduction}

\begin{figure}
   \begin{center}
     \includegraphics{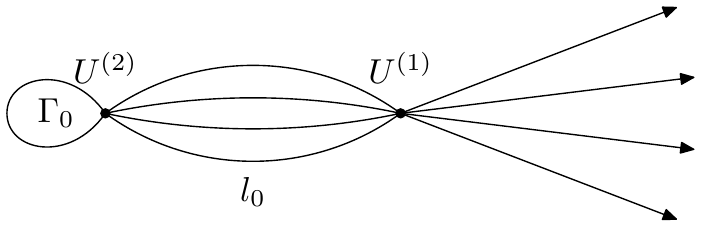}
     \caption{Graph with a balanced vertex}\label{mechanism}
   \end{center}
\end{figure}

A graph with a balanced vertex can be always thought as having the shape sketched in Fig.~\ref{mechanism}: with a vertex $\mathcal{X}_1$ which connects $p$ internal edges
of the same length $l_0$ and $p$ external edges; if the internal ones have different lengths we simply add a needed number of `dummy' vertices. We will suppose that the coupling at $\mathcal{X}_1$ is invariant with respect to edge permutations being described by a unitary matrix $U^{(1)} = a
J_{2p\times 2p} + b I_{2p\times 2p}$; the coupling of the other internal edge ends to the rest of the graph, denoted here as
$\Gamma_0$, is described by a $q\times q$ matrix $U^{(2)}$, where $q \geq p\,$ (which may express also the topology of $\Gamma_0$).

To find how the effective size of such a quantum graph may look like we employ the following property which can be derived  easily from coupling condition (\ref{rat-coup2}).

 \begin{proposition} \label{transfbal}
Let $\Gamma$ be the graph described above with the coupling given by arbitrary $U^{(1)}$ and $U^{(2)}$. Let further $V$ be an arbitrary unitary $p\times p$ matrix, $V^{(1)}:= \mathrm{diag\,}(V,V)$ and $V^{(2)}:=\mathrm{diag\,} (I_{(q-p)\times(q-p)},V)$ be $2p \times 2p$ and $q\times q$ block diagonal matrices, respectively. Then $H$ on $\Gamma$ is unitarily equivalent to the Hamiltonian $H_V$ on the graph with the same topology and  the coupling given by the matrices $[V^{(1)}]^{-1}U^{(1)}V^{(1)}$ and $[V^{(2)}]^{-1} U^{(2)} V^{(2)}$, respectively.
 \end{proposition}

Application to the couplings described by $U^{(1)} =
a J_{2p\times 2p} + b I_{2p\times 2p}$ at $\mathcal{X}_1$ is
straightforward. One has to choose the columns of $V$ as an
orthonormal set of eigenvectors of the corresponding $p\times p$ block $aJ_{p\times p} + b I_{p\times p}$ of $U^{(1)}$, the first one of them being $\frac {1}{\sqrt{p}}\,(1,1, \dots,1)^\mathrm{T}$. The transformed matrix $[V^{(1)}]^{-1} U^{(1)}V^{(1)}$ decouples then into blocks connecting only the pairs $(v_j,g_j)$. The first one of these, corresponding to a symmetrization of all the $u_j$'s and $f_j$'s, leads to the $2\times 2$ matrix $U_{2\times 2} = ap J_{2\times 2} + b I_{2\times 2}$, while the other lead to separation of the corresponding internal and external edges described by Robin conditions  $(b-1)v_j(0) + i (b+1)v_j'(0) = 0$ and $(b-1)g_j(0) + i (b+1)g_j'(0) = 0$ for $j=2,\dots,p$. We note that it resembles the reduction procedure of a tree graph due
to Solomyak \cite{SS02}.

It is easy to see that the `overall' Kirchhoff/anti-Kirchhoff condition at $\mathcal{X}_1$ is transformed to the `line' Kirchhoff/anti-Kirchhoff condition in the subspace of permutation-symmetric functions, leading to reduction of the graph effective size as mentioned above. In all the other
cases the point interaction corresponding to the matrix $ap
J_{2\times 2} + b I_{2\times 2}$ is nontrivial, and consequently, the graph size is preserved.

\smallskip

\subsubsection{Global character of non-Weyl asymptotics}\label{calculating}

The above reasoning might lead one to the conclusion that the effect discussed here is of a local character. We want to show now that while this is true concerning the occurrence of non-Weyl asymptotics, the effective size of a non-Weyl quantum graph is a global property because it may depend on the graph $\Gamma$ as a whole.

\begin{figure}
   \begin{center}
     \includegraphics[width=4cm]{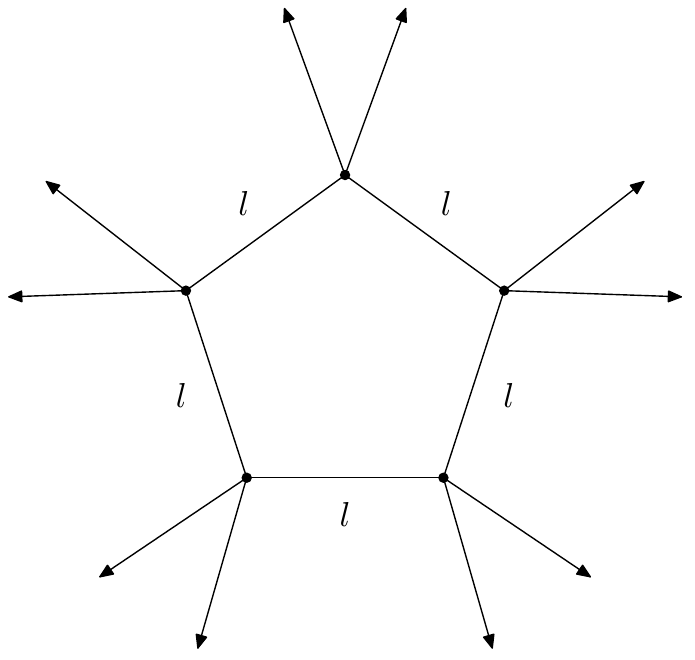}
     \caption{A polygonal balanced graph.}\label{polygon}
   \end{center}
\end{figure}

We will use an example to justify this claim. We shall consider the graph $\Gamma_n$ with an integer $n\geq 3$ which contains a regular $n$-gon, each edge of which has length $l$. To each of its vertices two semi-infinite leads are attached, cf.~Fig.~\ref{polygon}. Hence all the vertices of $\Gamma_n$ are balanced, and if the coupling in them is of Kirchhoff type the effective size $W_n$ of the graph is by Theorem~\ref{perm_result} strictly less than the actual size $V_n=n\ell$. It appears, however, that the actual value of the effective size depends in this case on the number $n$ of polygon vertices.

Since all the internal edges have the same length, the system has a rotational symmetry. One can thus perform a `discrete Floquet' analysis and investigate cells consisting of two internal and two external edges; the wave functions at the ends of the former have to differ by a multiplicative factor $\omega$ such that $\omega^n=1$. After a simple computation \cite{DEL10} we conclude that there is a resonance at $k^2$ \emph{iff}
\begin{equation} \label{poly-res}
-2(\omega^2+1)+4\omega \mathrm{e}^{-ik\ell}=0.
\end{equation}
The `Floquet component' $H_\omega$ of $H$ has thus effective size $W_\omega= \frac12 l$ if $\omega^2+1\not=0$ while for $\omega^2+1=0$ we have no resonances, $W_\omega=0$. Summing finally over all the $\omega$ with $\omega^n=1$ we arrive at the following conclusion.

\begin{theorem} \label{polygon-ex}
The effective size of the graph $\Gamma_n$ with Kirhhoff coupling is
$$
W_n=\left\{ \begin{array}{ll}
\frac12 nl&\;\mathrm{ if }\; n\not= 0\,\, \bmod 4\\[.3em]
\frac12(n\!-\!2)l&\;\mathrm{ if }\; n= 0\,\, \bmod 4
\end{array}\right.
$$
\end{theorem}

\medskip

\noindent Note that if one puts $\omega=\e^{i\theta}$ in
({\ref{poly-res}) the resonance poles behave according to
$$
k=\frac{1}{l}\Big(i\ln(\cos\theta)+2\pi n\Big)
$$
where $n\in \mathbb{Z}$ is arbitrary, hence they escape to imaginary infinity as $\theta\to\pm\frac12\pi$. Of course, the Floquet variable is discrete, $\theta= \frac{2\pi j}{n},\: j=0,1,\dots,n\!-\!1$, nevertheless, the limit still illustrates the mechanism of the resonance `disappearance'; it is illustrative to look at the behavior of the solutions for large values of $n$.

   \subsection{Non-Weyl graphs with non-balanced vertices}\label{sec-unbal}

Now we are going to show that there are many more graphs with non-Weyl asymptotics once we abandon the assumption of permutation symmetry of the vertex couplings. For the sake of brevity, we limit ourselves again to a simple example. In order to formulate it, however, we state first a general property of the type of Proposition~\ref{transfbal} above. Specifically, we will ask what happens if the coupling matrix $U$ of a guantum graph is replaced by $W^{-1} U W$, where $W$ is a block diagonal matrix of the form
$$
     W = \left(\begin{array}{cc}\mathrm{e}^{i\varphi} I_{p\times p}&0\\0&
     W_4\end{array}\right)
$$
and $W_4$ is a unitary $q\times q$ matrix. The following claim is obtained easily from (\ref{rat-coup2}).

 \begin{proposition}\label{sym1}
The family of resonances of a quantum-graph Hamiltonian $H_U$ does not change if the original coupling matrix $U$ is replaced by $W^{-1} U W$.
 \end{proposition}

Let us turn now to the example which concerns the graph investigated in Sec.~\ref{s:stub}, a line with a stub of length $l$, this time without a potential. Changing slightly the notation we use the symbols $f_j$ for wave function on the two halflines and $u$ for the stub. The function from the domain of any Hamiltonian $H_U$ are locally $W^{2,2}$ and satisfy the conditions $u(l)+c u'(l)=0$ with $c\in\R\cup\{\infty\}$ and
  $$
     (U-I)\, ( u(0), f_1 (0),f_2 (0))^\mathrm{T}+ i(U+I)\,(u'(0), f_1' (0),f_2' (0))^\mathrm{T} =0\,.
  $$
We split $\Gamma$ into two parts in a way different from Sec.~\ref{s:stub} choosing the coupling described by $U_0:=\mathrm{diag\,}\big( {0\;\;1 \choose 1\;\; 0},\, \e^{i\psi} \big)$ which gives two halflines with the conditions $u(l)+c u'(l)=0$ and $f_2(0)+ \cot\frac{\psi}{2}\, f'_2(0)=0$, respectively, at their endpoints; the first part consists of the halfline number one and the stub joined by Kirchhoff coupling. It is obvious that such a graph has at most two resonances, and thus a non-Weyl asymptotics. We now replace $U_0$ by $U_W = W^{-1}U_0 W$ with
  $$
     W  = \left(\begin{array}{ccc}
     1&0&0\\
     0& r \mathrm{e}^{i\varphi_1}&\sqrt{1-r^2}\,\mathrm{e}^{i\varphi_2}\\
     0&\sqrt{1-r^2}\,\mathrm{e}^{i\varphi_3}& -r \mathrm{e}^{i(\varphi_2 +\varphi_3-\varphi_1)}
     \end{array}\right)
 $$
for some $r\in[0,1]$ and obtain for every fixed value of $\psi$ and $c$ a three-parameter family of coupling conditions described by the unitary matrix
  $$
     U_W  = \left(\begin{array}{ccc}
     0& r \mathrm{e}^{i\varphi_1}&\sqrt{1-r^2}\mathrm{e}^{i\varphi_2}\\
     r \mathrm{e}^{-i\varphi_1}& (1-r^2)\mathrm{e}^{i\psi} & -r\sqrt{1-r^2}\mathrm{e}^{-i(-\psi+\varphi_1-\varphi_2)}\\
     \sqrt{1-r^2} \mathrm{e}^{-i\varphi_2}&-r\sqrt{1-r^2}
     \mathrm{e}^{i(\psi+\varphi_1-\varphi_2)}&r^2\mathrm{e}^{i\psi}
     \end{array}\right),
 $$
each of which has the same resonances as $U_0$ by Proposition~\ref{sym1}. The associated quantum graphs are thus of non-Weyl type despite the fact that their edges are connected in a single vertex of $\Gamma$ which is not balanced.

Note that among these couplings one can find, in particular, the one mentioned above in connection with relation
(\ref{stubfreefall}); choosing $\psi = \pi$ and $c = 0$, and furthermore, $\varphi_1 = \varphi_2 = 0$ and $r =
2^{-1/2}$, we get the conditions
 $$
     f_1 (0) = f_2(0),\quad  u(0) = \sqrt{2} f_1(0),\quad
     f_1' (0) - f_2'(0)= -\sqrt{2} u'(0)\,,
 $$
or (\ref{stub-bc}) with $b=\sqrt{2}$ and $c=d=0$. Similarly, such conditions with $b=-\sqrt{2}$ and $c=d=0$ correspond to $\varphi_1 = \varphi_2 = \pi$ and $r = 2^{-1/2}$; both these quantum graphs have no resonances at all. This fact is easily understandable, for instance, if we regard the line with the stub as a tree with the root at the end of the stub and apply the Solomyak reduction procedure \cite{SS02} mentioned above.

  \subsection{Magnetic field influence}

Let us finally look how can the high-energy asymptotics be influenced by a magnetic field. We have encountered magnetic quantum graphs already in the example of Sec.~\ref{s:lasso}, now we look at them in more generality. We consider a graph $\Gamma$ with a set of vertices $\{\mathcal{X}_j\}$ and set of edges $\{\mathcal{E}_j\}$ containing $N$ finite edges and $M$ infinite leads. We assume that it is equipped with the operator $H$ acting as $-\frac{\D^2}{\D x^2}$ on the infinite leads and as $-(\frac{\D}{\D x} + i A_j(x))^2$ on the internal edges, where $A_j$ is the tangent component of the vector potential; without loss of generality we may neglect it on external leads because one can always remove it there by a gauge transformation. The Hamiltonian domain consists of functions from $W^{2,2}_\mathrm{loc}(\Gamma)$ which satisfy $(U_j-I)\Psi_j + i(U_j+I)(\Psi_j' +i \mathcal{A}_j\Psi_j)=0$ at the vertex $\mathcal{X}_j$. As before it is useful to pass to a graph $\Gamma_0$ with a single vertex of degree $(2N+M)$ in which the coupling is described by the condition
 $$ 
  (U - I) \Psi + i (U+I)(\Psi'+ i \mathcal{A}\Psi)= 0\,,	
 $$ 
where the matrix $U$ consists of the blocks $U_j$ corresponding to the vertices of $\Gamma$ and the matrix $\mathcal{A}$ is composed of tangent components of the vector potential at the vertices, $\mathcal{A}= \mathrm{diag\,}(A_1(0),-A_1(l_1), \dots,A_N(0),-A_N(l_N),0,\dots,0)$.

Using the local gauge transformation $\psi_j (x) \mapsto \psi_j (x) \e^{-i \chi_j (x)}$ with $\chi_j(x)' = A_j(x)$ one can get rid of the explicit dependence of coupling conditions on the magnetic field and arrive thus at the Hamiltonian acting as $-\frac{\D^2}{\D x^2}$ with the coupling conditions given by a transformed unitary matrix,
 \begin{equation}
  (U_A - I) \Psi + i (U_A+I)\Psi' = 0\,,\quad U_A := \mathcal{F} U \mathcal{F}^{-1},	\label{coupl2}
 \end{equation}
with $\mathcal{F}= \mathrm{diag\,}(1,\mathrm{exp\,}(i\Phi_1), \dots,1,\mathrm{exp\,}(i\Phi_N),1,\dots,1)$ containing magnetic fluxes $\Phi_j = \int_0^{l_j} A_j(x) \,\mathrm{d}x$. Furthermore, one can reduce the analysis to investigation of the compact core of $\Gamma$ with an effective energy-dependent coupling described by the matrix $\tilde U_A(k)$ obtained from $U_A$ in analogy with (\ref{eff-res-condition}).

To answer the question mentioned above we employ another property of the type of Proposition~\ref{transfbal}. This time we consider replacement of $U$ by $V^{-1} U V$ where $V= \left(\begin{smallmatrix}  V_1& 0\\0& V_2  \end{smallmatrix}\right)$ is unitary block-diagonal matrix consisting of a $2N \times 2N$ block $V_1$ and an $M \times M$ block $V_2$; resonances are again invariant under this transformation. With respect to the relation between $U$ and $\tilde U_A$ we get the following result \cite{EL11}.

\begin{theorem}
A quantum graph with a magnetic field described by a vector potential $A$ is of non-Weyl type if and only if the same is true for $A=0$.
\end{theorem}

In other words, magnetic field alone cannot switch a graph with non-Weyl asymptotics into Weyl type and \emph{vice versa}. On the other hand, the magnetic field \emph{can} change the effective size of a non-Weyl graph. To illustrate this claim, let us return to the example discussed in Sec.~\ref{loop2leads}, a loop with two external leads Kirchhoff-coupled to a single point, now we add a magnetic field. It is straightforward to check \cite{EL11} that the condition determining the resonance pole becomes
$$
  -2\cos{\Phi} + \mathrm{e}^{-ikl} = 0\,,
$$
where $\Phi$ is the magnetic flux through the loop. The graph is non-Weyl as the term with $\e^{ikl}$ is missing on the left-hand side; if $\Phi  = \pm \pi/2\; (\mathrm{mod}\,\pi)$, that is, for odd multiples of a quarter of the flux quantum $2\pi$, the $l$-independent term disappears and the effective size of the graph becomes zero.

The conclusions of the example can be generalized \cite{EL11} to any graph with a single internal edge: if the elements of the effective $2\times2$ coupling matrix satisfy $|\tilde u_{12}(k)|  = |\tilde u_{21}(k)|$ for any $k>0$ there is a magnetic field such that the graph under its influence has at most finite number of resonances.

\section{Leaky graphs: a caricature of quantum wire and dots} \label{s: leaky}

A different class of quantum graph models employs Schr\"odinger operators which can be formally written as $-\Delta - \alpha \delta(x-\Gamma)$ where $\Gamma\subset\R^d$ is a graph; one usually speaks about \emph{`leaky' graphs}. Their advantage is that they can take into account tunneling between different parts of the graph as well as its geometry beyond just the edge lengths. A survey of results concerning these models can be found in \cite{Ex08}. In particular, even a simple $\Gamma$ like an infinite non-straight curve can give rise to resonances \cite{EN03}, however, one needs a numerical analysis to reveal them.

   \subsection{The model}

Instead we will describe here a simple model of this type which can be regarded as a caricature description of a system consisting of a quantum wire and one or several quantum dots. The state Hilbert space of the model is $L^2(\R^2)$ and the Hamiltonian can be formally written as
 $$ 
 -\Delta -\alpha \delta (x-\Sigma ) +\sum_{i=1}^n \tilde\beta_i
 \delta(x-y^{(i)})\,,
 $$ 
where $\alpha>0$, $\,\Sigma :=\{(x_1,0);\,x_1\in\R^2\}$, and
$\Pi:= \{y^{(i)}\}_{i=1}^n \subset \R^2 \setminus \Sigma$. The
formal coupling constants of the two-dimensional $\delta$
potentials are marked by tildes to stress they are not identical with the proper coupling parameters $\beta_i$ which we shall introduce below. Following the standard prescription \cite{AGHH} one can define the operator rigorously \cite{EK04} by introducing appropriated boundary conditions on $\Sigma \cup \Pi$. Consider functions $\psi\in W^{2,2}_{\mathrm{loc}} (\R^{2}\setminus (\Sigma \cup \Pi )) \cap L^{2}$ continuous on $\Sigma$. For a small enough $\rho>0$ the restriction $\psi\upharpoonright_{\mathcal{C}_{\rho ,i}}$ to the circle $\mathcal{C}_{\rho,i}:=\{q\in \R^2:|q-y^{(i)}|=\rho\}$ is well defined; we say that $\psi$ belongs to $D(\dot{H}_{\alpha , \beta })$ \emph{iff}
$(\partial^2_{x_1}+\partial^2_{x_2})\psi$ on $\R^{2}\setminus
(\Sigma \cup \Pi )$ belongs to $L^2$ in the sense of distributions and the limits
 \begin{eqnarray*} \label{}
 && \hspace{-1em} \Xi_{i}(\psi):=-\lim _{\rho \rightarrow 0}\frac {1}{\ln \rho
 }\,\psi \upharpoonright _{\mathcal{C}_{\rho ,i}}\,,\;
 \Omega_{i}(\psi):=\lim _{\rho \rightarrow 0}[\psi\upharpoonright
 _{\mathcal{C}_{\rho ,i}} +\Xi_{i}(\psi)\ln \rho ] \,,\;\:
 i=1,\dots,n\,, \\ && \hspace{-1em}
 \Xi_{\Sigma }(\psi)(x_{1}):=
 \partial_{x_{2}} \psi(x_{1},0+)-
 \partial_{x_{2}}\psi(x_{1},0-)\,, \quad
 \Omega_{\Sigma }(\psi)(x_{1}):= \psi(x_{1},0)
 \end{eqnarray*}
exist, they are finite, and satisfy the relations
\begin{equation}  \label{boucon}
2\pi \beta _i \Xi _{i}(\psi)=\Omega  _{i}(\psi)\,, \quad \Xi
_{\Sigma}(\psi)(x_{1})=-\alpha \Omega  _{\Sigma }(\psi)(x_{1})\,,
\end{equation}
where $\beta _{i}\in \R$ are the true coupling parameters; we put $\beta = (\beta _1 ,\dots,\beta _n )$ in the following. On
this domain we define the operator $\dot H_{\alpha, \beta}:D(\dot H_{\alpha, \beta})\rightarrow L^{2}(\R^2)$ by
$$
\dot H_{\alpha, \beta}\psi(x)=-\Delta \psi(x)\quad \mathrm{for}\quad
x\in \R^{2} \setminus (\Sigma \cup \Pi)\,. $$
It is now a standard thing to check that $\dot H_{\alpha, \beta}$ is essentially self-adjoint \cite{EK04}; we shall regard in the following its closure denoted as $H_{\alpha, \beta}$ as the rigorous counterpart to the above mentioned formal model Hamiltonian.

To find the resolvent of $H_{\alpha ,\beta }$ we start from $R(z)=(-\Delta -z)^{-1}$ which is for any $z\in \C\setminus[0,\infty)$ an integral operator with the kernel $G_{z}(x,x')=\frac{1}{2\pi}K_{0} (\sqrt{-z}|x-x'|)$, where $K_{0}$ is the Macdonald function and $z\mapsto \sqrt{z}$ has conventionally a cut along the positive halfline; we denote by $\mathbf{R}(z)$ the unitary operator with the same kernel acting from $L^{2}(\R^2)$ to $W^{2,2}(\R^2)$. We need two auxiliary spaces, $\mathcal{H}_0:= L^{2}(\mathbb{R})$ and $\mathcal{H}_1:=\mathbb{C}^n$, and the corresponding trace maps $\tau_j:W^{2,2}(\R^2)\to \mathcal{H}_j$ which act as
 $$
 \tau _0  \psi:=\psi\!\upharpoonright_{\,\Sigma }\,, \quad \tau_1
 \psi:=\psi\!\upharpoonright_{\,\Pi}=\big(\psi\!\upharpoonright _{\,\{y^{(1)}\}},
 \dots,\psi\!\upharpoonright _{\,\{y^{(n)}\}}\big)\,,
 $$
respectively; they allow us to define the canonical embeddings of $\mathbf{R} (z)$ to $\mathcal{H}_{i}$, i.e.
 $$ 
 \mathbf{R}_{iL}(z)=\tau _{i}R(z):L^{2}\rightarrow \mathcal{H}
 _{i}\,,\quad \mathbf{R}_{Li}(z)=[\mathbf{R}_{iL}(z)]^{\ast
 }:\mathcal{H} _{i}\rightarrow L^{2}\,,
 $$ 
and $\mathbf{R}_{ji}(z)=\tau _{j}\mathbf{R}_{Li}(z):\mathcal{H}_{i}\rightarrow \mathcal{H}_{j}$, all expressed naturally through the free Green's function in their kernels, with the variable range corresponding to a given $\mathcal{H}_{i}$. The operator-valued matrix $\Gamma (z)=[\Gamma_{ij}(z)]:\mathcal{H}_{0} \oplus\mathcal{H}_{1}\rightarrow \mathcal{H}_{0}\oplus\mathcal{H}_{1}$ is defined by
 \begin{eqnarray*} \label{forg22}
 \Gamma _{ij}(z)g &\!:=\!& -\mathbf{R}_{ij}(z)g \qquad
 \mathrm{for}\quad i\neq j \quad \mathrm{and }\;\; g\in
 \mathcal{H}_{j}\,, \\ \Gamma_{00}(z)f &\!:=\!& \left[\alpha^{-1}
 -\mathbf{R}_{00}(z)\right] f \qquad \mathrm{if} \;\;
 f\in \mathcal{H}_0\,, \\ \Gamma _{11}(z)\varphi &\!:=\!&
 \left[ s_{\beta _{l}}(z) \delta_{kl} - G_{z}(y^{(k)},y^{(l)}) (1\!-\!
 \delta_{kl}) \right]_{k,l=1}^{n} \varphi \quad \mathrm{for}
 \;\; \varphi \in \mathcal{H}_1\,,
 \end{eqnarray*}
where $s_{\beta_{l}}(z)= \beta_{l}+s(z):=\beta_{l}+\frac{1}{2\pi}(\ln \frac{\sqrt{z}} {2i}-\psi(1))$ and $-\psi(1)$ is the Euler number. For $z$ from $\rho(H_{\alpha ,\beta })$ the operator $\Gamma (z)$ is boundedly invertible. In particular, $\Gamma _{00}(z)$ is invertible which makes it possible to employ the Schur reduction procedure one more time and to define the map $D(z):\mathcal{H}_{1}\rightarrow \mathcal{H}_{1}$ by
 \begin{equation} \label{Gamhat}
 D(z)=\Gamma _{11}(z)-\Gamma _{10}(z)\Gamma _{00}(z)^{-1}\Gamma_{01}(z)\,.
 \end{equation}
We call it the \emph{reduced determinant} of $\Gamma $; it allows us to write the inverse of $\Gamma (z)$ as $[\Gamma(z)]^{-1}: \mathcal{H}_{0}\oplus \mathcal{H}_{1}\rightarrow\mathcal{H}_{0}\oplus\mathcal{H}_{1}$ with the `block elements' defined by
 \begin{eqnarray*}
 \left[\Gamma(z)\right]_{11}^{-1} &=& D(z)^{-1}\,, \\
 \left[\Gamma(z)\right]_{00}^{-1} &=&
 \Gamma_{00}(z)^{-1} + \Gamma_{00}(z)^{-1} \Gamma_{01}(z)D(z)^{-1}
 \Gamma_{10}(z)\Gamma_{00}(z)^{-1}\,, \\
 \left[\Gamma(z)\right]_{01}^{-1} &=& -\Gamma_{00}(z)^{-1}
 \Gamma _{01}(z) D(z)^{-1}\,, \\
 \left[\Gamma(z)\right]_{10}^{-1} &=& -D(z)^{-1}
 \Gamma_{10}(z)\Gamma_{00}(z)^{-1}\,;
\end{eqnarray*}
in the above formul{\ae} we use the notation $\Gamma _{ij}(z)^{-1}$ for the inverse of $\Gamma _{ij}(z)$ and $[\Gamma (z)]_{ij}^{-1}$ for the matrix element of $[\Gamma (z)]^{-1}$.

Before using this to express the resolvent $R_{\alpha ,\beta }(z):= (H_{\alpha ,\beta }-z)^{-1}$ we introduce another notation which allow us to write $R_{\alpha ,\beta }(z)$ through a perturbation of the `line only' Hamiltonian $\tilde{H}_{\alpha }$ describing the system without the point interactions, i.e. $\beta_i=\infty,\: i=1,\dots,n$. By \cite{BEKS94} the resolvent of $\tilde{H}_{\alpha }$ is equal to
 $$ 
 R_{\alpha }(z)=R(z)+R_{L0}(z)\Gamma ^{-1}_{00}R_{0L}(z)
 $$ 
for $z\in \C\setminus [-\frac{1}{4}\alpha ^{2},\infty )$. We define the map $\mathbf{R}_{\alpha ; L1}(z):\mathcal{H}_{1}\rightarrow
L^{2}(\R^2)$ by $\mathbf{R}_{\alpha ; 1L}(z)\psi:=R_{\alpha }(z) \psi\!\upharpoonright_{\Pi}$ and  $\mathbf{R}_{\alpha ;1L}(z):L^2(\R^2) \rightarrow \mathcal{H}_{1}$ as its adjoint, $\mathbf{R}_{\alpha ;L1}(z):=\mathbf{R}^{\ast }_{\alpha ;1L}(z)$. The resolvent difference between $H_{\alpha ,\beta }$ and
$\tilde{H}_{\alpha }$ is then given by Krein's formula \cite{AGHH}. A straightforward computation \cite{EK04} yields now the following result.

 \begin{theorem} \label{resoth}
 For any $z\in \rho (H_{\alpha ,\beta })$ with $\mathrm{Im \,}z>0$
 we have
 $$ 
 R_{\alpha ,\beta }(z) =R(z)+\sum_{i,j=0}^{1} \mathbf{R}
 _{Li}(z)[\Gamma(z)]_{ij}^{-1}\mathbf{R}_{jL}(z) =R_{\alpha}(z)
 + \mathbf{R}_{\alpha;L1}(z) D(z)^{-1} \mathbf{R}_{\alpha;1L}(z)\,.
 $$ 
\end{theorem}

\noindent The obtained resolvent expressions allow us to investigate various spectral properties of the operator $H_{\alpha ,\beta }\,$ \cite{EK04}; here we concentrate only on those related to the subject of the paper, namely to perturbations of embedded eigenvalues.

\subsection{Resonance poles} \label{ss: respole}

The mechanism governing resonance and decay phenomena in this model is the tunneling between the points and the line. This interaction can be `switched off' if the line is removed, in other words, put to infinite distance from the points. Consequently, the `free' Hamiltonian $\tilde{H}_{\beta}:= H_{0,\beta}$ has the point interactions only. It has $m$ eigenvalues, $1\le m\le n$ of which we assume
 \begin{equation} \label{hypoth1}
 -\frac14 \alpha^2< \epsilon_1< \cdots< \epsilon_m <0\,,
 \end{equation}
i.e. that the discrete spectrum of $\tilde{H}_{\beta}$ is simple and contained in (the negative part of) $\sigma(\tilde{H}_{\alpha})= \sigma_{ac}(\Hab) =[-\frac14 \alpha^2,\infty )$; this can be always achieved by an appropriate choice of the configuration of the set $\Pi$ and the coupling parameters $\beta$. Let us specify the interactions sites by their Cartesian coordinates, $y^{(i)}=(c_i ,a_i )$. It is also useful to introduce the notations $a=(a_1 ,...,a_n )$ and $d_{ij}=|y^{(i)}-y^{(j)}|$ for the distances between the point interactions.

Resolvent poles will be found through zeros of the
operator-valued function (\ref{Gamhat}), more exactly, through the analytical continuation of $D(\cdot )$ to a subset $\Omega_-$ of the lower halfplane across the segment $(-\frac14 \alpha^2,0)$ of the real axis, in a similar way to what we did for Friedrichs model using formula (\ref{Fcontin}). For the sake of definiteness we employ the notation $D(\cdot)^{(l)}$, where $l=-1,0,1$ refers to the argument $z$ from $\Omega_-$, the segment $(-\frac14 \alpha^2,0)$, and the upper
halfplane, $\mathrm{Im\,}z>0$, respectively. Using the resolvent formula of the previous section we see that the first component of the operator-valued function $D(\cdot)^{(l)}$ is an $n\times n$ matrix with the elements
 $$
 \Gamma _{11;jk}(z)^{(l)}=-(1-\delta_{jk})\, \frac{1}{2\pi}
 K_{0}\big(d_{jk}\sqrt{-z}\big) + \delta_{jk}
 \big(\beta_{j}+1/2\pi (\ln \sqrt{-z}-\psi (1))\big)
 $$
for all the $l$. To find an explicit form of the second component let us introduce
 $$
 \mu_{ij}(z,t):=\frac{i\alpha }{2^{5}\pi }\frac{(\alpha
 -2i(z-t)^{1/2})\, \e^{i(z-t)^{1/2}(|a_{i}|+|a_{j}|)}}{t^{1/2}(z-t)^{1/2}}
 \,\e^{it^{1/2}(c_i -c_j )}
 $$
and $\mu_{ij}^{0}(\lambda ,t):= \lim_{\eta\to0+} \mu_{ij} (\lambda+i\eta,t)$. Using this notation we can rewrite the matrix elements of $(\Gamma _{10}\Gamma _{00}^{-1}\Gamma _{01})^{(l)}(z)$ appearing in (\ref{Gamhat}) in the following form,
 \begin{eqnarray*}
 \theta_{ij}^{(0)}(\lambda ) &\!=\!& \mathcal{P}\int_{0}^{\infty}
 \frac{\mu_{ij}^0 (\lambda,t) }{t-\lambda -\frac14 \alpha^{2}}\,\mathrm{d}t
 +g_{\alpha,ij}(\lambda )\,, \qquad\;\: \lambda \in \big(\! - \textstyle{\frac14 }\alpha^{2},0\big)
 \\
 \theta_{ij}^{(l)}(z) &\!=\!& l \int_{0}^{\infty}\frac{\mu _{ij}(z,t)}
 {t-z -\frac14 \alpha^{2}}\,\mathrm{d}t+(l-1) g_{\alpha,ij}(z )\quad
 \mathrm{for} \;\; l=1,\, -1
 \end{eqnarray*}
where $\mathcal{P}$ indicates again principal value of the integral, and
 $$
 g_{\alpha,ij}(z):= \frac{i\alpha }{(z+\alpha ^2/4)^{1/2}}\,
 \e^{-\alpha(|a_{i}|+|a_j |)/2}\, \e ^{i(z+\alpha^{2}/4)^{1/2}
 (c_i -c_j )}\,.
 $$
Using these formul{\ae} one has to find zeros of $\det D(\cdot)^{(-1)}$; we shall sketch the argument referring to \cite{EK04, EIK07} for details. We have mentioned that resonances are caused by tunneling between the parts of the interaction support, hence it is convenient to introduce the following reparametrization,
 $$
 b(a)= (b_1(a),\dots, b_n (a))\quad \text{with} \quad b_{i}(a)
 :=\e^{-|a_i|\sqrt{-\epsilon_i}}
 $$
and to put $\eta(b(a),z):=\det D(z)^{(-1)}$. Since the
absence of the line-supported interaction can be regarded as putting the line to an infinite distance from the points,
it corresponds to $b=0$ in which case we have $\eta(0,z)= \det
\Gamma_{11}(z)$ and the zeros are nothing else than
the eigenvalues of the point-interaction Hamiltonian $\tilde
H_\beta$, in other words, $\eta(0,\epsilon_i)=0\,, \; i=1,\dots,m$. Then, in analogy with Sec.~\ref{s:friedrichs}, we have to check that the hypotheses of the implicit-function theorem are satisfied which makes it possible to formulate the following conclusion.

\begin{proposition} \label{linedot-res}
The equation $\eta(b,z)=0$ has for all the $b_i$ small enough exactly $m$ zeros which admit the following weak-coupling asymptotic expansion,
 $$ 
z_i(b)=\epsilon_i +\mathcal{O}(|b|)+i\mathcal{O}(|b|)\quad
\mathrm{where}\quad  |b|:=\max_{1\leq i\leq m}b_i \,.
 $$ 
\end{proposition}

This result is not very strong, because it does provides just a bound on the asymptotic behavior and it does not guarantee that the interaction turns embedded eigenvalues of $\tilde H_\beta$ into true resonances. This can be checked in the case $n=1\:$ \cite{EK04} but it may not be true already for $n=2$. The simplest example involves a pair of point interactions with the same coupling placed in a mirror-symmetric way with respect to $\Sigma$. The Hamiltonian can be then decomposed according to parity, its part acting on functions even with respect to $\Sigma$ has a resonance, exponentially narrow in terms of the distance between the points and the line, while the odd one has a embedded eigenvalue independently of the distance. On the other hand, if the mirror symmetry is violated, be it by changing one of the point distances or one of the coupling constants, the latter turns into a resonance and one can derive a weak-perturbation expansion \cite{EK04} in a way similar to those of Sec.~\ref{ss:noeckel}.

Let us also note that the explicit form of the resolvent given in Theorem~\ref{resoth} makes it possible to find the on-shell S-matrix from energies from the interval $(-\frac14 \alpha^2,0)$, that is, for states traveling along the `wire', and to show that their poles coincide with the resolvent poles; for $m=1$ this is done in \cite{EK04}.

   \subsection{Decay of the `dot' states}

The present model gives us one more opportunity to illustrate relations between resonances and time evolution of unstable systems, this time on bound states of the quantum `dots' decaying due to tunneling between them and the `wire'. By assumption (\ref{hypoth1}) there is a nontrivial discrete spectrum of $\tilde H_\beta$ embedded in $(-\frac14 \alpha^2,0)$, and the respective eigenfunctions are
 $$ 
 \psi_j(x) =\sum_{i=1}^m d_i^{(j)} \phi_i^{(j)}(x)\,,\;j=1,\dots,m\,, \quad
 \phi_i^{(j)}(x):= \sqrt{-\frac{\epsilon_j}{\pi}}\,
 K_{0} \big(\sqrt{-\epsilon_j}|x-y^{(i)}|\big)\,,
 $$ 
where in accordance with \cite[Sec.~II.3]{AGHH} the coefficient vectors $d^{(j)}\in\C^m$ solve the equation $\Gamma_{11} (\epsilon_j)d^{(j)}=0$ and the normalization condition $\|\phi_i^{(j)}\|=1$ gives
 $$ 
 |d^{(j)}|^2 +2\mathrm{Re\,} \sum_{i=2}^m \sum_{k=1}^{i-1}\,
 \overline{d_i^{(j)}} d_k^{(j)} (\phi_i^{(j)},\phi_k^{(j)})=1\,.
 $$ 
In particular, $d^{(1)}=1$ if $n=m=1$; if $m>1$ and the distances between the points of $\Pi$ are large, the natural length scale being given by $(-\epsilon_j)^{-1/2}$, the cross terms are small and the vector lengths $|d^{(j)}|$ are close to one.

Let us now identify the unstable system Hilbert space $\HH_\mathrm{u}=E_\mathrm{u} L^2(\R^2)$ with the span of the vectors $\psi_1,\dots,\psi_m$. The decay law of the system prepared at the initial instant $t=0$ at a state $\psi\in \HH_\mathrm{u}$ is according to (\ref{decaylaw}) given by the formula
 $$ 
 P_\psi(t) = \| E_\mathrm{u}\, \e^{-iH_{\alpha, \beta}t}\psi\|^2.
 $$ 
We are particularly interested in the \emph{weak-coupling
situation} which in the present case means that the distance between $\Sigma$ and $\Pi$ is a large at the scale given by $(-\epsilon_m)^{-1/2}$. Let us denote by $E_j$ the one-dimensional projection associated with the eigenfunction $\psi_j$, the one can make the following claim \cite{EIK07}.

\begin{theorem}\label{tdecay1}
Suppose that $\Hab$ has no embedded eigenvalues. Then, with the notation introduced above, we have in the limit $|b|\to 0$, i.e. $\mathrm{dist\,}(\Sigma,\Pi)\to\infty$
 $$
 \|E_j\, \e^{-i\Hab t}\psi_j -\e^{-iz_j t}\psi_j \|\to 0\,,
 $$
pointwise in $t\in(0,\infty)$, which for $n=1$ implies $|P_{\psi_1}(t)- \e^{2\im z_1 t}|\to 0$ as $|b|\to 0$.

\end{theorem}

Let us add a couple of remarks. The result implies more generally that for large values of $\mathrm{dist\,} (\Sigma,\Pi)$ the reduced evolution can be approximated by a semigroup. On the other hand, despite the approximately exponential decay in the case $n=1$ the lifetime defined as $T_{\psi_1} = \int_0^\infty P_{\psi_1}(t)\,\D t$ diverges; the situation is similar to those mentioned is Sections~\ref{s:friedrichs} and \ref{ss: 2channel}: the operator $\Hab$ has a bound state which is not exactly orthogonal to $\psi_1$ for $b\ne 0$, cf.~\cite{EK04}, hence $\lim_{t\to\infty} P_{\psi_1}(t) \ne 0$. Furthermore, the decay of the `dot' states in this model offers a possibility to compare the `stable' dynamics, i.e. evolution of vector in $\HH_\mathrm{u}$ governed $\e^{-i\tilde H_\beta t}$, with the Zeno dynamics obtained from $\e^{-i\Hab t}$ by permanent observation. cf. \cite{EIK07} for details. Finally, let us finally mention that a related model with a singular interaction in $\R^3$ supported by a line and a circle and resonances coming from a symmetry violation has been investigated recently in \cite{Ko12}.

\section{Generalized graphs}

In the closing section we will mention another class of solvable models in which resonances can be studied, which may be regarded as another generalization of the quantum graphs discussed in Section~\ref{s:qgraphs}. What they have in common is that the configuration space consists of parts connected together through point contacts. In the present case, however, we consider parts of different dimensions; for simplicity we limit ourselves to the simplest situation when the dimensions are one and two.

   \subsection{Coupling different dimensions}

To begin with we have to explain how such a coupling can be constructed. The technique is known since \cite{ES87}, we demonstrate it on the simplest example in which a halfline lead is coupled to a plane. In this case the state Hilbert space is  $L^2({\R}^-) \oplus L^2({\R}^2)$ and the Hamiltonian acts on its elements $\psi_\mathrm{lead} \choose \psi_\mathrm{plane}\,$ (belonging locally to $W^{2,2}$) as $\,-\psi''_\mathrm{lead} \choose -\Delta\psi_\mathrm{plane}$; to make such an operator self-adjoint one has to impose suitable boundary conditions which couple the wave functions at the junction.

The boundary values to enter such boundary condition are obvious on the lead side being the columns of the values $\psi_\mathrm{lead}(0+)$ and $\psi'_\mathrm{lead} (0+)$. On the other hand, in the plane we have to use generalized ones analogous to those appearing in the first relation of (\ref{boucon}). If we restrict two-dimensional Laplacian to functions vanishing at the origin and take an adjoint to such an operator, the functions in the corresponding domain will have a logarithmic singularity at the origin and the generalized boundary values will be the coefficients in the corresponding expansion,
 $$ 
\psi_\mathrm{plane}(x) = -\frac{1}{2\pi}\,
L_0(\psi_\mathrm{plane})\,\ln |x| + L_1(\psi_\mathrm{plane}) +
o(|x|)\,;
 $$ 
using them we can write the sought coupling conditions as
 \begin{equation} \label{geng-bc}
\begin{array}{rcl}
\psi'_\mathrm{lead} (0+) &=& A\psi_\mathrm{lead} (0+) +
2\pi \bar C L_0(\psi_\mathrm{plane})\,, \\ [.3em] L_1(\psi_\mathrm{plane})
&=& C\psi_\mathrm{lead} (0+) + DL_0(\psi_\mathrm{plane})\,,
\end{array}
 \end{equation}
where $A,D\in\R$ and $C$ is a complex number, or more generally
 $$
\AAA {\psi_\mathrm{lead}(0+) \choose L_0(\psi_\mathrm{plane})} +
\BB {\psi'_\mathrm{lead}(0+) \choose L_1(\psi_\mathrm{plane})}
=0
 $$
with appropriately chosen matrices $\AAA,\BB$ in analogy with (\ref{rat-coup2}), however, for our purpose here the generic conditions (\ref{geng-bc}) are sufficient.

As in the case of quantum graphs the choice of the coupling based on the probability current conservations leaves many possibilities open and the question is which ones are physically plausible. This is in general a difficult problem. A natural strategy would be to consider leads of finite girth  coupled to a surface and the limit when the transverse size tends to zero. While for quantum graphs such limits are reasonably well understood nowadays \cite{Gr08, EP09, CET10}, for mixed dimensions the current knowledge is limited to heuristic results such as the one in \cite{ES97} which suggests that an appropriate parameter choice in (\ref{geng-bc}) might be
\begin{equation} \label{natcoupl}
A =\, {1\over 2\rho}\,,\quad\, B = \sqrt{{2\pi\over \rho}}\,, \quad\; C = {1\over\sqrt{2\pi\rho}}\,, \quad\; D = -\ln\rho\,,
\end{equation}
where $\rho$ is the contact radius. At the same time, other possibilities have been considered such as the simplest choice keeping just the coupling term, $A=D=0$, or an indirect approach based on fixing the singularity of the Hamiltonian Green's function at the junction which avoids using the coupling conditions explicitly \cite{Ki97}.

While the example concerned a particular case, the obtained coupling conditions are of a local character and can be employed whenever we couple a one-dimensional lead to a locally smooth surface. In  this way one can treat a wide class of such systems, in particular to formulate the scattering theory on configuration spaces consisting of a finite numbers of manifolds, finite and infinite edges --- one sometimes speaks about `hedgehog manifolds' --- cf.~\cite{BG03}.

Before turning to an example of resonances on such a `manifold' let us mention that while the system of a plane and a halfline lead considered above has at most two resonances coming from the coupling, one can produce an infinite series of them if the motion in the plane is under influence of a magnetic field. The same is true even if Laplacian is replaced by a more complicated Hamiltonian describing other physical effects such as spin-orbit interaction --- cf.~\cite{CE11}.

   \subsection{Transport through a geometric scatterer}

Let us look in more detail into an important example in which we have a `geometric scatterer' consisting of a compact and connected manifold $\Omega$, which may or may not have a boundary, to which two semi-infinite leads are attached at two different points $x_1,x_2$ from the interior of $\Omega$. One may regard such a system as a motion on the line which is cut and the loose ends are attached to a black-box object which can be characterized by the appropriate transfer matrix, ${u(0+) \choose u'(0+)} = L{u(0-) \choose u'(0-)}$. To find the latter one has to fix the dynamics: we suppose that the motion on the line is free being described by the negative Laplacian, while the manifold part of the Hamiltonian is Laplace-Beltrami operator on the state Hilbert space $L^2(\Omega)$ of the scatterer; they are coupled by conditions \eqref{geng-bc} with the coefficients indexed by $j=1,2$ referring to the `left' and `right' lead, respectively.

We need the Green function $G(.,.;k)$ of the Laplace-Beltrami operator which exists whenever the $k^2$ does not belong to the spectrum. Its actual form depends on the geometry of $\Omega$ but the diagonal singularity does not: the manifold $\Omega$ admits in the vicinity of any point a local Cartesian chart and the Green's function behaves with respect to those variables as that of Laplacian in the plane,
 $$ 
G(x,y;k)= -\,\frac{1}{2\pi}\,\ln |x\!-\!y| +\OO(1)\,,
\qquad |x\!-\!y|\to 0\,.
 $$ 
Looking for transient solutions to the Schr\"odinger equation at energy $k^2$, we note that its manifold part can be written as $u(x) = a_1G(x,x_1;k) +a_2G(x,x_2;k)$, cf.~\cite{Ki97}, which allows us to find the generalized boundary values
 $$ 
L_0(x_j) = -\frac{a_j}{2\pi}\,,\qquad L_1(x_j) =
a_j\xi(x_j,k)+a_{3-j}G(x_1,x_2;k)
 $$ 
for $j=1,2$, where we have employed the regularized Green's function at $x_j$,
\begin{equation} \label{xi1}
\xi(x_j;k) := \lim_{x\to x_j} \left\lbrack G(x,x_j;k)+
\frac{\ln|x\!-\!x_j|}{2\pi} \right\rbrack\,.
\end{equation}
Let $u_j$ be the wave function on the $\,j$-th lead; using the abbreviations $u_j, \,u'_j$ for its boundary values we get from the conditions (\ref{geng-bc}) a linear system which can be easily solved \cite{ETV01}; it yields the transfer matrix in terms of the quantities $\,Z_j:= {d_j\over 2\pi} +\xi_j\,$ and $\,\Delta := g^2\!-Z_1Z_2\,$, where $\xi_j:= \xi(x_j;k)$ and $g:=G(x_1,x_2;k)$. The expression simplifies if the couplings are the same at the two junctions; then $\det L=1$ and the transfer matrix is given by
 $$ 
L = \frac{1}{g}\left(\begin{array}{cc}
Z_2+\frac{A}{C^2}\Delta & -2\frac{\Delta}{C^2} \\[.5em]
C^2-A(Z_1\!+\!Z_2)-\frac{A^2}{C^2}\Delta &
\frac{A}{C^2}\Delta +Z_1
\end{array}\right)\,.
 $$ 
From here one can further derive the on-shell scattering matrix \cite{ETV01}, in particular, the reflection and transmission amplitudes are
 $$
r = -\,\frac{L_{21} + ik(L_{22} \!-\! L_{11}) + k^2L_{12}}
{L_{21} - ik(L_{22} \!+\! L_{11}) - k^2L_{12}}\,, \quad
t = -\,\frac{2ik}{L_{21} - ik(L_{22} \!+\! L_{11}) -
k^2L_{12}}\,;
 $$
they naturally depend on $k$ through $\xi$ and $g$, and satisfy $|r|^2+|t|^2=1$. To find these quantities for a particular $\Omega$ we may use the fact that it is compact by assumption, hence the Laplace-Beltrami operator on it has a purely discrete spectrum. We employ the eigenvalues, $\{\lambda_n \}_{n=1}^{\infty}\,$, numbered in ascending order and with the multiplicity taken into account, corresponding to eigenfunctions $\{\phi_n\}_{n=1}^{\infty}\,$ which form an orthonormal basis in $L^2(\Omega)$. The common Green's function expression then gives
 $$
g(k)=\sum_{n=1}^{\infty}\, \frac{\phi_n(x_1)
\overline{\phi_n(x_2)}}{\lambda_n\!-k^2}\,,
 $$
while the regularized value (\ref{xi1}) can be expressed \cite{ES97} as
 $$ 
\xi (x_j,k) \,=\, \sum_{n=1}^{\infty}\, \left(\frac{|\phi_n(x_j)|^2}
{\lambda_n\!-k^2}\,-\,\frac{1}{4\pi n}\right) \,+ c(\Omega)\,,
 $$ 
where the series is absolutely convergent and the constant $c(\Omega)$ depends on the manifold $G$. Note that a nonzero value of $c(\Omega)$ amounts in fact just to a coupling parameter renormalization: $\,D_j$ has to be changed to $\,D_j\!+2\pi c(\Omega)\,$.

Several examples of such a scattering has been worked out in the literature, mostly for the case when $\Omega$ is a sphere. If the coupling is chosen according to (\ref{natcoupl}) and the leads are attached at opposite poles, the transmission probability has resonance peaks around the values $\lambda_n$ where the transmission probability is close to one, and a background, dominating at high energies, which behaves as $\OO(k^{-2} (\ln k)^{-1})$, cf.~\cite{ETV01}. Similar behavior can be demonstrated for other couplings at the junctions \cite{Ki97}; the background suppression is faster if the junctions are not antipolar \cite{BGMP02}. Recall also that this resonance behavior is manifested in conductance properties of such systems as a function of the electrochemical potential given by the Landauer-B\"uttiker formula, see e.g. \cite{BGMP02}.

  \subsection{Equivalence of the resonance notions}

Let us return finally to a more general situation\footnote{Considerations of this section follow the paper \cite{EL12}. Similarly one can treat `hedgehogs' with three-dimensional manifolds, just replacing logarithmic singularities by polar ones.} and consider a `hedgehog' consisting of a two-dimensional Riemannian manifold $\Omega$, compact, connected, and for simplicity supposed to be embedded into $\R^3$, endowed with a metric $g_{rs}$, to which a finite number $n_j$ of halfline leads is attached at points $x_j,\: j=1, \dots, n$ belonging to a finite subset $\{x_j\}$ of the interior of $\Omega$; by $M = \sum_j n_j$ we denote the total number of the leads. The Hilbert space will be correspondingly $\mathcal{H} = L^2 (\Omega, \sqrt{|g|} \,\mathrm{d}x)\oplus \bigoplus_{i=1}^M L^2 (\mathbb{R}_+^{(i)})$.

Let $H_0$ be the closure of the Laplace-Beltrami operator $- g^{-1/2} \partial_r (g^{1/2} g^{rs} \partial_s)$ defined on functions from $C_0^\infty(\Omega)$; if $\partial\Omega \ne \emptyset$ we require that they satisfy there appropriate boundary conditions, either Neumann/Robin, $(\partial_n +\gamma) f|_{\partial \Omega} = 0$, or Dirichlet, $f|_{\partial \Omega} = 0$. The restriction $H_0'$ of $H_0$ to the domain $\{f\in D(H_0):\: f(x_j) = 0,\; j=1,\dots,n\}$ is a symmetric operator with deficiency indices $(n,n)$. Furthermore, we denote by $H_i$ the negative Laplacian on $L^2(\mathbb{R}_+ ^{(i)})$ referring to the $i$-th lead and by $H_i'$ its restriction to functions which vanish together with their first derivative at the halfline endpoint. The direct sum $H' = H_0' \oplus H_1' \oplus \cdots \oplus H_M'$ is obviously a symmetric operator with deficiency indices $(n+M,n+M)$.

As before admissible Hamiltonians are identified with self-adjoint extensions of the operator $H'$ being described by the conditions (\ref{rat-coup2}) where $U$ is now an $(n+M)\times (n+M)$ unitary matrix, $I$ the corresponding unit matrix, and furthermore, $\Psi = (L_{1,1}(f), \dots , L_{1,n}(f), f_1(0),\dots , f_n(0))^\mathrm{T}$ and is $\Psi'$ the analogous column of (generalized) boundary values with $L_{1,j}(f)$ replaced by $L_{0,j}(f)$ and $f_1(0)$ by $f'_j(0)$, respectively. The first $n$ entries correspond to the manifold part being equal to the appropriate coefficients in the expansion of functions $f \in D(H_0^*)$ the asymptotic expansion near $x_j$, namely $f(x) = L_{0,j}(f) F_0 (x, x_j) + L_{1,j}(f) + \mathcal{O}(r(x,x_j))$, where
  $$ 
   F_0 (x, x_j) =  -\frac{q_2(x,x_j)}{2 \pi}\, \ln{r(x,x_j)}
 $$ 
with $r(x,x_j)$ being the geodetic distance on $\Omega$; according to Lemma~4 in \cite{BG03} $\,q_2$ is a continuous functions of $x$ with $q_i(x_j, x_j) = 1$. The extension described by such conditions will be denoted as $H_U$. We are naturally interested in \emph{local} couplings; in analogy with considerations of Section~\ref{ss:graphs} we can work with one `large' matrix $U$ and encode the junction geometry in its block structure.

Another useful thing we can adopt from the previous discussion is the possibility to employ conditions $(\tilde U_j(k) - I) d_j(f) + i (\tilde U_j(k) + I) c_j(f) = 0$ on the manifold $\Omega$ itself with the effective, energy-dependent coupling described by the matrix
  $$
   U_j(k) = U_{1j} -(1-k) U_{2j} [(1-k) U_{4j} - (k+1) I]^{-1} U_{3j}
 $$
at the $j$-th lead endpoint, where $U_{1j}$ denotes top-left entry of $U_j$, $U_{2j}$ the rest of the first row, $U_{3j}$ the rest of the first column and $U_{4j}$ is $n_j\times n_j$ part corresponding to the coupling between the leads attached to the manifold at the same point.

To find the on-shell scattering matrix at energy $k^2$ one has to couple solutions $a_j(k) \mathrm{e}^{-ikx}+b_j(k) \mathrm{e}^{ikx}$ on the leads to solution on the manifold and to look at the continuation of the result to the complex plane. On the other hand to find the resolvent singularities, we can again employ the complex scaling and to find complex eigenvalues of the resulting non-selfadjoint operator. In both cases we need the solution on the manifold; modifying the conclusions of \cite{Ki97} mentioned above we can infer that it has to be of the form $f(x,k) =  \sum_{j=1}^{n} c_j G(x,x_j;k)$,

Consider first the scattering resonances. Denoting the coefficient vector of $f(x,k)$ as $\mathbf{c}$ and using similar abbreviations $\mathbf{a}$ for the vector of the amplitudes of the incoming waves, $(a_1(k), \dots, a_M(k))^\mathrm{T}$, and $\mathbf{b}$ for the vector of the amplitudes of the outgoing waves, one obtains in general a system of equations,
 $$ 
  A(k)  \mathbf{a} + B(k)  \mathbf{b} + C(k)\mathbf{c}  = 0\,,
 $$ 
in which $A$ and $B$ are $(n+M) \times M$ matrices and $C$ is $(n+M) \times n $ matrix the elements of which are exponentials and Green's functions, regularized if necessary; what is important that all the entries of the mentioned matrices allow for an analytical continuation which makes it possible to ask for complex $k$ for which the above system is solvable. For $k_0^2 \not \in \mathbb{R}$ the columns of $C(k_0)$ are linearly independent and one can eliminate $\mathbf{c}$ and rewrite the above system as
 $$ 
   \tilde A(k_0)  \mathbf{a} + \tilde B(k_0)  \mathbf{b} = 0\,,
 $$ 
where $\tilde A(k_0)$ and $\tilde B(k_0)$ are $M\times M$ matrices the entries of which are rational functions of the entries of the previous ones. If $\det\tilde A(k_0) = 0$ there is a solution with $\mathbf{b} = 0$, and consequently, $k_0$ should be an eigenvalue of $H$ since $\mathrm{Im\,}k_0<0$ and the corresponding eigenfunction belongs to $L^2$, however, this contradicts to the self-adjointness of $H$. Next we notice that the S-matrix analytically continued to the point $k_0$ equals $-\tilde B(k_0)^{-1} \tilde A(k_0)$ hence its singularities must solve $\det \tilde B(k) = 0$.

On the other hand, for resolvent resonances we use exterior complex scaling with $\mathrm{arg\,}\theta > \mathrm{arg\,} k_0$, then the solution $a_j(k) \mathrm{e}^{-ikx}$ on the $j$-th lead, analytically continued to the point $k=k_0$, is after the transformation by $U_\theta$ exponentially increasing, while $b_j(k) \mathrm{e}^{ikx}$ becomes square integrable. This means that solving in $L^2$  the eigenvalue problem for the complex-scaled operator one has to find solutions of the above system with $\mathbf{a} = 0$ which leads again to the condition  $\det\tilde B(k) = 0$. This allows us to make the following conclusion.

 \begin{theorem}
In the described setting, the hedgehog system has a scattering resonance at $k_0$ with $\mathrm{Im\,}k_0<0$ and $k_0^2 \not \in \mathbb{R}$ \emph{iff} there is a resolvent resonance at $k_0$. Algebraic multiplicities of the resonances defined in both ways coincide.
 \end{theorem}

\bibliographystyle{amsalpha}

\end{document}